\documentclass[twocolumn,times,tighten,twocolappendix]{aastex631}

\usepackage{float}
\usepackage{subcaption}
\usepackage{threeparttable}
\usepackage{colortbl}
\usepackage{graphicx}
\usepackage{tikz}
\usetikzlibrary{shapes,arrows,positioning}

\usepackage{soul}
\newcommand{\add}[1]{\textbf{\textcolor{blue}{#1}}}

\shorttitle{Spike subpulses and quasi-periodic subpulses of 25 pulsars by FAST}
\shortauthors{Wang et al.}


\begin{document}

\title{FAST Pulsar Database IV. Spike subpulses and quasi-periodic subpulses of 25 pulsars observed by FAST}

\author[0000-0002-0786-7307]{Tao Wang}
\author{N. N. Cai}
\affiliation{National Astronomical Observatories, Chinese Academy of Sciences, Jia-20 Datun Road, ChaoYang District, Beijing 100012, China}
\affiliation{State Key Laboratory of Radio Astronomy and Technology, Beijing 100101, China }

\author[0000-0002-9274-3092]{J.~L. Han}\thanks{E-mail: hjl@nao.cas.cn}
\affiliation{National Astronomical Observatories, Chinese Academy of Sciences, Jia-20 Datun Road, ChaoYang District, Beijing 100012, China}
\affiliation{School of Astronomy and Space Science, University of Chinese Academy of Sciences, Beijing 100049, China}
\affiliation{State Key Laboratory of Radio Astronomy and Technology, Beijing 100101, China }

\author[0000-0002-1056-5895]{W.~C. Jing} 
\author{Z.~L. Yang}
\author{Yi Yan}
\author{D.~J. Zhou} 
\affiliation{National Astronomical Observatories, Chinese Academy of Sciences, Jia-20 Datun Road, ChaoYang District, Beijing 100012, China}
\affiliation{State Key Laboratory of Radio Astronomy and Technology, Beijing 100101, China }

\author{P.~F. Wang}
\author{C. Wang}
\affiliation{National Astronomical Observatories, Chinese Academy of Sciences, Jia-20 Datun Road, ChaoYang District, Beijing 100012, China}
\affiliation{State Key Laboratory of Radio Astronomy and Technology, Beijing 100101, China }
\affiliation{School of Astronomy and Space Science, University of Chinese Academy of Sciences, Beijing 100049, China}


\begin{abstract}
Fine structures of individual pulses can be detected when observations are conducted with a high time resolution and a great sensitivity. We examined pulsar data observed by the Five-hundred-metre Aperture Spherical radio Telescope (FAST) with a time resolution of 49~{\textmu}s, and detected a large number of spike subpulses of 21 pulsars and quasi-periodic subpulses from 13 pulsars. These spike subpulses cannot be or are marginally resolved by the FAST observation time resolution, and are generally strongly linearly polarized, which {may be} primary emission elements of subpulses. 
For the quasi-periodic subpulses from 13 pulsars, we measured their characteristic periods, 
generally a few tenths of a millisecond, and examined their possible correlation with pulsar rotation period. 
\end{abstract}


\section{Introduction} \label{sec:intro}

Pulsars are rotating neutron stars that emit periodic radio pulses. The mean profiles of these pulses stabilize after averaging over many periods. The mean polarization profiles of a pulsar are closely linked to the emission geometry in the pulsar magnetosphere, and polarization angles, in particular, are related to the sweeping of the curved magnetic field lines across the line of sight, which can be described by the Rotation Vector Model \citep[RVM,][]{rc+1969}.

High-sensitivity observations enable the detection of detailed structures of individual pulses with excellent signal-to-noise ratio (S/N). Generally, individual pulses typically consist of multiple sub-pulses. For instance, \citet{wwh+2024} reported numerous highly polarized isolated sub-pulses in PSR B1916+14, inferring that the emission region in the pulsar magnetosphere has a scale length of about 120~m. With improved time resolution, structures narrower than sub-pulses, such as microstructures, were first identified in PSR B0950+08 at 430 MHz \citep{ccd+1968}. Microstructures have durations as short as a few microseconds \citep{Bartel+1978} and even nanoseconds \citep{hb+1978}, and are considered elemental emission features often characterized by high linear polarization \citep{mar+2015}. In some pulsars, extremely bright giant pulses have been observed \citep{hb+1978, Hankins+1996, hkw+2003, spb+2004, jpk+2010, kp+2018}, and in contrast, narrow dwarf pulses have recently also been detected from a dozen pulsars \citep{cyh23, yhz24}.

Consecutive subpulse sequences within some individual pulsar pulses exhibit quasi-periodic behavior, a phenomenon first noted in PSR B0950+08 \citep{Hankins+1971, Hankins+1972}. Quasi-periodic subpulses (hereafter, QPS) manifest as broadband emission \citep{Boriakoff+1983, bsf+1981, kjs+2002, dgs+2016, lab+2022}, with a characteristic quasi-period of $P_\mu$ typically 0.3 ms \citep{rhc+1975}. To date, quasi-periodic subpulses have been detected from 38 pulsars, including very long-period pulsars \citep[PSR J1901$-$4046, $P \simeq 76$ s;][]{chr+2022}, magnetars \citep{kld+2024}, interpulse pulsars \citep{LDW+2025}, and millisecond pulsars \citep{dgs+2016, lab+2022}. 
A correlation between $P_\mu$ and pulsar spin period $P$ has been found from observational data \citep{kjs+2002, mar+2015, dgs+2016, lab+2022, kld+2024}.
%

\begin{figure}[t]
    \centering
    \includegraphics[width=0.36\textwidth]{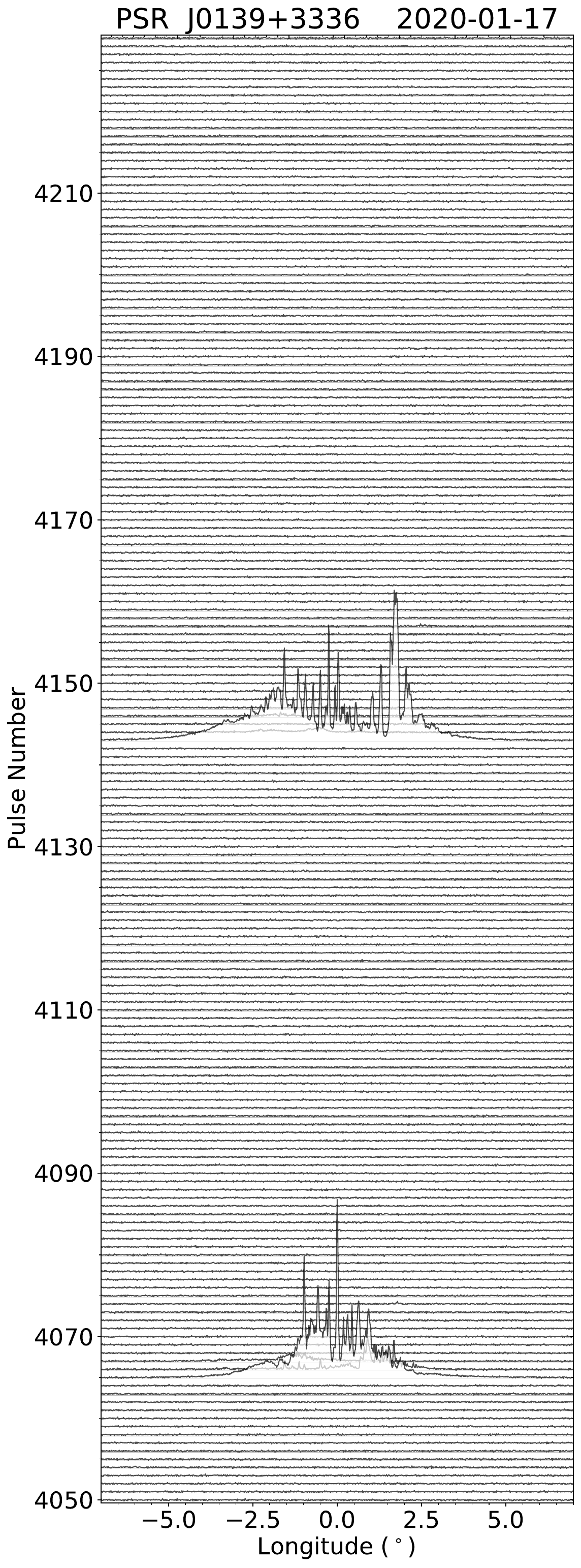}
    \caption{An example for fine subpulse structures of two individual pulses among many nulling periods of PSR J0139+3336 detected by FAST on January 17th, 2020 (hereafter in the form of 2020-01-17).
    }
    \label{fig:J0139exam}
\end{figure}

The proposed physical origin mechanisms for quasi-periodic subpulses are broadly categorized into geometric and temporal (intrinsic) classes. Geometric mechanisms rely on longitudinally aligned flux tubes that channel particle streams, thereby imprinting periodic structures on the emitted radiation \citep{Benford+1977}.
Temporal models can be classified into three main scenarios: (1) neutron star vibrations that modulate particle acceleration \citep{van+1980}; (2) magnetospheric heating–cooling cycles that generate quasi-periodic emission intervals \citep{Benford+1977}; and (3) radiation transfer near the light cylinder, where shearing plasmas induce multiple reflections to produce periodic signals \citep{ht+1981, cjd+2004}.
Recent studies have proposed additional intrinsic mechanisms, including Cerenkov-like instabilities of current-carrying Alfvén waves in relativistic plasma \citep{t+2022b, kld+2024}, and two-beam instabilities. The latter satisfies the emission coherence condition and can form the periodically bunched plasma along magnetic field lines, which may be responsible for quasi-periodic subpulses generation \citep{pts+2020, bbl+2023, Jones+2024}.

\begin{table*}[tbh!]
  \caption{Basic pulsar parameters and observational parameters for 25 pulsars observed by FAST. Columns (1-4) list the sequence number (No.), pulsar name (Jname), pulsar period (${\rm P_0}$) and dispersion measurement (DM) with the uncertainty; Columns (5-7) list FAST observation date (Date), observation duration (${\rm T_{Obs}}$, in minutes) and number of observed periods (${\rm N_{P}}$); Columns (8-9) list the bin number per period (N$_{\rm BIN}$), the 10\% pulse width of the mean pulse ${\rm W_{10}}$; Columns (10-11) list the number of spike subpulses detected (${\rm N_{sp}}$) and the number of sets of quasi-periodic subpulses (${\rm N_{QPS}}$). The last column gives the references for previous discussions of fine pulse structures. 
  } 
\label{tab:PSRlist}
  \footnotesize
  \renewcommand{\arraystretch}{0.82} 
  \begin{tabular}{clrrcrrrrrrc}
    \hline
    No. & PSR name & ${\rm Period}$  & DM & Obs. Date  & ${\rm T_{obs}}$ & ${\rm N_{\rm P}}$ & N$_{\rm bin}$ & ${\rm W_{10}}$ & ${\rm N_{sp}}$ & ${\rm N_{QPS}}$ & Ref. \\
    & & (ms)   & (pc cm$^{-3}$) & (yyyy-mm-dd) & (Min.)  &  &  & (${^\circ}$)  &  &  &  \\ 
    (1)  &  (2) & (3) & (4) & (5) & (6) & (7) & (8) & (9) & (10) & (11) & (12)\\ 
    \hline
 1 &J0139$+$3336  &1248.0  &21.203$\pm$0.026  &2020-01-17  &200.0  &9615  &25344  &6.8  &1  &1  & 1\\
 2 &J0304$+$1932  &1387.6  &15.679$\pm$0.031  &2021-09-13  &59.2  &2559  &28224  &15.71  &-  &1  & 2\\
 3 &J0426$+$4933  &922.5   &84.325$\pm$0.012  &2021-11-17  &19.3  &1258  &18760  &9.0  &17  &- & 0\\
 4 &J0454$+$5543  &340.7   &14.596$\pm$0.025  &2021-05-02  &26.0  &4581  &6928  &30.2  &12  &-  & 0\\
 5 &J0612$+$3721  &298.0  &27.144$\pm$0.022  &2021-01-09  &9.9  &1987  &6062  &19.2   &2  &-  & 0\\
 6 &J0614$+$2229  &335.0  &96.935$\pm$0.026  &2020-08-25  &88.7  &15892  &6800  &14.8  &52  &-  & 0 \\
 7 &J0627$+$0706  &475.9  &138.193$\pm$0.024  &2019-11-26  &60.0  &7567  &9664  &6.7  &9  &-  & 3\\
 8 &J0631$+$1036  &287.8  &125.253$\pm$0.024  &2021-05-01  &30.0  &6256  &5840   &24.0  &23  &- & 0\\
 9 &J0659$+$1414  &384.9  &14.064$\pm$0.020  &2023-04-30  &80.0  &12471 &7824  &31.0 & 1947 & - & 2\\
10 &J0826$+$2637  &530.7  &19.471$\pm$0.022  &2019-11-26  &60.0  &6787  &10784  &6.6  &53  &5  & 2,3,4\\
11 &J1136$+$1551  &1187.9  &4.843$\pm$0.022  &2019-11-21  &60.0  &3032  &24128  &10.0 &78  &24  & 2,5,6,7\\
12 &J1239$+$2453  &1382.4  &9.258$\pm$0.022  &2020-01-18  &120.0  &5210  &28096  &13.2  &11  &3  & 2,8\\
13 &J1607$-$0032  &421.8  &10.677$\pm$0.026  &2021-10-19  &58.0  &8246  &8576  &14.1  &29  &-  & 0 \\
14 &J1645$-$0317  &387.7  &35.745$\pm$0.030  &2022-09-07  &10.2  &1577  &7872  &7.3  &1  &3 & 9 \\
15 &J1844$+$1454  &375.5  &41.487$\pm$0.032  &2021-05-01  &30.0  &4797  &7632  &14.9   &2  &-  & 0\\
16 &J1913$-$0440  &825.9  &89.323$\pm$0.026  &2022-10-22  &23.8  &1729  &16800  &7.5   &-  &1  & 0 \\
17 &J1917$+$1353  &194.6  &94.611$\pm$0.027  &2020-11-22  &10.0  &3075  &3952  &14.9   &1  &-  &  0\\
18 &J1921$+$2153  &1337.3  &12.427$\pm$0.029  &2023-09-30  &48.9  &2195  &27200  &11.4  &-  &8  & 2 \\
19 &J1932$+$1059  &226.5  &3.187$\pm$0.020  &2019-11-22  &60.0  &15896 &4608  &17.8 &70  &35 & 2,10,11 \\
20 &J1935$+$1616  &358.7  &158.521$\pm$0.023  &2019-09-19  &52.8  &8829  &7296  &9.1  &-  &151  & 0\\
21 &J1946$+$1805  &440.6   &16.152$\pm$0.029  &2019-12-04  &60.0  &8172  &8960  &41.6  &2  &22  & 3,12\\
22 &J1955$+$5059  &518.9  &31.983$\pm$0.020  &2020-12-20  &58.0  &6703  &10544  &7.5  &67  &2 & 0 \\
23 &J2022$+$2854  &343.4  &24.630$\pm$0.022  &2021-07-11  &30.0  &5238  &6976  &15.5   &7  &-  & 2,13\\
24 &J2043$+$2740  &96.1  &21.021$\pm$0.022  &2021-10-10  &38.0  &23702  &1952  &18.4   &10  &-  & 0\\
25 &J2219$+$4754  &538.5  &43.478$\pm$0.026 &2023-08-15  &49.0  &5465  &10944  &10.5   &49  &1  & 0 \\
    \hline
  \end{tabular}\\
  \textbf{References}: (0): only this work; (1): \cite{dys+2024}; (2): \cite{mar+2015}; (3): \cite{LDW+2025}; (4): \cite{lkw+1998}; (5): \cite{Hankins+1972}; (6): \cite{fs+1978}; (7): \cite{Popov2024}; (8): \cite{yhz24}; (9): \cite{sgd+2024}; (10): \cite{lkw+1998}; (11): \cite{pbc+2002}; (12): \cite{cwh+1990}; (13): \cite{cordes+1979}.
\end{table*}

The Five-hundred-metre Aperture Spherical Radio Telescope \citep[FAST,][]{NRD+2006, jth+2020} enables highly sensitive pulsar observations using a modern digital backend with precise polarization calibration. Data {from} the FAST Galactic Plane Pulsar Snapshot (GPPS) survey \citep{HWW+2021}\footnote{http://zmtt.bao.ac.cn/GPPS/} and targeted observations of known pulsars have been recorded in the pulsar searching format. A large number of known pulsars were serendipitously detected during the survey observations. 
In this work, we analyze data from the FAST GPPS survey and other publicly released FAST projects to investigate the emission of spikes and quasi-periodic subpulses in individual pulsar pulses. Section~\ref{data} briefly describes the FAST observations. Section~\ref{result} presents the results of identifying fine subpulse structures for 25 pulsars, with detailed analysis of spike subpulses and quasi-periodic subpulses. 
We do statistical analyses of fine pulse widths and characteristic quasi-periods, and examine their correlations with pulsar rotation periods. Section~\ref{discussion} summarizes our key results and discusses their implications for coherent emission processes in the pulsar magnetosphere.

\section{FAST observations}
\label{data}

The data used for this study were mainly extracted from the publicly released archive FAST data, supplemented by some FAST GPPS survey observations \citep{HWW+2021}. All observations were carried out by using the 19-beam receiver, which is mounted to FAST at the main focus \citep{jth+2020} and works in a radio band at a central frequency of 1.25~GHz with a bandwidth of 500~MHz. In general, the GPPS survey recorded data for 2048 frequency channels covering the band, with a sampling time of $49.152~\mu$s. The follow-up pulsar verification observations of the GPPS survey generally last for 15 minutes, with the four polarization products ($XX$, $YY$, Re[$X^{*}Y$], and Im[$X^{*}Y$]) recorded for each frequency channel. The 1-K on–off noise signals are injected at the beginning or the end of observation sessions for calibration. 

Offline data processing has several steps. First, we dedispersed the data with the pulsar
ephemeris initially obtained from the ATNF pulsar catalog \citep{mht+2005} and formed single-pulse sequences using DSPSR \citep{dspsr}. A better DM value was obtained for each pulsar (see Table~\ref{tab:PSRlist}) by using our high time resolution data for the spike subpulses (see below) {through maximizing their sharpness \citep{hss+19}}. Frequency channels severely affected by instrumental performance or radio frequency interference (RFI) were excised by using the package PSRCHIVE \citep{hvm04}. Polarization calibration, gain calibration \citep[$G=16$ K/Jy,][]{jth+2020}, and the bandpass corrections were then made with the injected calibration signals. 
To get the mean polarization profiles of pulsars, we folded the FAST observation data with 1024 phase bins per rotation period, following the procedures given in \citet{whx+2023}. To satisfy the requested high time resolution for fine subpulse structures and quasi-periodic subpulses, only bright pulsars with a period of $P_0 \gtrsim 50$~ms are included in this work.  Table~\ref{tab:PSRlist} lists basic pulsar parameters and observational parameters for 25 pulsars.

\begin{figure*}[t]
\centering
\includegraphics[width=0.600\textwidth]{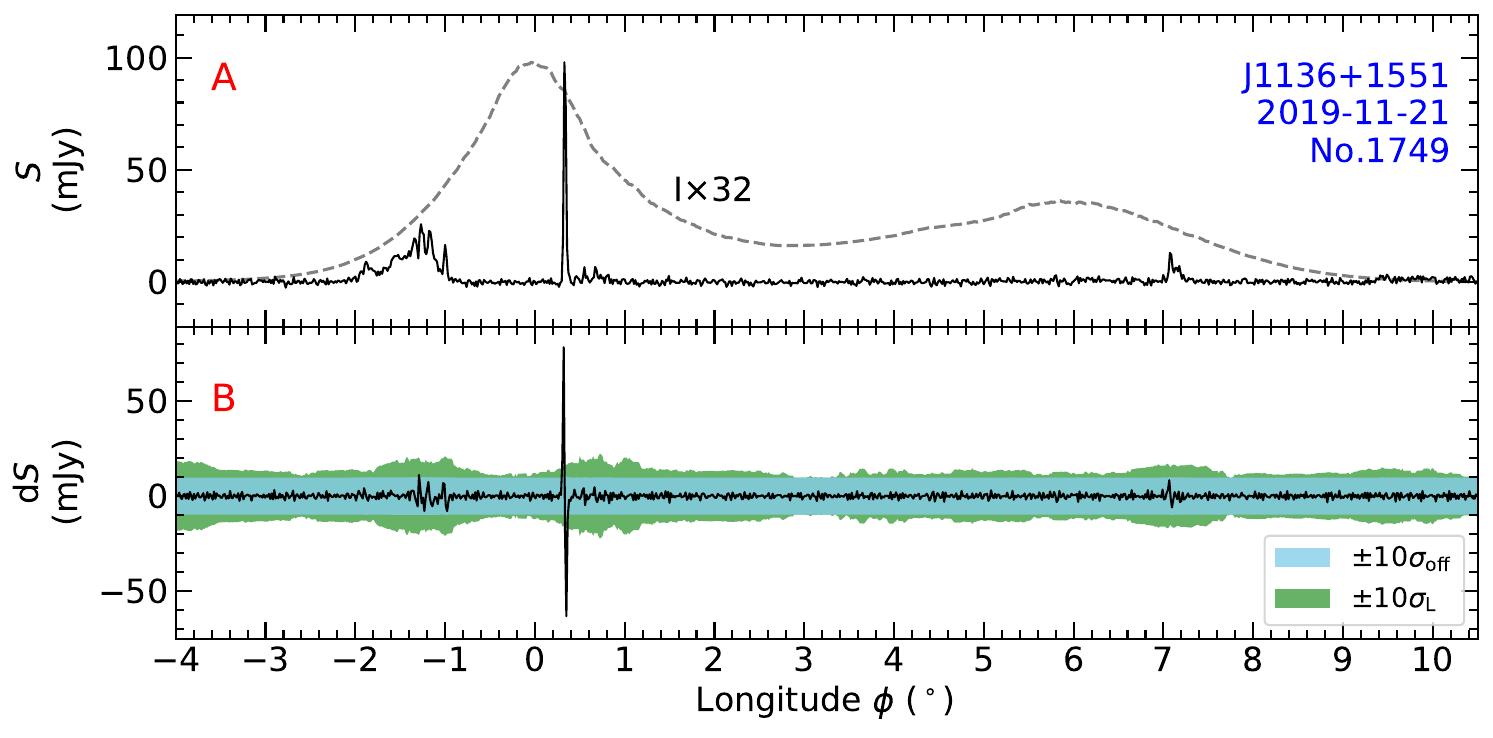}
\includegraphics[width=0.303\textwidth]{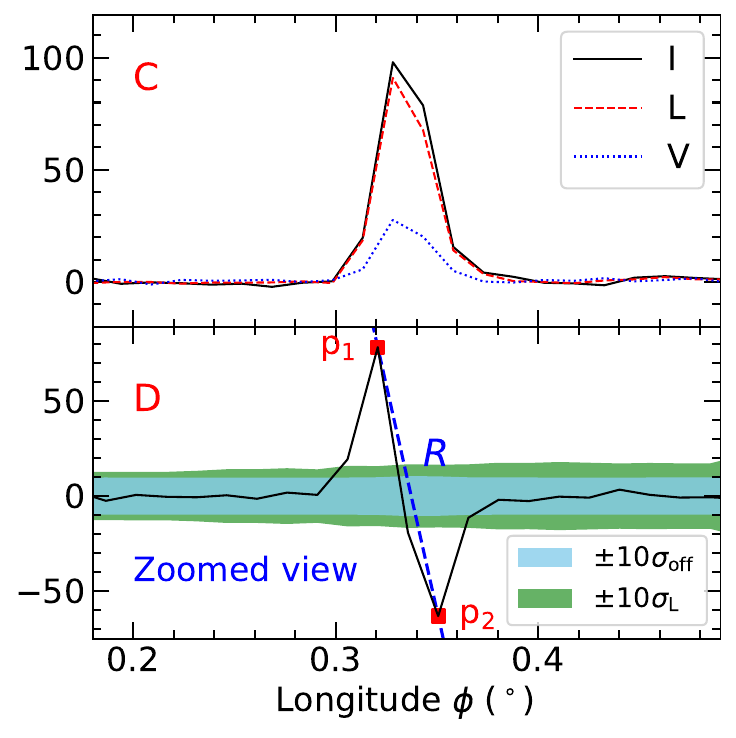}
%
%
%
\caption{A spike subpulse appears in the individual pulse of the period number 1749 of PSR J1136+1551 observed on 2019-11-21. %
%
Subpanel (A) shows the total intensity profile of the individual pulse of this period (solid line), compared with the mean pulse profile (light dashed line for $I$, amplified by a factor of 32); Subpanel (B) shows the differential profile (d$S = S_{i+1}-S_i$; here $i$ is the bin number along the pulse longitude $\phi$), together with marked $\pm10\sigma_{\rm Loc}$ and $\pm10\sigma_{\rm off}$ regions. Here, $\sigma_{\rm off}$ is the standard deviation of the off-pulse bins, and $\sigma_{\rm Loc}$ is the standard deviation estimated by 1.4826 times the median absolute deviation \citep{llk+2013} of the $\pm$30 phase bins for a given phase. The spike subpulse is identified at the longitude phase of $\phi = 0.33^\circ$ in Subpanel (A), and it causes the positive and negative slopes in Subpanel (B). The polarization profiles of this sharp subpulse are shown in a zoomed view in Subpanel (C), and the zoomed view of the differential profile is shown in Subpanel (D). We define the sharpness $R$ of this spike subpulse as being the slope between the positive and the negative peaks $p_1$ and $p_2$.
}
\label{fig:J1136N1749}
\end{figure*}

\begin{table*}[t]
  \caption{A full list of detected spike subpulses of 21 pulsars. Columns list the period number, longitude phase $\phi$, peak flux $S_{\rm peak}$, local flux fluctuation $\sigma_{\rm Loc}$, pulse width $\Delta\phi$ (in bins), sharpness $R$,  linear polarization percentage ($L/I$), circular polarization percentage ($V/I$), and polarization position angle (PA). The digits in brackets are the uncertainties. Here are the parameters of 5 example spike subpulses detected from individual pulses of PSR J1136+1551 observed by FAST on 2019-11-21. {\it A full list is presented in electronic version {\footnotesize [temporarily a long table behind the text of this paper for referee process.]}}. 
  }
  \label{table:78spikeJ1136}
    \centering
    \renewcommand{\arraystretch}{0.85}
    \begin{tabular}{crrcccrrr}
    \hline
No. of &  Longitude $\phi$ &  $S_{\rm peak}$ &  $\sigma_{\rm Loc}$  & $\Delta \phi$&   R          &   L/I&   V/I  & PA\\ 
Period & ($^\circ$)        &  (mJy)          &  (mJy)    & (bin)        &              &    (\%)&   (\%) & ($^\circ$)\\
    \hline

... & ... & ... & ... & ... & ... &  ... & ... & ... \\
1687      &      0.00&           20.6&      1.30&         1&      32.3&      95.3(5.0)&       5.3(4.4)&      -5.8(0.9)\\
1729      &      5.23&           55.8&      2.50&         2&      20.3&      17.6(0.7)&      -5.2(0.9)&     -42.3(0.4)\\
1744      &      5.63&           64.1&      2.10&         1&      60.9&      18.1(0.6)&      -6.8(0.9)&     -48.3(1.0)\\
1749      &      0.33&           90.0&      1.60&         2&      45.6&      89.8(0.8)&      16.6(0.7)&      -9.5(0.2)\\
1754      &      5.35&          397.6&      6.00&         2&      58.1&      48.0(0.1)&     -36.5(0.2)&     -76.9(0.1)\\
... & ... & ... & ... & ... & ... &  ... & ... & ... \\
    \hline
    \end{tabular}
\end{table*}

\section{Results}
\label{result}

As shown in Fig.~\ref{fig:J0139exam}, we see unprecedented details of individual pulses.  Many spike subpulses or quasi-periodic structures are revealed by 
FAST observations, even though some of them are superimposed on gradually varying wide pulses. Given that the time resolution of FAST data is $49.152~\mu$s, we fold the FAST data to keep the highest time resolution after considering the period of a given pulsar. For example, PSR J1136+1551 has a spin period of 1.188~s, and we fold the data with 24128~bins. PSR J2219+4754 has a period of 0.538~s; we fold the data with 10944 bins. The bin numbers are set according to the pulsar period, and listed in Column (8) in Table~\ref{tab:PSRlist}. From the FAST data of 25 pulsars, we identify spike subpulses, find the quasi-periodic subpulses, and then perform statistical studies.

\subsection{Identifying spike subpulses}
\label{method}

It is impossible to manually inspect all individual pulses to identify these spike subpulses. Therefore, we have to develop an automatic method to detect such spikes from the dedispersed FAST data. 
As illustrated in Figure~\ref{fig:J1136N1749}, spike subpulses are characterized by a steep rise and rapid decay. We obtained the differential profile (d$S$) by subtracting the amplitude of the preceding bin from that of each given bin. The spike subpulses can be clearly identified by a tilted `Z' from such a differential pulse profile, see the zoomed panels (C and D) in Figure~\ref{fig:J1136N1749}. A few key parameters can be derived: (1) peak flux density, $S_{\rm peak}$; (2) width of a spike subpulse, $\Delta\phi$ (in bins), which is defined as the longitude phase bin number between the maximum rising slope ($p_1$, ahead of the peak, scaled by the local $\sigma_{\rm Loc}$ defined below) and the maximum decaying slope ($p_2$, behind the peak); (3) pulse sharpness, $R = (p_1 - p_2)/\Delta\phi$, defined as the difference of the maximum rising and decaying slopes ($p_1 - p_2$) divided by their phase difference $\Delta\phi$. A few supplementary parameters are also derived from data: (4) off-pulse fluctuation $\sigma_{\rm off}$ estimated as being the standard deviation of intensities of off-pulse phase-bins outside the pulse window;  and (5) the local flux fluctuation $\sigma_{\rm Loc}$ for a given phase bin, which is the standard deviation locally within $\pm 30$ phase bins estimated from the median absolute deviation multiplied by a factor of 1.4826 \citep{llk+2013}. In addition, we also got the polarization measurements for spike subpulses, including linear polarization percentage ($L/I$) and circular polarization percentage ($V/I$) calculated from the polarized and total intensities within the phase bins between $p_1$ and $p_2$. When a spike subpulse lies upon the gradual ``background'' pulsar emission, the mean flux $S$ of the pulse is the average between the bins where the rising and decaying slopes reach one third of the maximums, respectively, and also the flux difference between the two bins ($\Delta S$) which indicate if the spike pulse is isolated or on a wide profile slop. The peak flux density $S_{\rm peak}$ of the spike pulse is then obtained after subtracting the gradual background emission.

\begin{figure}
\centering
\includegraphics[width=0.9\columnwidth]{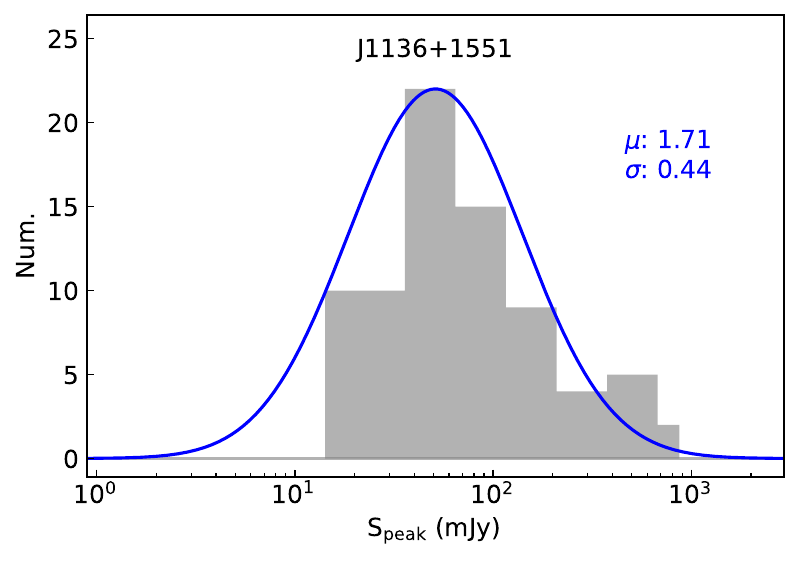}
\caption{The distribution of the peak intensities ($S_{\rm peak}$, in mJy) of 78 spike subpulses of PSR J1136+1551. It can be fitted with a log-normal function with the peak $ {\rm \mu} (\log S_{\rm peak}) =  1.71$ and $\sigma({\log S_{\rm peak})=0.44}$, corresponding to a flux density of $51^{+90}_{-33}$ mJy.
}
\label{fig:J1136spikePeak}
\end{figure}

\begin{figure}
\centering
\includegraphics[width=0.9\columnwidth]{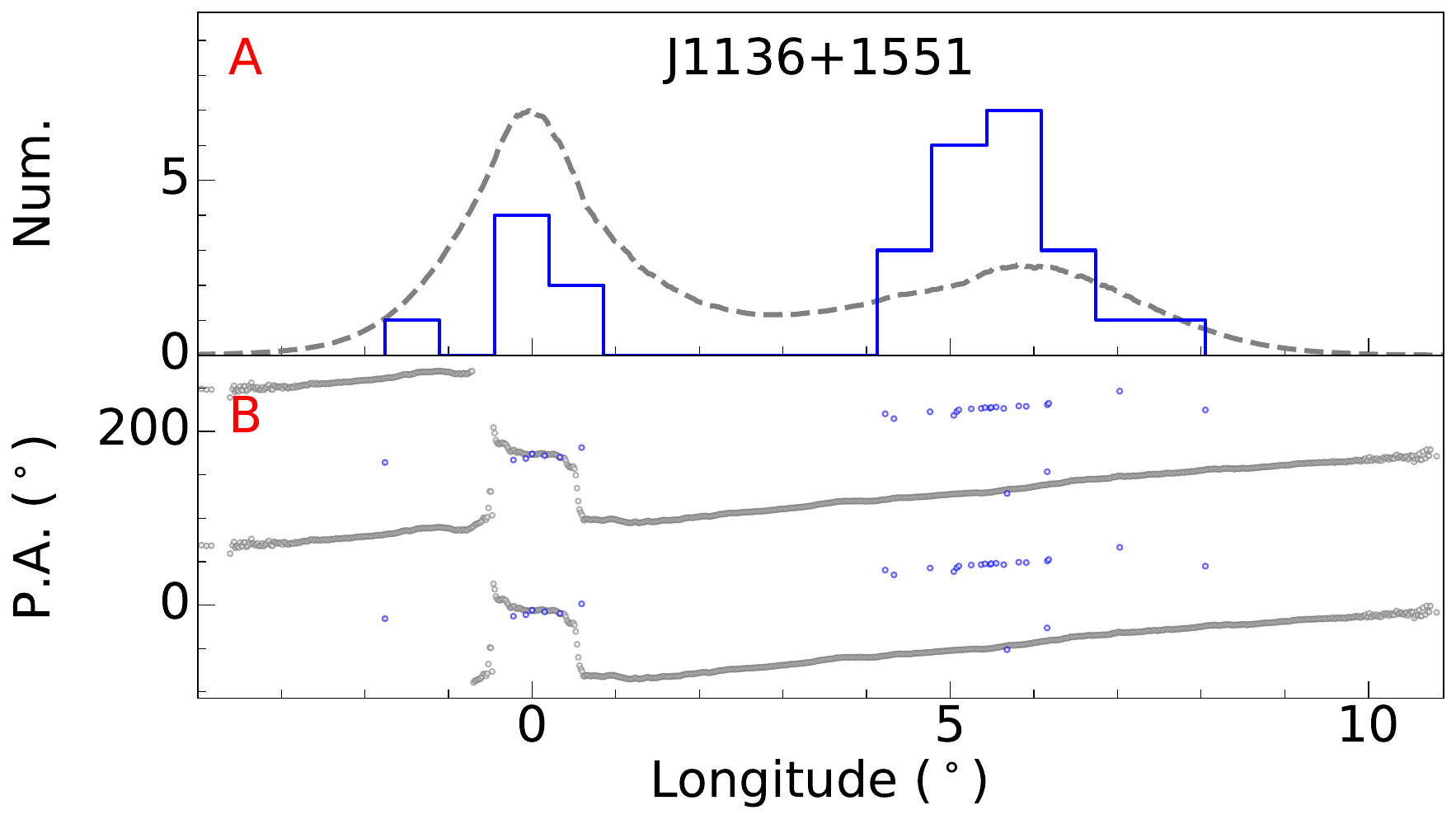}
\includegraphics[width=0.9\columnwidth]{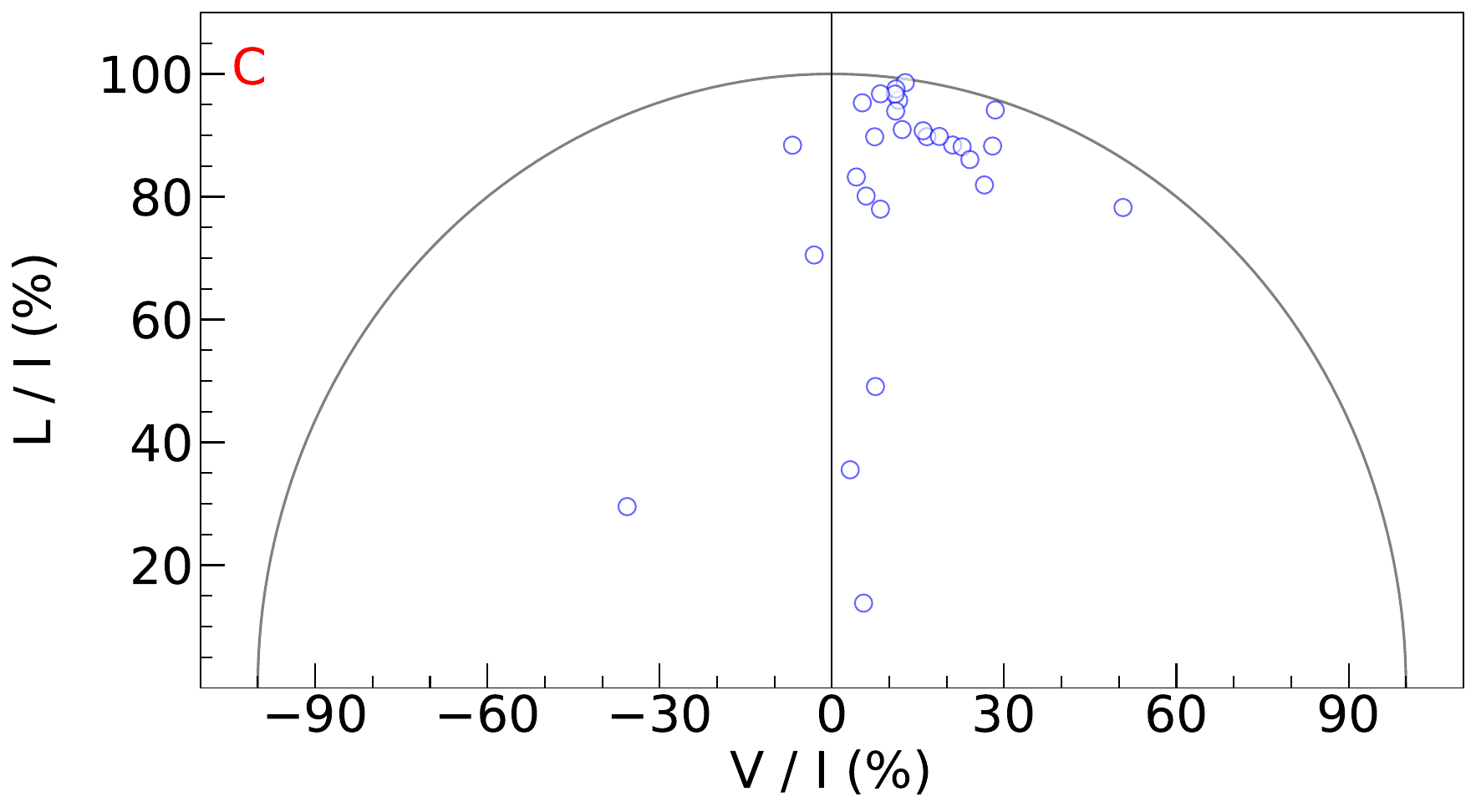}
%
\caption{
 Subpanel A shows the longitude distribution of 28 spike subpulses of PSR J1136+1551 with their linearly polarization detected with high confidence ($L>\sigma_{\rm Loc}$). Subpanel B shows the PA values of these spike subpulses which are generally in the orthogonal mode relative to the grey PA curve for the mean polarization profile. Subpanel C shows their linear and circular polarization percentages, and most of the spike subpulses are highly linearly polarized.
}
\label{fig:J1136spikePoln}
\end{figure}

Taking the one-hour FAST data for PSR J1136+1551 observed on 2019-11-21, for example, we detected 78 spike subpulses (Table~\ref{table:78spikeJ1136}) that satisfy four conditions: \\ 
(1) significance: ${\rm d}S_{p1} \geq 10 \times \sigma_{\rm Loc}$ and $|{\rm d}S_{p2}| \geq 10 \times \sigma_{\rm Loc}$; \\ 
(2) sharpness: $R \geq 20 $; \\ 
(3) symmetry:  Min\{${\rm d}S_{p1}$,${\rm d}S_{p2}$\} $\geq$ 0.8 $\times$ Max\{${\rm d}S_{p1}$,${\rm d}S_{p2}$\}; \\
(4) outstanding: $S_{\rm peak} \geq 5 \times \Delta S$. \\
%
%
%
So-selected spike subpulses in general appear as an isolated spike subpulse with only one or very few bins, as shown in Figure~\ref{fig:J1136N1749}. 
{Note that the measured widths of these spike subpulses, including possible residual dispersion smearing and scattering, are mainly limited by the sampling time of 49~$\mu$s. The intrinsic subpulse widths should be narrower than these values, and are only measurable by future higher-time-resolution observations.}
%
%
We noticed that the peak flux densities of these 78 spike subpulses follow a log-normal distribution, as shown in Fig.~\ref{fig:J1136spikePeak}.
%
Those spike subpulses are detected both in the two emission components of the mean pulse profile, as shown in the top panel of Figure~\ref{fig:J1136spikePoln}. 
%
%
%
%
We also examined the polarization properties of the detected spike subpulses. To diminish the  ``background'' emission, we focused on 28 isolated spike subpulses with peak intensities exceeding 10 times the local ``background'' emission and linear polarized intensity exceeding 10$\sigma_{\rm Loc}$, and plotted the distribution of the linear and circular polarization percentage in Fig.~\ref{fig:J1136spikePoln}. One can see that most of them are highly linearly polarized. It is intriguing to see that the polarization angles of these spike subpulses are of the orthogonal mode relative to the PA curve of the mean polarization profile.




\begin{table}[t]
  \caption{Parameters of example quasi-periodic subpulses of 13 pulsars. 
   Columns list the pulse number for quasi-periodic subpulses detection, the longitude phase range (in degrees), the number of subpulses for a set of quasi-periodic subpulses (${\rm Num}$), and the quasi-period of the subpulses (in degrees). 
   Here we present 4 sets of quasi-periodic subpulses of PSR J1136+1551 detected by FAST on 2019-11-21 and the only one set of quasi-periodic subpulses of PSR J2219+4754 detected by FAST on 2023-08-15.
   {\it A full list of detected quasi-periodic subpulses of 13 pulsars 
   is presented in 
   electronic form. [temporarily a long table behind the text of this paper for referee process.]}
  }
  \label{table:QPSJ1136J2219}
    \centering
    \small
    \renewcommand{\arraystretch}{0.8}
    \begin{tabular}{cccc}
    \hline 
Period    &        Longitude range &    Num. of    &  Quasi-period  \\
Number    &      ($^\circ$)  &   subpulses     &       ($^\circ$)     \\
    \hline
   ... &  ... & ... & ... \\    
    \hline 
    \multicolumn{4}{c}{24 QPS of PSR J1136+1551 observed by FAST on 2019-11-21} \\  
    \hline 
   ... &  ... & ... & ... \\
      1927&     (6.97$^\circ$~,~8.41$^\circ$)&        12&     0.119\\
      2232&     (-1.27$^\circ$~,~0.94$^\circ$)&        30&     0.074\\
      2582&     (-0.75$^\circ$~,~1.22$^\circ$)&        13&     0.149\\
      2661&     (-0.65$^\circ$~,~0.42$^\circ$)&        12&     0.089\\
    \hline 
       \multicolumn{4}{c}{1 QPS of PSR J2219+4754 observed by FAST on 2023-08-15} \\  
       \hline
      1920&     (-1.39$^\circ$~,~2.98$^\circ$)&        27&         0.164\\
    \hline
   ... &  ... & ... & ... \\    
   \hline
    \end{tabular}
\end{table}

\begin{figure*}[tb]
\centering 
\includegraphics[width=0.495\textwidth]{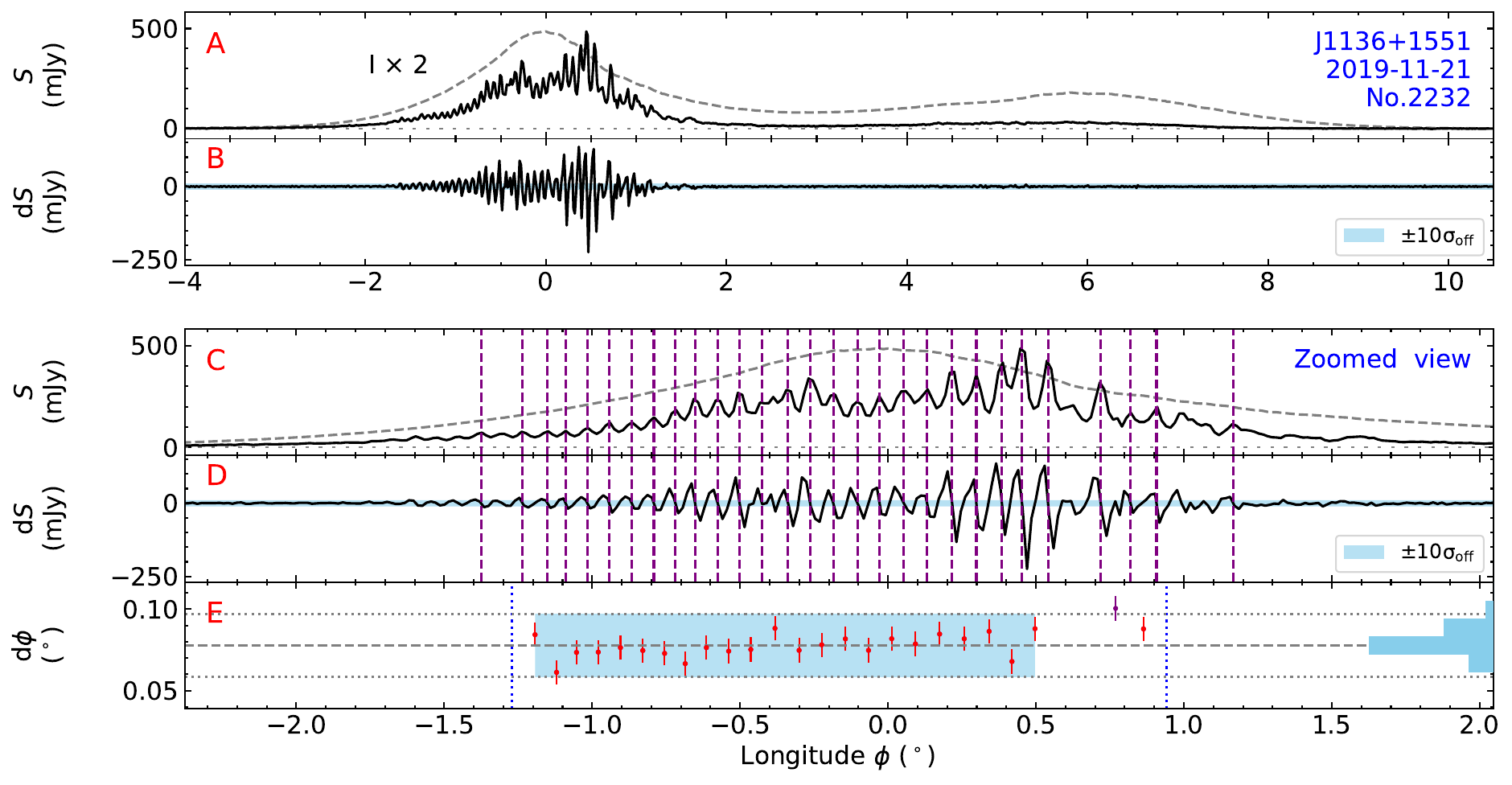}
\includegraphics[width=0.495\textwidth]{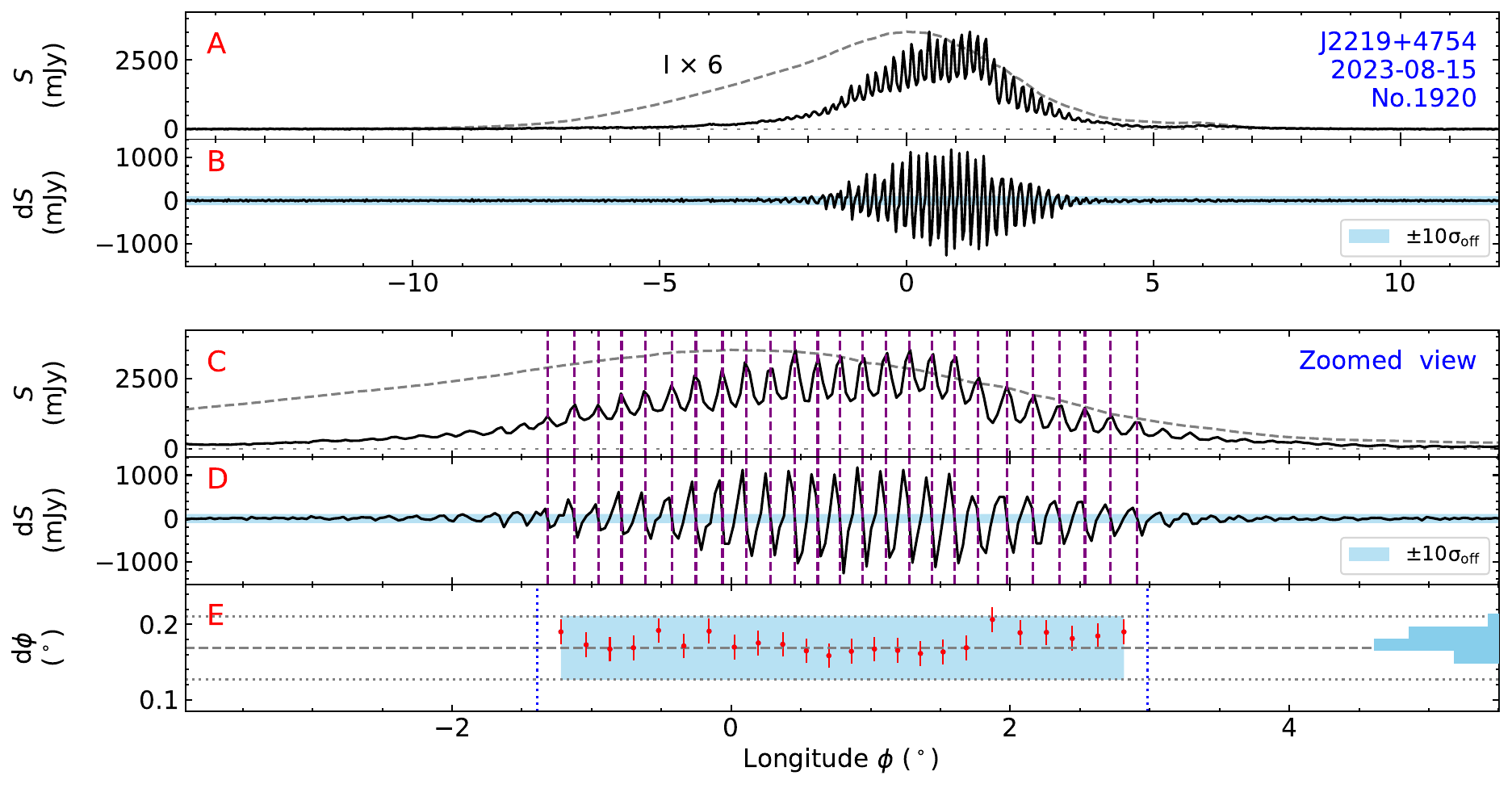}
\caption{Two examples of quasi-periodic subpulses. Left panels for quasi-periodic subpulses of PSR J1136+1551 observed by FAST on 2019-11-21 at the period number 2232, and right panels for PSR J2219+4754 on 2023-08-15 at the period number 1920. 
Subpanel (A) shows the total intensity profile ($S$) by the solid line, compared to the scaled mean pulse profile in the light dashed line with a multiplied factor;  Subpanel (B) shows the differential profile (d$S$) together with the shaded area for $\pm10\sigma_{\rm off}$; the zoomed views of $S$ and d$S$ of the longitude range for the quasi-periodic subpulses are shown in subpanels (C, D), with peaks marked by vertical dashed lines; the phase differences ($d\phi$) between every two subpulse peaks are plotted in subpanel (E), together with an indication of the $\pm$25\% deviation from the mean quasi-period and the histogram distribution in the right. 
%
%
%
}
\label{fig:QPSJ1136J2219}
\end{figure*}

\subsection{Identifying quasi-periodic subpulses from individual pulses}
\label{A:periodicMicroPs}

Quasi-periodic subpulses (QPS) have sometimes been detected from highly sensitive observations of individual subpulses \citep{kjs+2002, mar+2015, dgs+2016, lab+2022, kld+2024}, and the periodicity was often obtained by the autocorrelation of pulse emission along the pulse longitude. We examined the high-time-resolution subpulses by using the FAST pulsar data and identified a large number of quasi-periodic subpulses for 13 pulsars (see Table~\ref{table:QPSJ1136J2219}). 

Specifically, for example, we identified 24 sets of quasi-periodic subpulses from one hour of FAST observation of PSR J1136+1551  observed on 2019-11-21 and one set of quasi-periodic subpulses from PSR J2219+4754 observed on 2023-08-15. 
Two examples of quasi-periodic subpulses are shown in Fig.~\ref{fig:QPSJ1136J2219}. 
%
%
The details of quasi-periodic subpulses of the other pulsars are presented in the section~\ref{otherPSRs}.

Let us explain how a set of quasi-periodic subpulses is identified with the help of differential profiles (d$S$) which can diminish the gradual emission profiles. 
For a given pulsar, many outstanding subpulses may be detected within each period, as shown for the period number 1920 of PSR J2219+4754 in Figure~\ref{fig:QPSJ1136J2219}. Here, the subpulses are detected with more relaxed criteria than the spike subpulses: (1) d$S_{p1} \geq 10 \times \sigma_{\rm off}$; and (2) the separations of subpulses are less than 3 times the largest width of subpulses. In general, quasi-periodic subpulses emerge with nearly equally separated longitudinal phases, 
as marked by the {vertical} dashed lines in the subpanels C and D of  Figure~\ref{fig:QPSJ1136J2219}. Therefore, we get the phase separations ${\rm d\phi}$ for adjacent subpulses with an uncertainty of half a bin width. If the distribution of the separations remains approximately constant and shows an outstanding peak, then a set of quasi-periodic subpulses is identified. We plotted in Figure~\ref{fig:QPSJ1136J2219} the mean of separations ${\rm d\phi}$ together with a shadowed area for $\pm$25\%, and the subpulses with the phase-separations in the range are identified as one set. 
%
The phase separation of the subpulses, which is the quasi-period, is obtained by the 
auto-correlation of the individual pulse intensity in the phase range. 

\begin{figure}
\centering
\includegraphics[width=0.9\columnwidth]{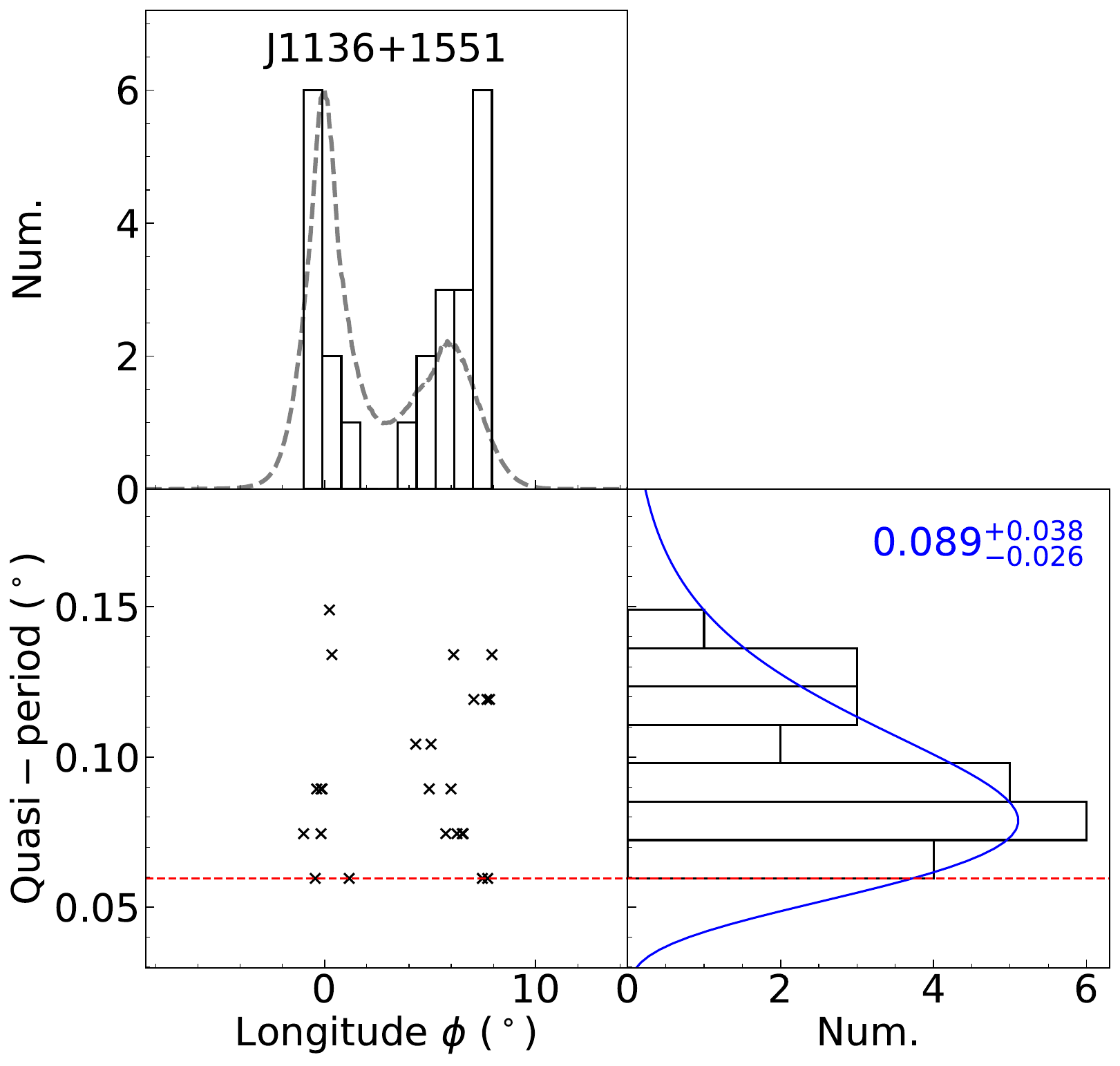}
\caption{Distribution of the phase quasi-period and phase longitudes of the 24 sets of quasi-periodic subpulses of PSR J1136+1551 observed on 2019‑11‑21.
}
\label{fig:QPSJ1136distr}
\end{figure}

We checked the phase range of 24 sets of quasi-periodic subpulses of PSR J1136+1551 observed on 2019-11-21, and found that they can emerge in the phase range of both mean pulse profile components, as shown in the top panel of Figure~\ref{fig:QPSJ1136distr}. The phase quasi-periods are distributed by following a log-normal distribution with the typical phase quasi-period of $0.089^{+0.038}_{-0.026}$ degree, corresponding to $0.294^{+0.124}_{-0.087}$ ms.

{We do have polarization data of these quasi-periodic subpulses; some have sufficient signal-to-noise ratio for polarization, and some do not. In general, the modulations of polarized emission are consistent with those of the total intensity, as shown by \citet{mar+2015}. Two sets of quasi-periodic subpulses exhibit orthogonal polarization modes, which we will analyze in a separate paper. }

    
    \subsection{Detailed Results for 25 pulsars}
    \label{otherPSRs}

    In this subsection, we introduce the 25 pulsars one by one and present the results with one example for spike subpulses and/or quasi-periodic subpulses. 
    
    \begin{figure*}[tbh]
    \centering
    \includegraphics[width=0.600\textwidth]{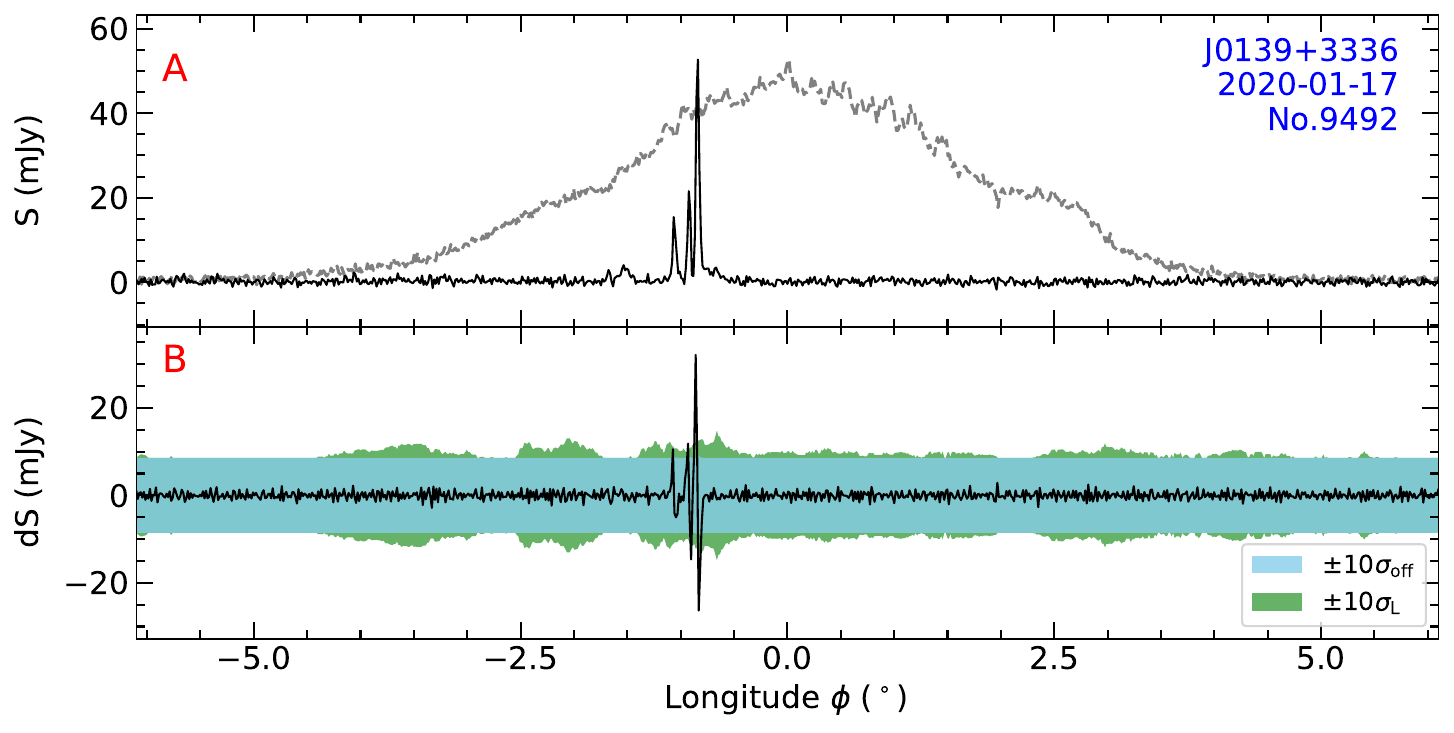}
    \includegraphics[width=0.303\textwidth]{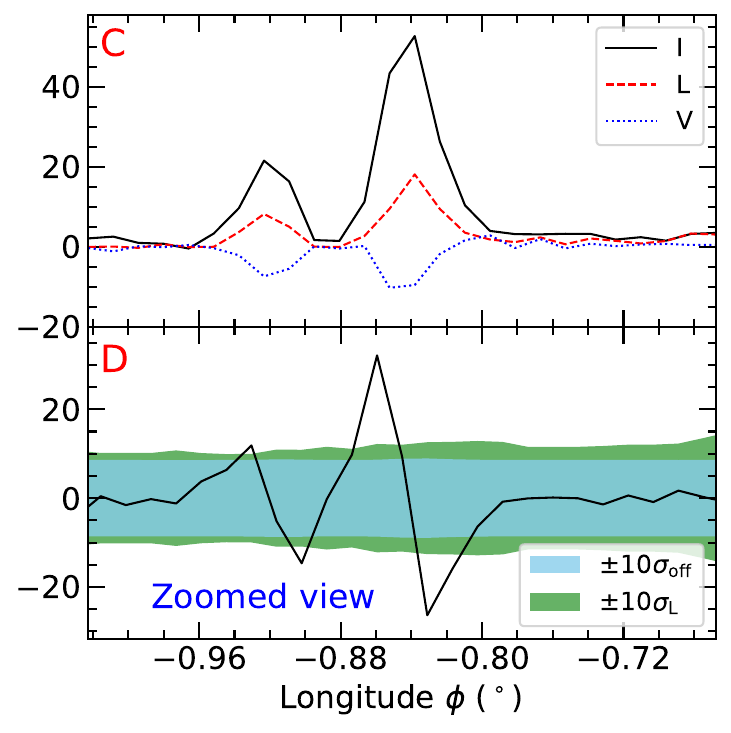}
    \caption{Similar to figure~\ref{fig:J1136N1749}, but for a spike subpulse  of PSR J0139+3336 appearing at the period number 9492 observed on 2020-01-17.}
    \label{fig:J0139N9492}
    \end{figure*}
    
    \begin{figure}[tbh]
    \centering
    \includegraphics[width=0.98\columnwidth]{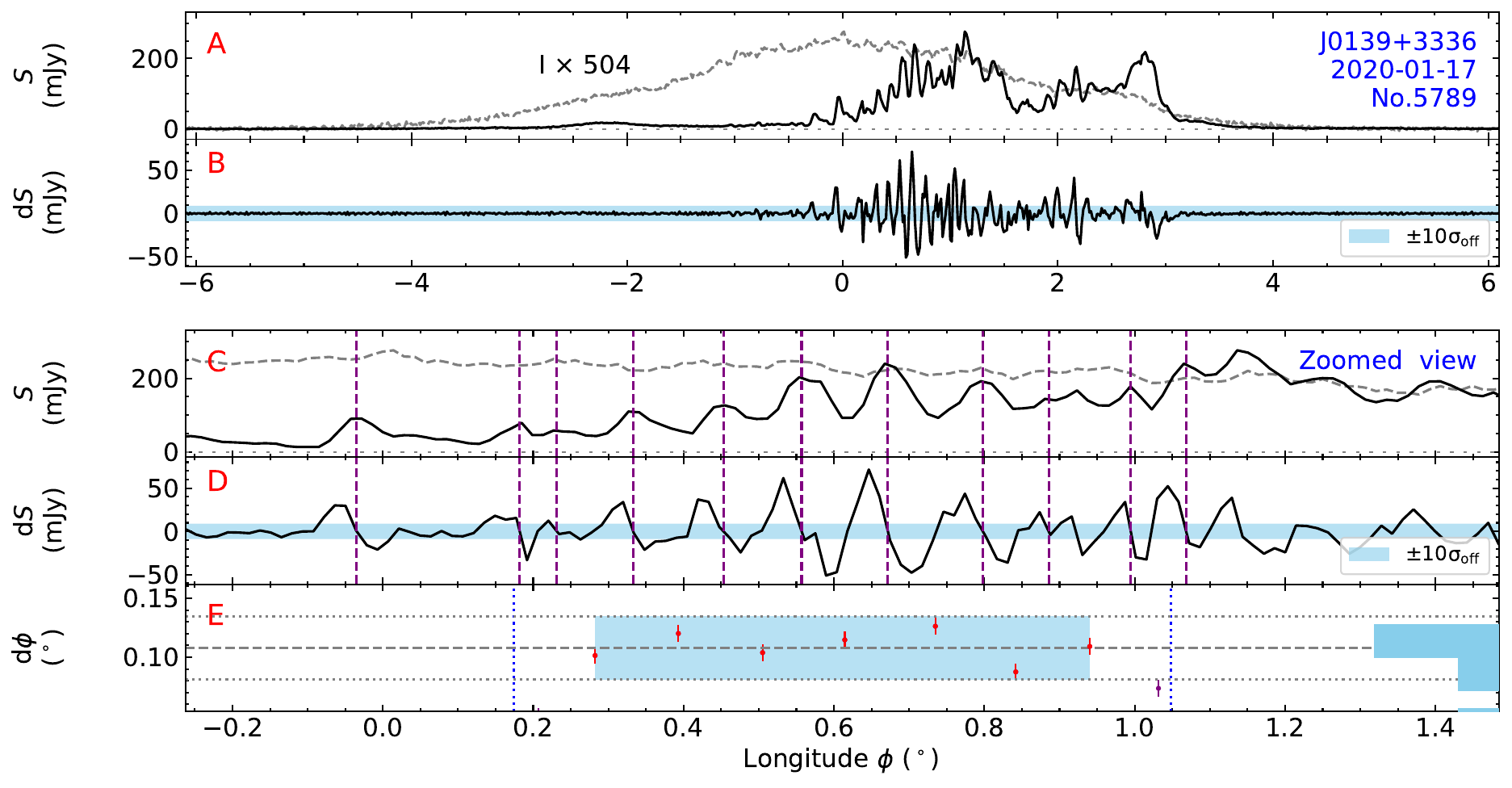}
    \caption{Similar to Figure~\ref{fig:QPSJ1136J2219}, but for the quasi-periodic subpulses of PSR J0139+3336 in the period number 5789 observed on 2020-01-17.}
    \label{fig:QPSJ0139}
    \end{figure}
    
    \subsubsection{J0139+3336}
    \label{sec:J0139+3336}
    
    PSR J0139+3336 was discovered as an RRAT by 
    Pushchino Radio Astronomy Observatory at 111 MHz and a time resolution of 12.5 ms \citep{tt+2017}. 
    Follow-up observations at 149 MHz and 1532 MHz with LOFAR and the Lovell telescope got one pulse 
    every 5 minutes \citep{mbc+2020}. 
    
    FAST observed this RRAT for 3.3 hr and detected 
    45 pulses per hour \citep{xww+2022}. \citet{dys+2024} identified two quasi-periodic microstructures at a time resolution of $\sim 152~\mu{\rm s}$ from FAST data, with a typical microstructure width of ${\rm \tau_\mu}\approx 460~\mu{\rm s}$ and a quasi-period of ${\rm P_\mu}\approx 910~\mu{\rm s}$.
    We analysed the FAST observation data obtained on 2020-01-17, which covers 9615 periods (${\rm P_0}=1.248~{\rm s}$, Table~\ref{tab:PSRlist}). We folded data with 25344 bins per period. 
    As shown in Figure~\ref{fig:J0139exam}, 
    this pulsar is nulling for most periods. 
    We identified only one spike subpulse in period No.~9492 (Figure~\ref{fig:J0139N9492}) with parameters in Table~\ref{table:SpikeAllPSR}, and detected one set of quasi-periodic subpulses in period No.~5789 satisfying the conditions above, as shown in Figure~\ref{fig:QPSJ0139}. The phase quasi-period is $0.113^\circ$ (corresponding to 0.392 ms), and these parameters of this set of quasi-periodic subpulses are given in Table~\ref{table:QPSallPSR}.

    \begin{figure}[tbh]
    \centering
    \includegraphics[width=0.98\columnwidth]{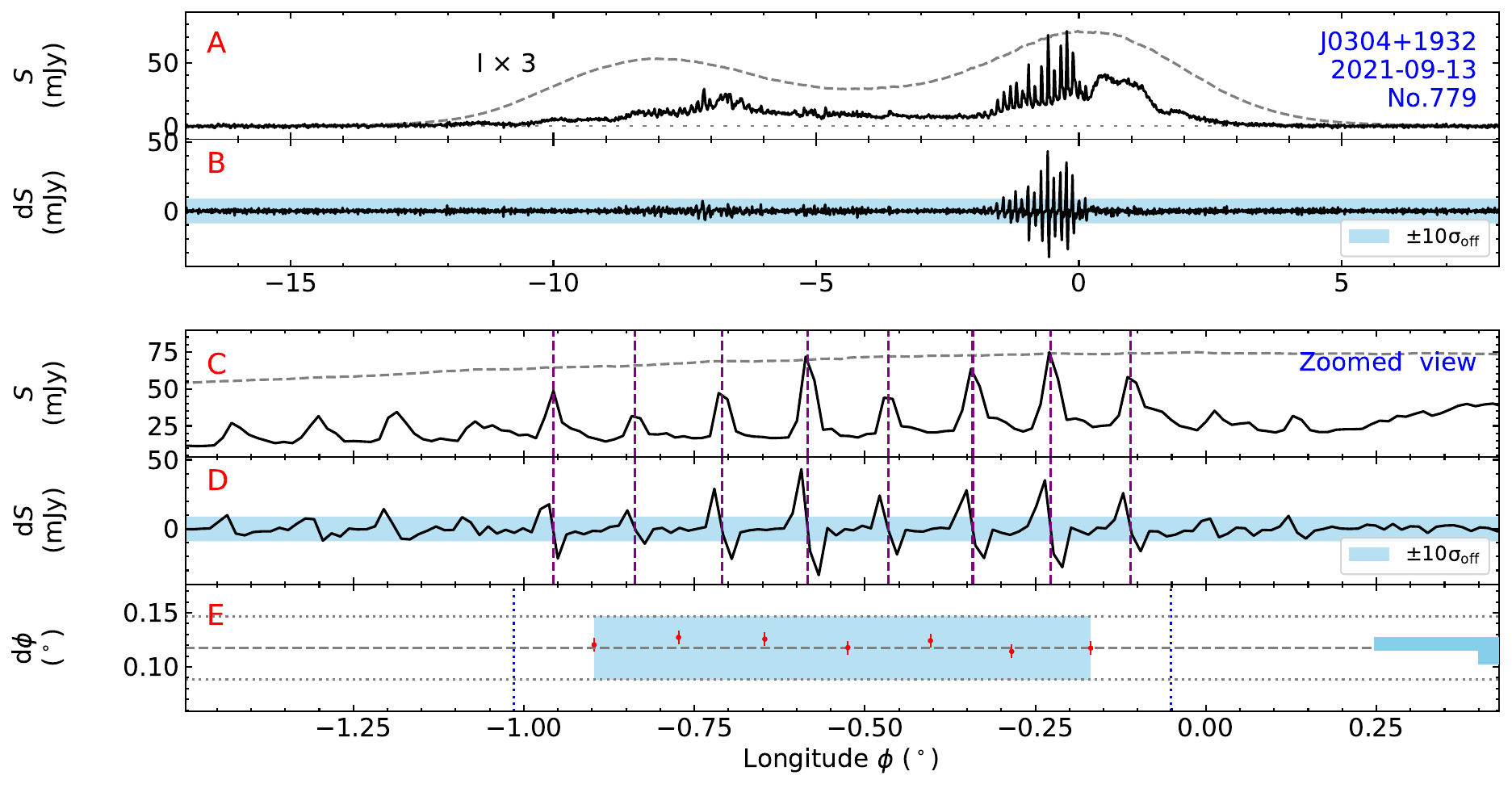}
    \caption{Similar to Figure~\ref{fig:QPSJ1136J2219}, but for the quasi-periodic subpulses of PSR J0304+1932 in the period number 779 of PSR J0304+1932 observed on 2021-09-13.}
    \label{fig:QPSJ0304}
    \end{figure}
    
    \subsubsection{J0304+1932}
    \label{subsec:J0304+1932}
    
    PSR J0304+1932 is a bright pulsar first detected at 408 MHz with the Bologna Cross telescope \citep{fss+1973}. It is also a nulling pulsar, with a nulling fraction in the range of $6.1\%$--$17.5\%$ at different observing frequencies \citep{r+1986, rr+2009, bmm+2017, whh+2020}. The nulls recur quasi-periodically every $\sim 100$ periods in some observations \citep{bmm+2017}, but appear random at 2.25 GHz \citep{whh+2020}. Giant pulses with a mean width of 2.8~ms have also been reported at 111 MHz \citep{kps+2019}. Microstructures were detected by \citet{mar+2015}, and \citet{kld+2024} reported a typical microstructure quasi-period of $1340^{+268}_{-223}~\mu{\rm s}$.
    
    We analyzed the FAST observation data obtained on 2021-09-13. The observation lasted for about 1 hour, and the data covered 2559 periods. We folded the data with 28224 phase bins per period (${\rm P_0}=1.39~{\rm s}$). 
    No spike subpulse was identified during this observation, but we detected one set of quasi-periodic subpulses in period No.~779, as shown in Figure~\ref{fig:QPSJ0304}, with parameters listed in Table~\ref{table:QPSallPSR}. The phase quasi-period is $0.115^\circ$, corresponding to 0.444 ms. 

    \begin{figure*}[tb!]
    \centering
    \includegraphics[width=0.600\textwidth]{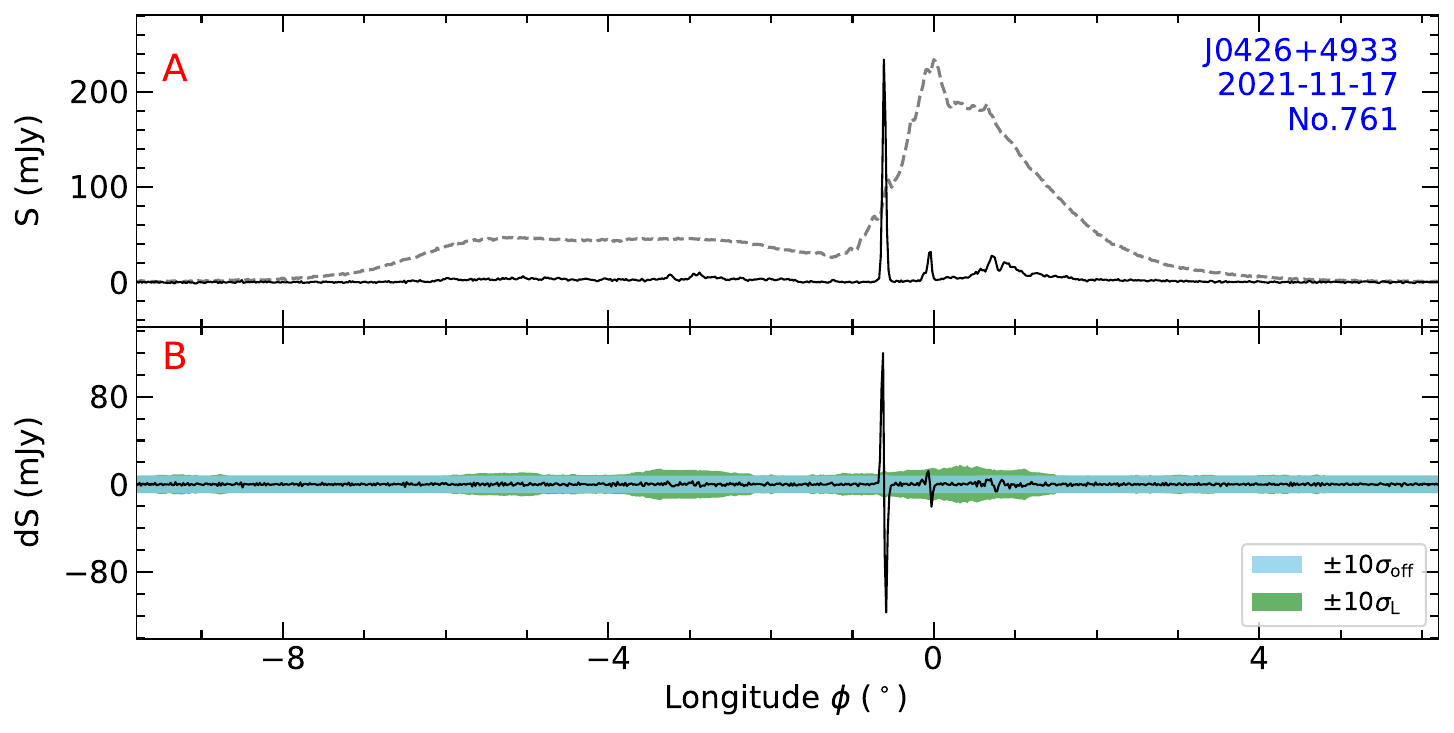}
    \includegraphics[width=0.303\textwidth]{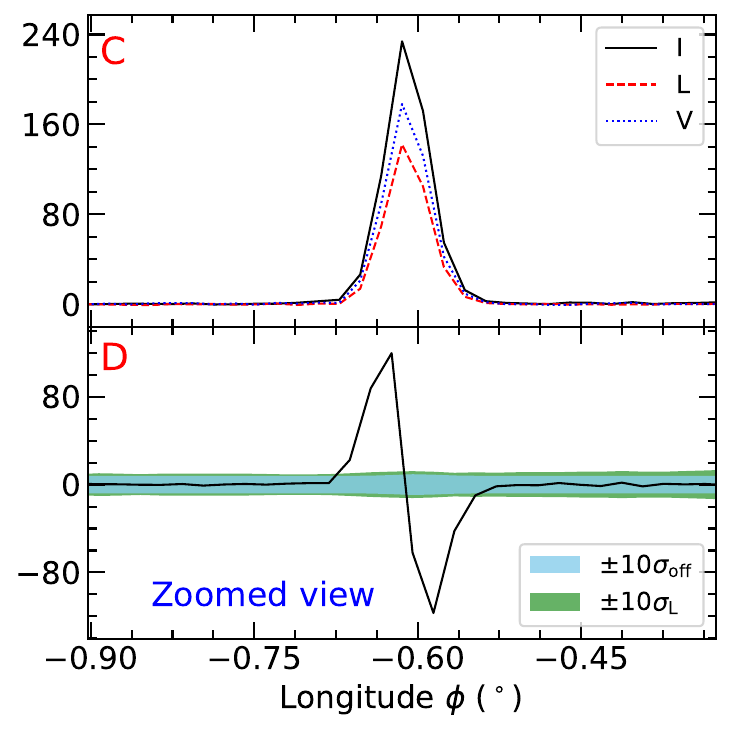}
    \caption{Similar to figure~\ref{fig:J1136N1749}, but for a spike subpulse appearing in the period number 761 of PSR J0426+4933 observed on 2021-11-17.}
    \label{fig:J0426N633}
%
    \centering
    \includegraphics[width=0.600\textwidth]{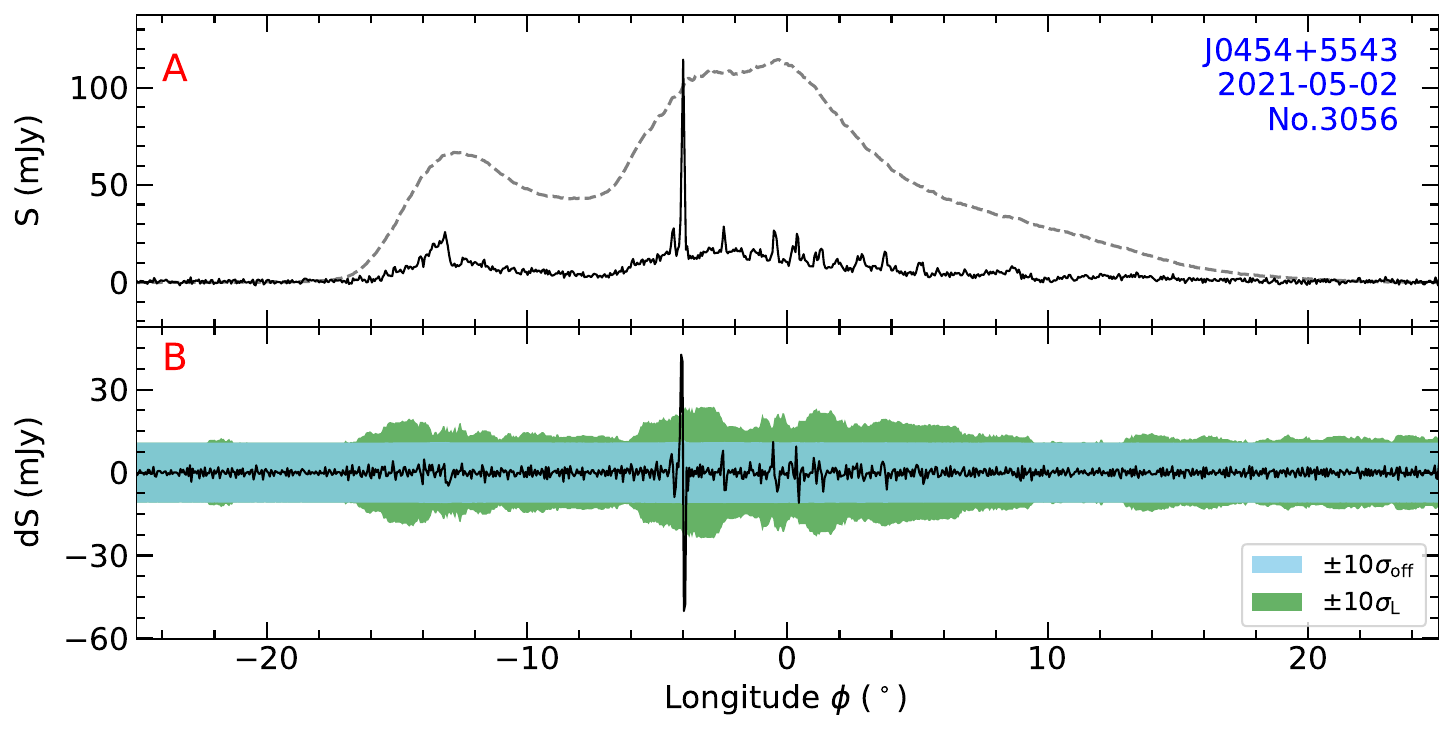}
    \includegraphics[width=0.303\textwidth]{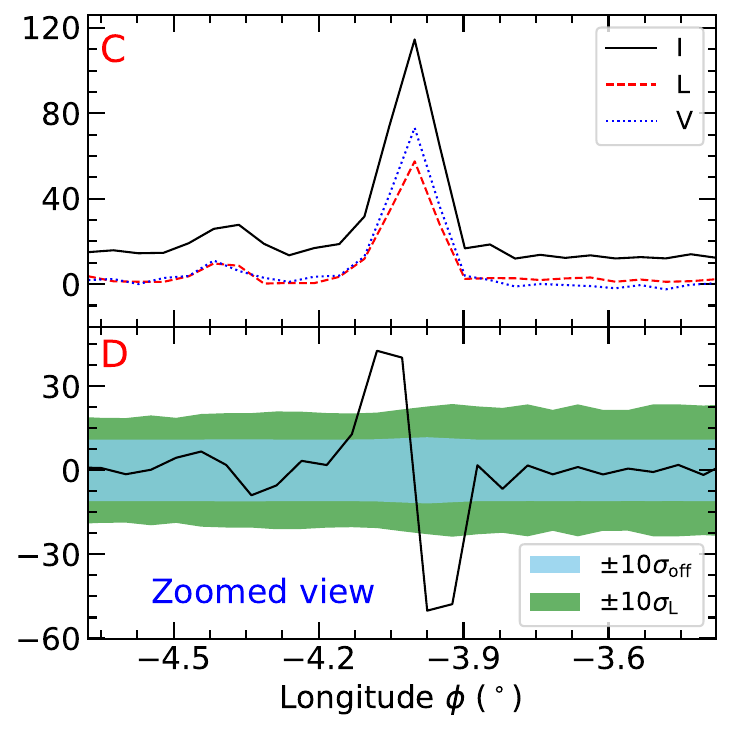}
    \caption{Similar to figure~\ref{fig:J1136N1749}, but for a spike subpulse appearing in the period number 3056 of PSR J0454+5543 observed on 2021-05-02.}
    \label{fig:J0454N3056}
  %
    \centering
    \includegraphics[width=0.600\textwidth]{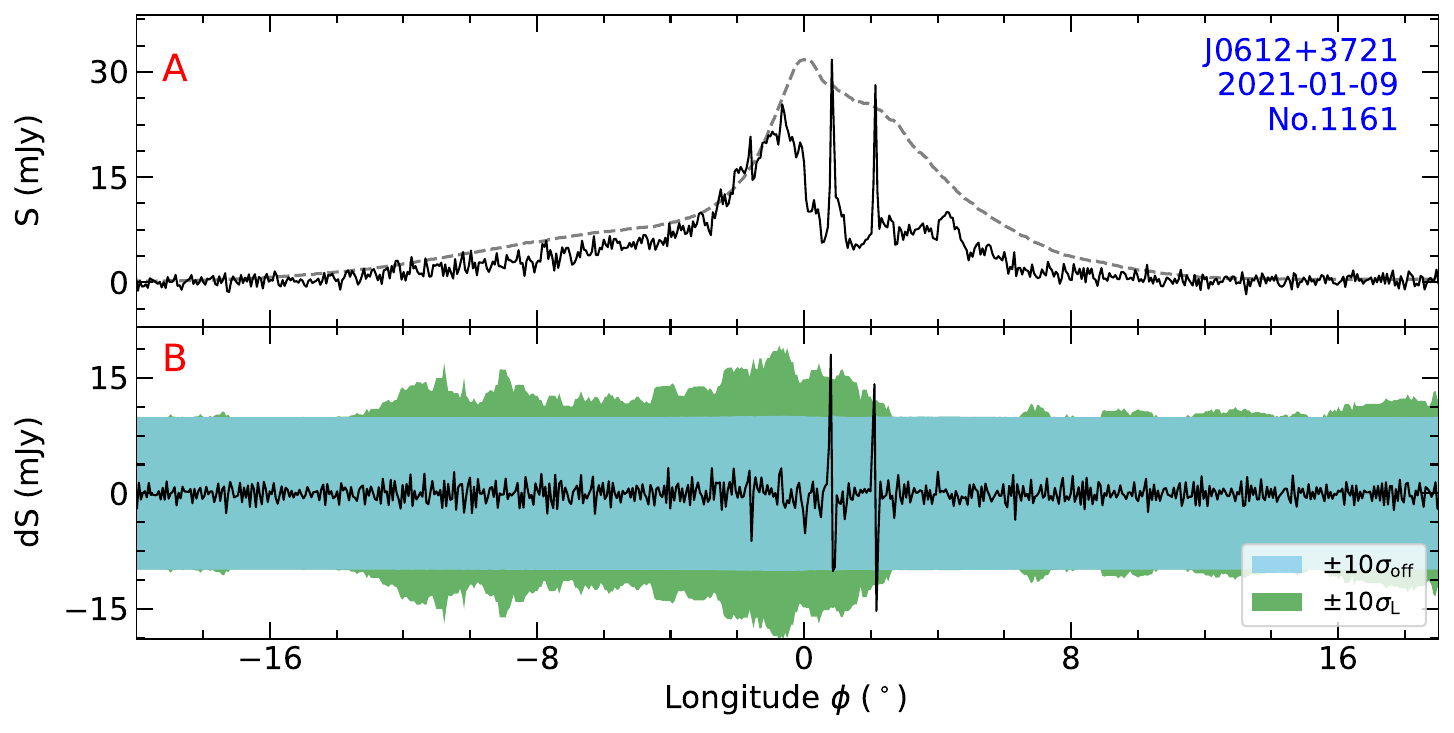}
    \includegraphics[width=0.303\textwidth]{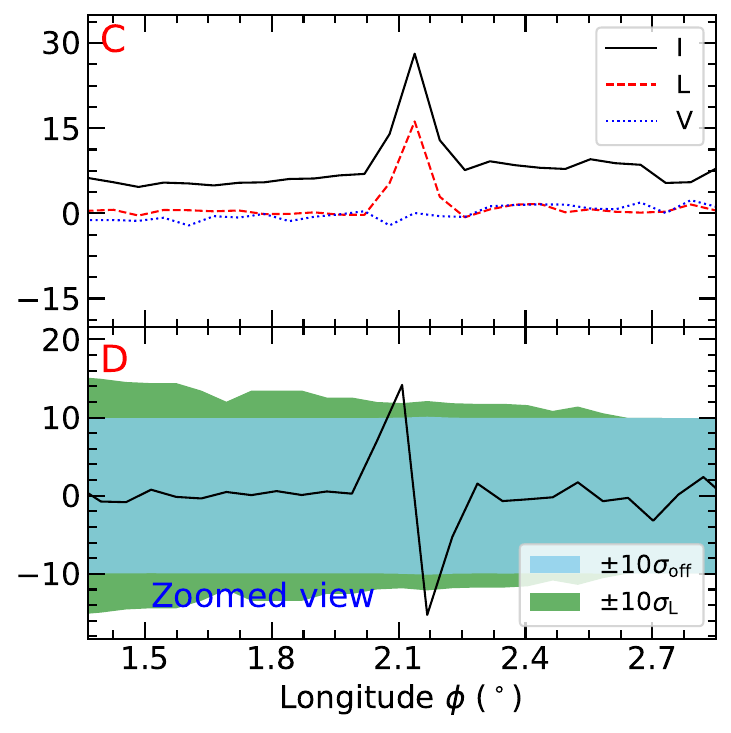}
    \caption{Similar to figure~\ref{fig:J1136N1749}, but for a spike subpulse appearing in the period number 1161 of PSR J0612+3721 observed on 2021-01-09.}
    \label{fig:J0612N1161}
    \end{figure*}

    \begin{figure*}[tb!]
    \centering
    \includegraphics[width=0.600\textwidth]{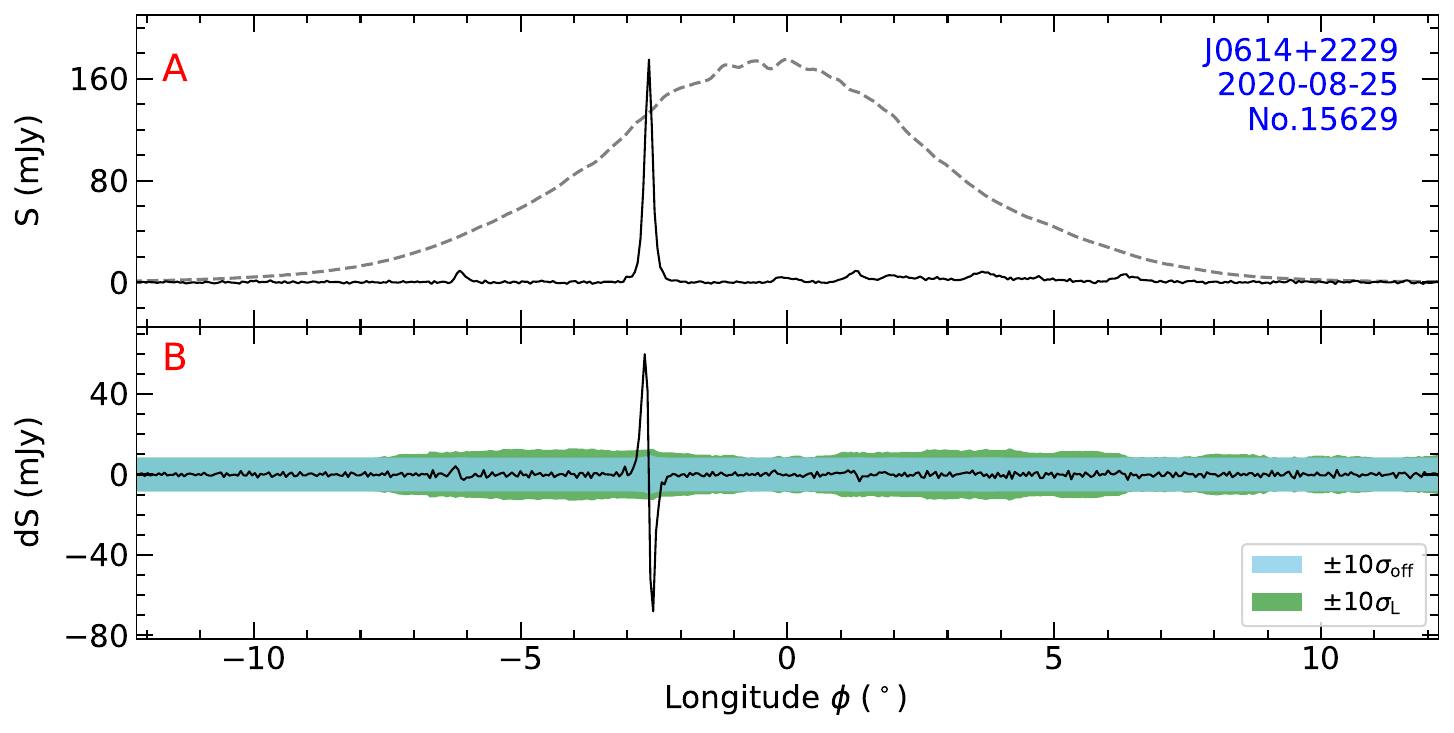}
    \includegraphics[width=0.303\textwidth]{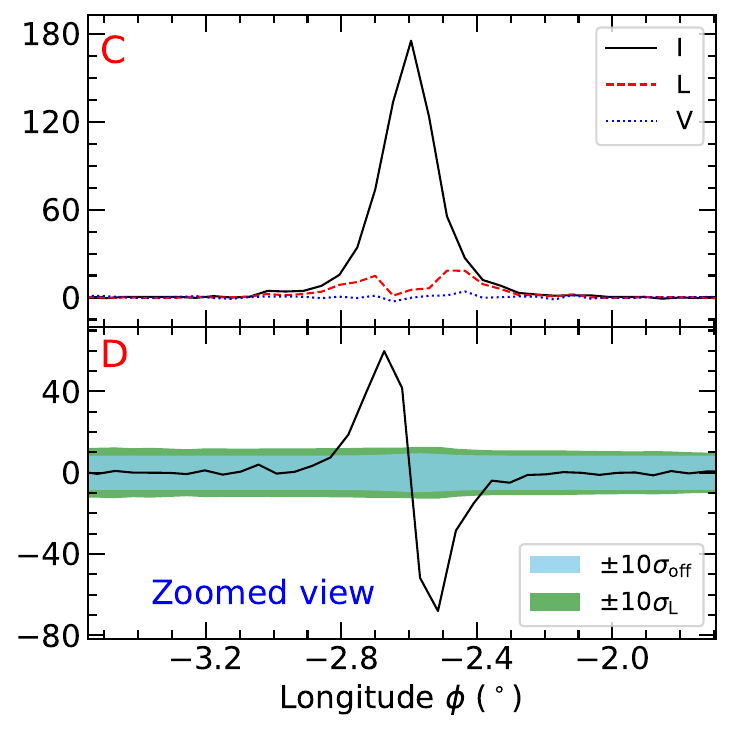}
    \caption{Similar to figure~\ref{fig:J1136N1749}, but for a spike subpulse appearing in the period number 15629 of PSR J0614+2229 observed on 2020-08-25.}
    \label{fig:J0614N15629}
    %
    \centering
    \includegraphics[width=0.600\textwidth]{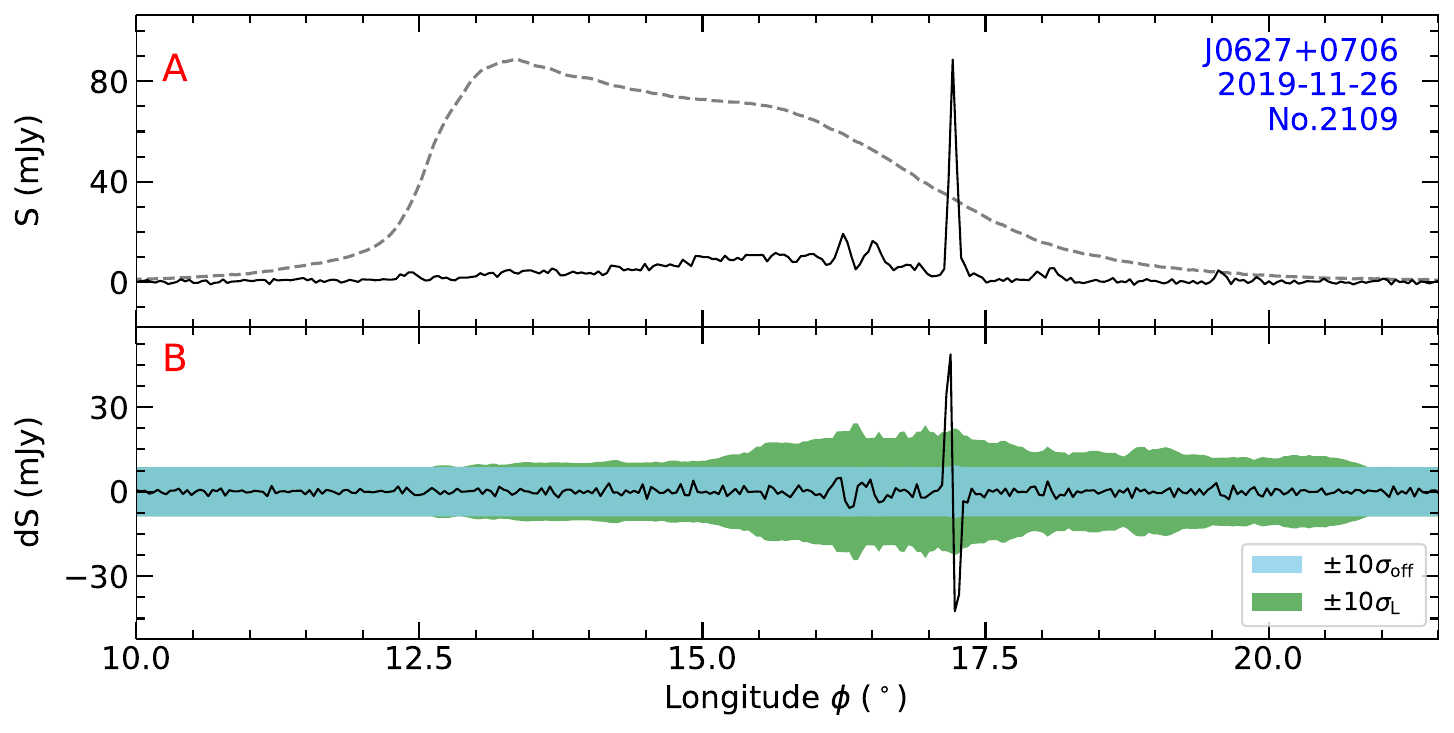}
    \includegraphics[width=0.303\textwidth]{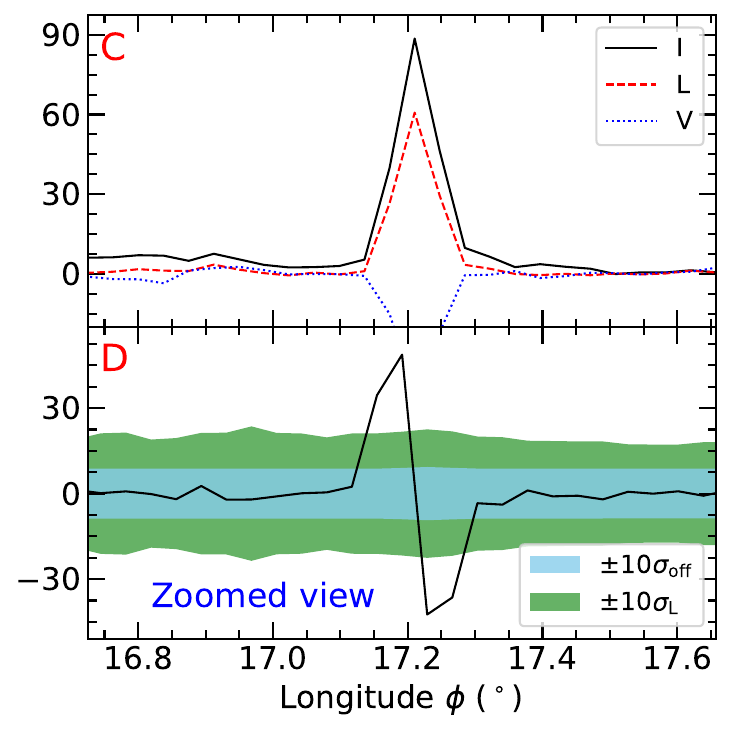}
    \caption{Similar to figure~\ref{fig:J1136N1749}, but for a spike  subpulse from the period No.~2109 of PSR J0627+0706 observed on 2019-11-26.}
    \label{fig:J0627N2109}
    %
    \centering
    \includegraphics[width=0.600\textwidth]{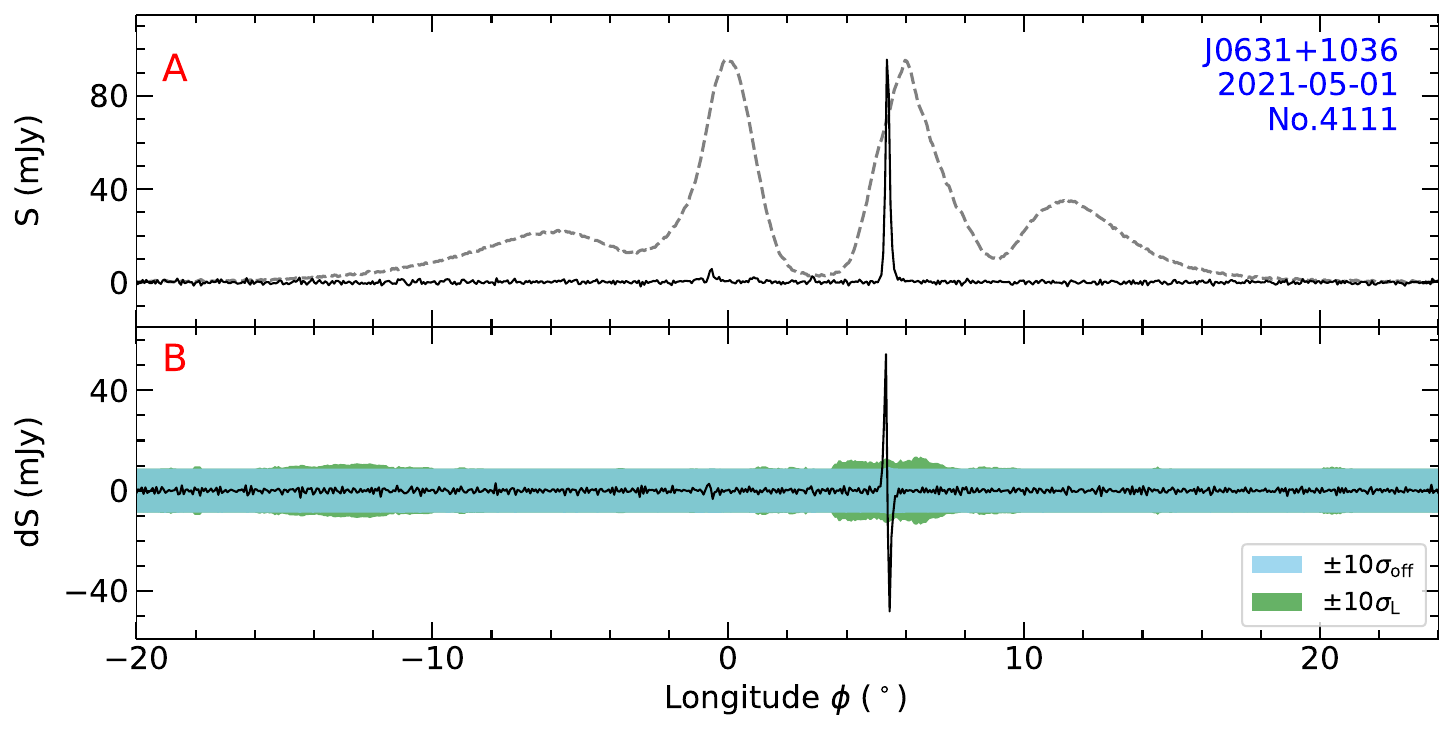}
    \includegraphics[width=0.303\textwidth]{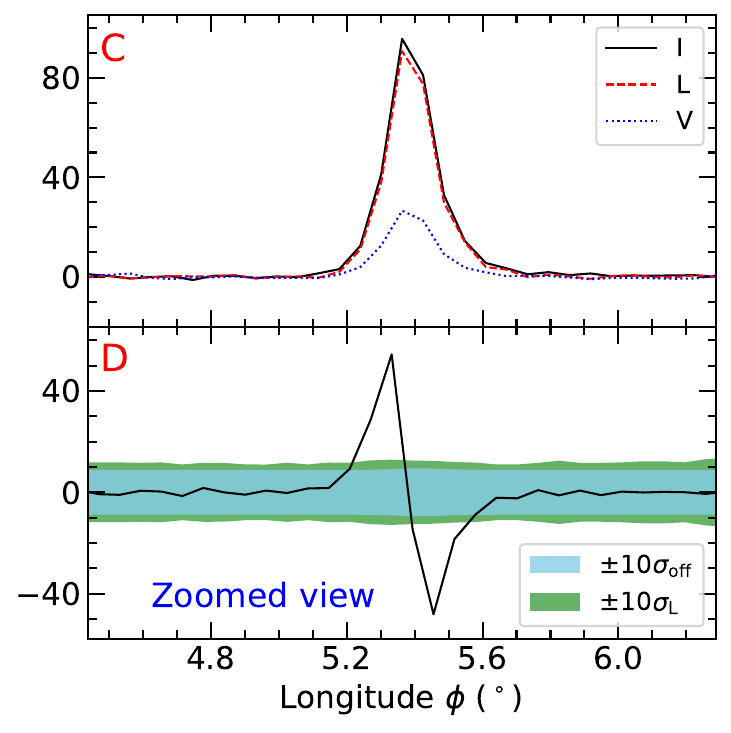}
    \caption{Similar to figure~\ref{fig:J1136N1749}, but for a spike subpulse appearing in the period number 4111 of PSR J0631+1036 observed on 2021-05-01.}
    \label{fig:J0631N4111}
    \end{figure*}

    \begin{figure*}[tb!]
    \centering
    \includegraphics[width=0.600\textwidth]{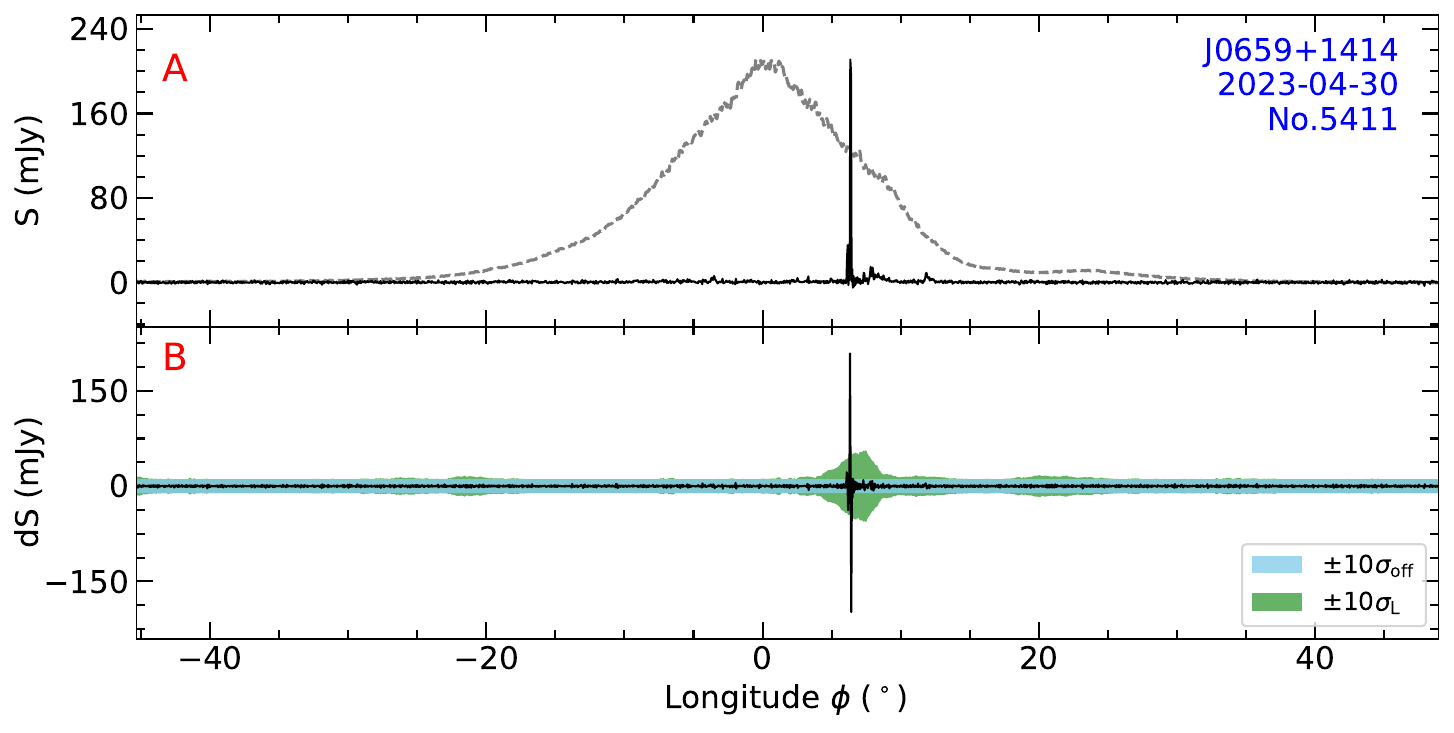}
    \includegraphics[width=0.303\textwidth]{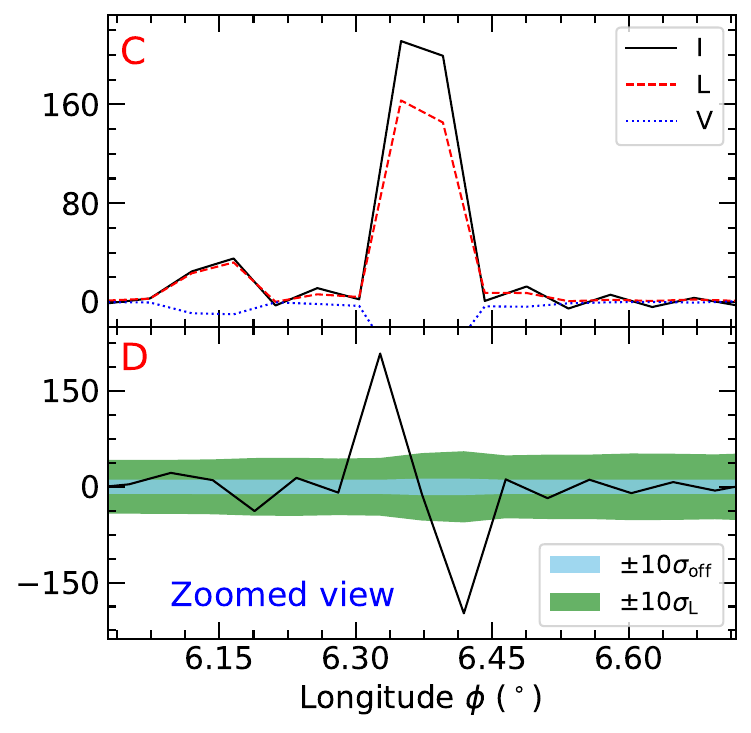}
    \caption{Similar to figure~\ref{fig:J1136N1749}, but for a spike subpulse appearing in the period number 5411 of PSR J0659+1414 observed on 2023-04-30.}
    \label{fig:J0659N5411}
  %
    \centering
    \includegraphics[width=0.600\textwidth]{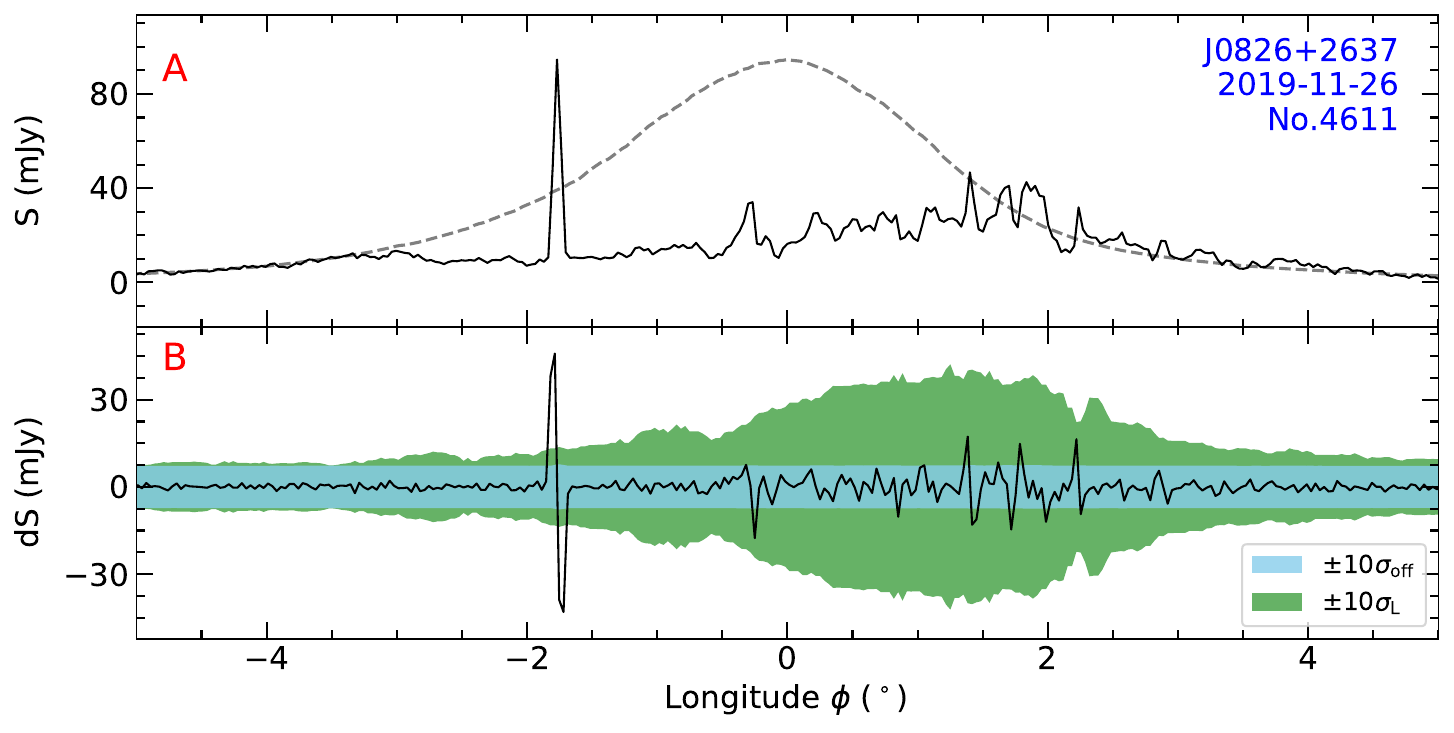}
    \includegraphics[width=0.303\textwidth]{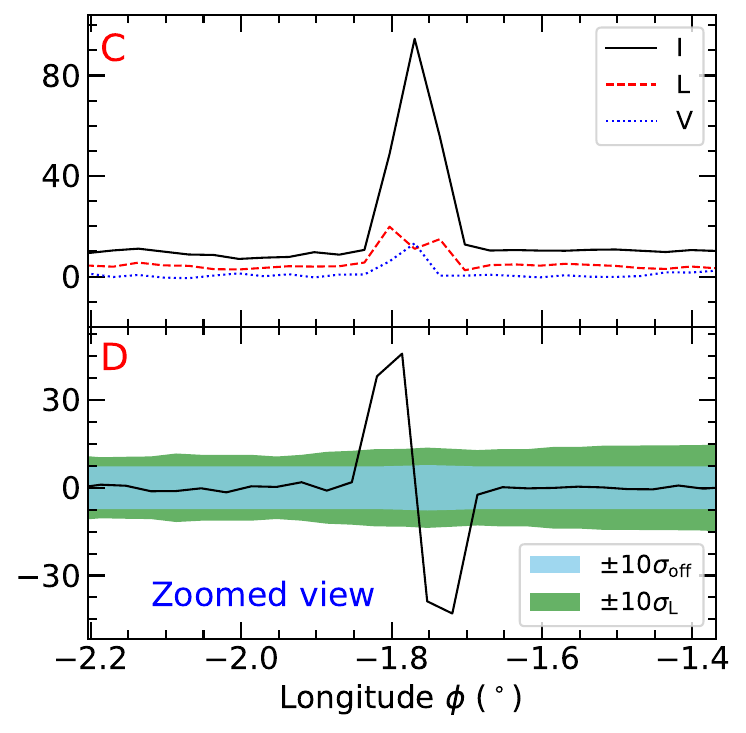}
    \caption{Similar to figure~\ref{fig:J1136N1749}, but for a spike subpulse appearing in the period number 4611 of PSR J0826+2637 observed on 2019-11-26.}
    \label{fig:J0826N4611}
    \end{figure*}
    
    \subsubsection{J0426+4933}
    \label{sec:J0426+4933}
            
    PSR J0426+4933 was discovered in the GBT Northern Galactic Plane survey for radio pulsars and transients \citep{hrk+2008}. A later high-sensitivity FAST observation revealed subpulse drifting at 1.4 GHz, with ${\rm P_2}=6.2\pm1.0^\circ$ from the ACF and ${\rm P_3}=3.5\pm0.3\,{\rm P_0}$ from the two-dimensional fluctuation spectrum \citep{zlh+2023}. 
    
    We analysed the FAST data observed for 19.3 minutes on 2021-11-17, 
    covering 1258 periods (${\rm P_0}=0.922~{\rm s}$). We folded the data with 18760 phase bins per period, and  
    identified 17 spike subpulses (Table~\ref{table:SpikeAllPSR}). One example from period No.~761 is shown in Figure~\ref{fig:J0426N633}. 
    We got a better DM value than in the pulsar catalog from the spike subpulse, as listed in Table~\ref{tab:PSRlist}.

    \subsubsection{J0454+5543}
    \label{subsec:J0454+5543}
       
    PSR J0454+5543 was discovered in a systematic survey of the northern sky with the 300-foot (91 m) NRAO transit telescope at Green Bank \citep{dth+1978}. Giant pulses were later detected with the Large Phased Array at Pushchino at 111 MHz with a time resolution longer than $1~{\rm ms}$ \citep{Kazantsev2021}. To our knowledge, no microstructure has been reported previously for this pulsar.
    
    We analyzed the FAST observation obtained on 2021-05-02. The observation lasted for  26.0 minutes, covering 4581 periods. We folded data with the period  of ${\rm P_0}=0.340~{\rm s}$ with 6928 phase bins per period (Table~\ref{tab:PSRlist}).
    
    In these data, we identified 12 spike subpulses. One example, from period No.~3056, is shown in Figure~\ref{fig:J0454N3056}, and the measured parameters are listed in Table~\ref{table:SpikeAllPSR}. 
        
    \subsubsection{J0612+3721}
    \label{subsec:J0612+3721}
        
    PSR J0612+3721 was discovered by the 
    92 m Green Bank telescope \citep{stwd+1985}. Subpulse drifting at 21 cm was detected by \citet{wes+2006}, with  the phase-shift ${\rm P_2}=-20^{+4}_{-18}$ degrees and the repeating period of ${\rm P_3}=32\pm4{\rm P_0}$. 
    
    We analysed the FAST data obtained on 2021-01-09. The observation lasted for 10 minutes, and we got data for 1987 periods (${\rm P_0}=0.298~{\rm s}$). We folded the pulses with 6062 phase bins per period, and 
    identified two spike subpulses. One example from the period No.~1161 is shown in Figure~\ref{fig:J0612N1161}, with parameters listed in Table~\ref{table:SpikeAllPSR}. 


    \subsubsection{J0614+2229}
    \label{subsec:J0614+2229}
    
    PSR J0614+2229 is a bright pulsar discovered by the Jodrell Bank Mark IA telescope \citep{dls+1972}. We analysed the FAST data obtained on 2020-08-25. The observation lasted for  88.7 minutes, and we obtained data covering 15892 periods. We folded data with the period of ${\rm P_0}=0.335~{\rm s}$ with 6800 bins per period (Table~\ref{tab:PSRlist}), and identified 52 spike subpulses. One example (from period No.~15629) is shown in Figure~\ref{fig:J0614N15629}, and the measured parameters of 52 spike subpulses are listed in Table~\ref{table:SpikeAllPSR}. The peak flux densities of these spike subpulses follow a log-normal distribution with $ {\rm \mu} (\log S_{\rm peak}) =  1.88$ and $\sigma({\log S_{\rm peak})=0.09}$, corresponding to a flux density of $76^{+17}_{-14}$ mJy.
    

    \subsubsection{J0627+0706}
    \label{subsec:J0627+0706}
      
    
    PSR J0627+0706 was discovered in the Caltech--Arecibo drift-scan search \citep{chandler2011}. 
    \citet{LDW+2025} reported microstructures in both the main pulse and the interpulse using FAST at a time resolution of $\sim115~\mu{\rm s}$. For the main pulse they measured a typical width of ${\rm \tau_\mu}\approx460~\mu{\rm s}$ and a quasi-period of ${\rm P_\mu}=820^{+170}_{-120}~\mu{\rm s}$, while for the interpulse they obtained ${\rm \tau_\mu}\approx460~\mu{\rm s}$ and ${\rm P_\mu}=940^{+220}_{-240}~\mu{\rm s}$.
    
    We analyzed the FAST L-band observation data obtained on 2019-11-26. The observation lasted for  60.0 minutes, covering 7567 periods. We folded the data with the period of ${\rm P_0}=0.476~{\rm s}$ with 9664 phase bins per period (Table~\ref{tab:PSRlist}), and identified 9 spike subpulses. One example, from period No.~2109, is shown in Figure~\ref{fig:J0627N2109}, and the measured parameters are listed in Table~\ref{table:SpikeAllPSR}. 
    We did not find quasi-periodic subpulses satisfying our conditions.


    \subsubsection{J0631+1036}
    \label{subsec:J0631+1036}
    
    PSR J0631+1036 was discovered by the Arecibo telescope \citep{zcwl+1996}. 
    No previous studies have ever been made on single-pulse behavior or microstructures.
    
    We analyzed the FAST data obtained on 2021-05-01. The observation lasted for  30.0 minutes, and we got 6256 single pulses. We folded data with the period of ${\rm P_0}=0.288~{\rm s}$ with 5840 phase bins per period   (Table~\ref{tab:PSRlist}) and identified 23 spike subpulses. One example from period No.~4111 is shown in Figure~\ref{fig:J0631N4111}, and the measured parameters of 23 spike subpulses are listed in Table~\ref{table:SpikeAllPSR}. 
    The peak-flux distribution 
    has 
    a mean value of $54\pm19$ mJy. 



    \begin{figure}[tb]
    \centering
    \includegraphics[width=0.98\columnwidth]{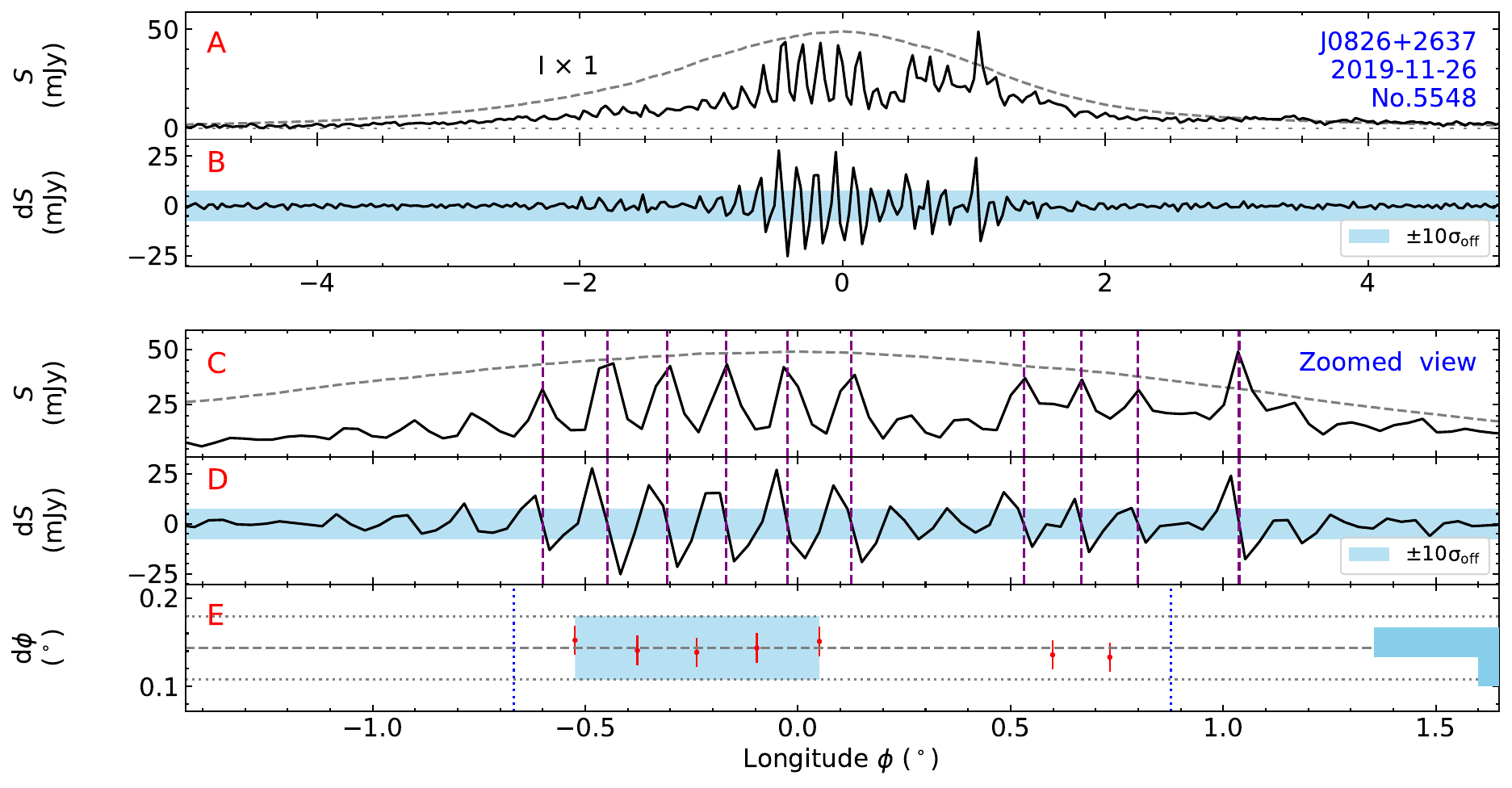}
    \caption{Similar to Figure~\ref{fig:QPSJ1136J2219}, but for the quasi-periodic subpulses in the period number 5548 of PSR J0826+2637 observed on 2019-11-26.}
    \label{fig:QPSJ0826}
    \end{figure}

    \begin{figure*}[tb]
    \centering
    \includegraphics[width=0.600\textwidth]{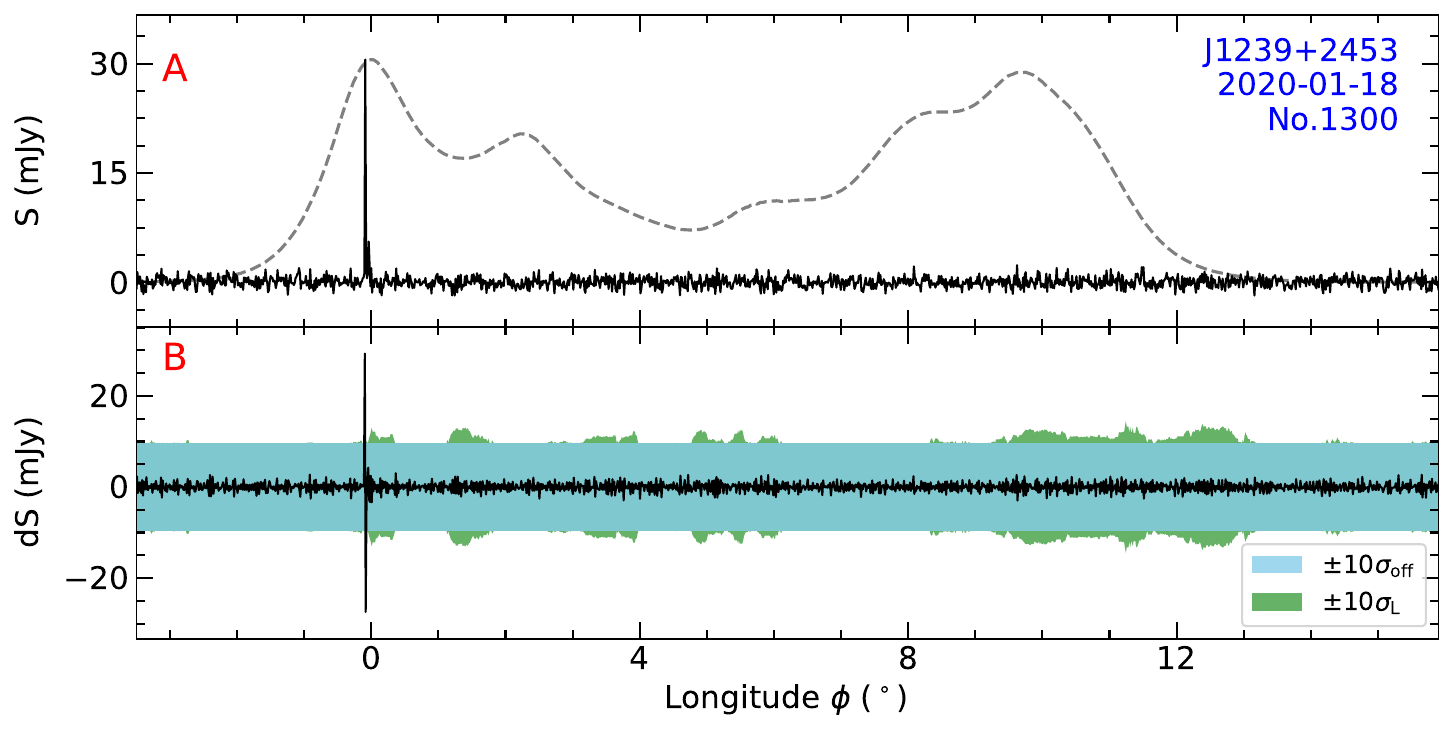}
    \includegraphics[width=0.303\textwidth]{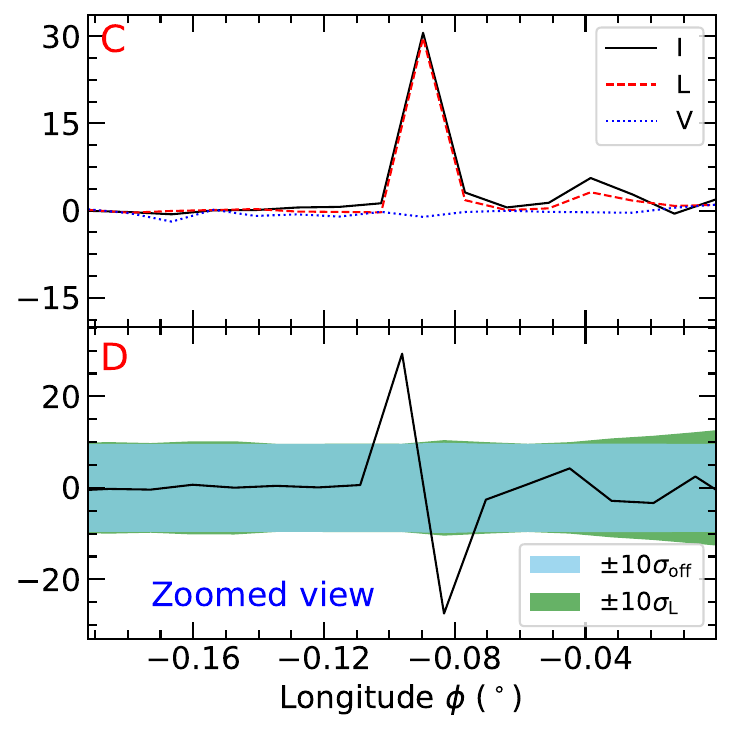}
    \caption{Similar to figure~\ref{fig:J1136N1749}, but for a spike subpulse appearing in the period number 1300 of PSR J1239+2453 observed on 2020-01-18.}
    \label{fig:J1239N1300}
    \end{figure*}

    \subsubsection{J0659+1414}
    \label{subsec:J0659+1414}
    
    The bright pulsar PSR J0659+1414 was discovered in the Second Molonglo Pulsar Survey \citep{mlt+1978}. Strong single pulses were detected 
    at 327~MHz with the Arecibo telescope \citep{wws+2006}. Long-term observations with the Urumqi 25-m radio telescope also detected such strong pulses at 1540~MHz \citep{teh+2012}. Subsequent Arecibo observations with a time resolution of $59.5~\mu{\rm s}$ revealed quasi-periodic microstructures at 1.2~GHz, with a typical period of $420\pm90~\mu{\rm s}$ \citep{mar+2015}.
    
    We analyzed the FAST data obtained on 2023-04-30. The observation lasted for 80.0 minutes and covered 12471 single pulses. We folded data with the period of ${\rm P_0}=0.385~{\rm s}$ with 7824 phase bins per period (Table~\ref{tab:PSRlist}), and identified 1947 spike subpulses. One example from period No.~5411 is shown in Figure~\ref{fig:J0659N5411}, and the measured parameters of 1947 spike subpulses are listed in Table~\ref{table:SpikeAllPSR}. A full list is presented in electronic version. Their peak flux densities follow a log-normal distribution with $ {\rm \mu} (\log S_{\rm peak}) =  1.82$ and $\sigma({\log S_{\rm peak})=0.31}$, corresponding to a flux density of $66^{+69}_{-33}$ mJy. 

    \subsubsection{J0826+2637}
    \label{subsec:J0826+2637}


    PSR J0826+2637 is a bright pulsar discovered by \citet{cls+1968}. 
    It exhibits diverse single-pulse behavior, including multiple emission modes, namely Bright, Quiet, Null, and Q-bright, as well as occasional giant pulses \citep{wes+2006, wse+2007, ysw+2012, syh+2015, bm+2019}. 
    Microstructures were detected with the Effelsberg telescope at 4.85 GHz and a time resolution of $60~\mu{\rm s}$,  
    the microstructure periods were found to span $360$--$660~\mu{\rm s}$ with a peak near $500~\mu{\rm s}$ \citep{lkw+1998}. Arecibo observations at 1.2 GHz and a time resolution of $59.5~\mu{\rm s}$ found a similar microstructure period of $720\pm45~\mu{\rm s}$ \citep{mar+2015}. More recently, \citet{kld+2024} reported a typical microstructure width of ${\rm \tau_\mu}=530^{+53}_{-48}~\mu{\rm s}$ and a quasi-period of ${\rm P_\mu}=660^{+66}_{-60}~\mu{\rm s}$. Using FAST at $\sim115~\mu{\rm s}$ sampling, \citet{LDW+2025} found numerous quasi-periodic microstructures in the main pulse but none in the interpulse, with ${\rm \tau_\mu}=260^{+60}_{-70}~\mu{\rm s}$ and ${\rm P_\mu}=490^{+130}_{-100}~\mu{\rm s}$.
    
    We analyzed the FAST data obtained on 2019-11-26. The observation lasted for  60.0 minutes, covering 6787 single pulses. We folded data with the period of ${\rm P_0}=0.530~{\rm s}$ with 10784 phase bins per period (Table~\ref{tab:PSRlist}), and identified 53 spike subpulses. One example from period No.~4611 is shown in Figure~\ref{fig:J0826N4611}. The measured parameters of 53 spike subpulses are listed in Table~\ref{table:SpikeAllPSR}. Their peak flux densities follow a log-normal distribution with a typical value of $71^{+180}_{-51}$ mJy. 
    We also detected 5 sets of quasi-periodic subpulses. One example shown in Fig.~\ref{fig:QPSJ0826} from the period No.~5548 has a phase quasi-period of $0.133^\circ$, corresponding to 0.196 ms. The parameters of other sets of quasi-periodic subpulses are listed in Table~\ref{table:QPSallPSR}. 

    \begin{figure}[tbh]
    \centering
    \includegraphics[width=0.98\columnwidth]{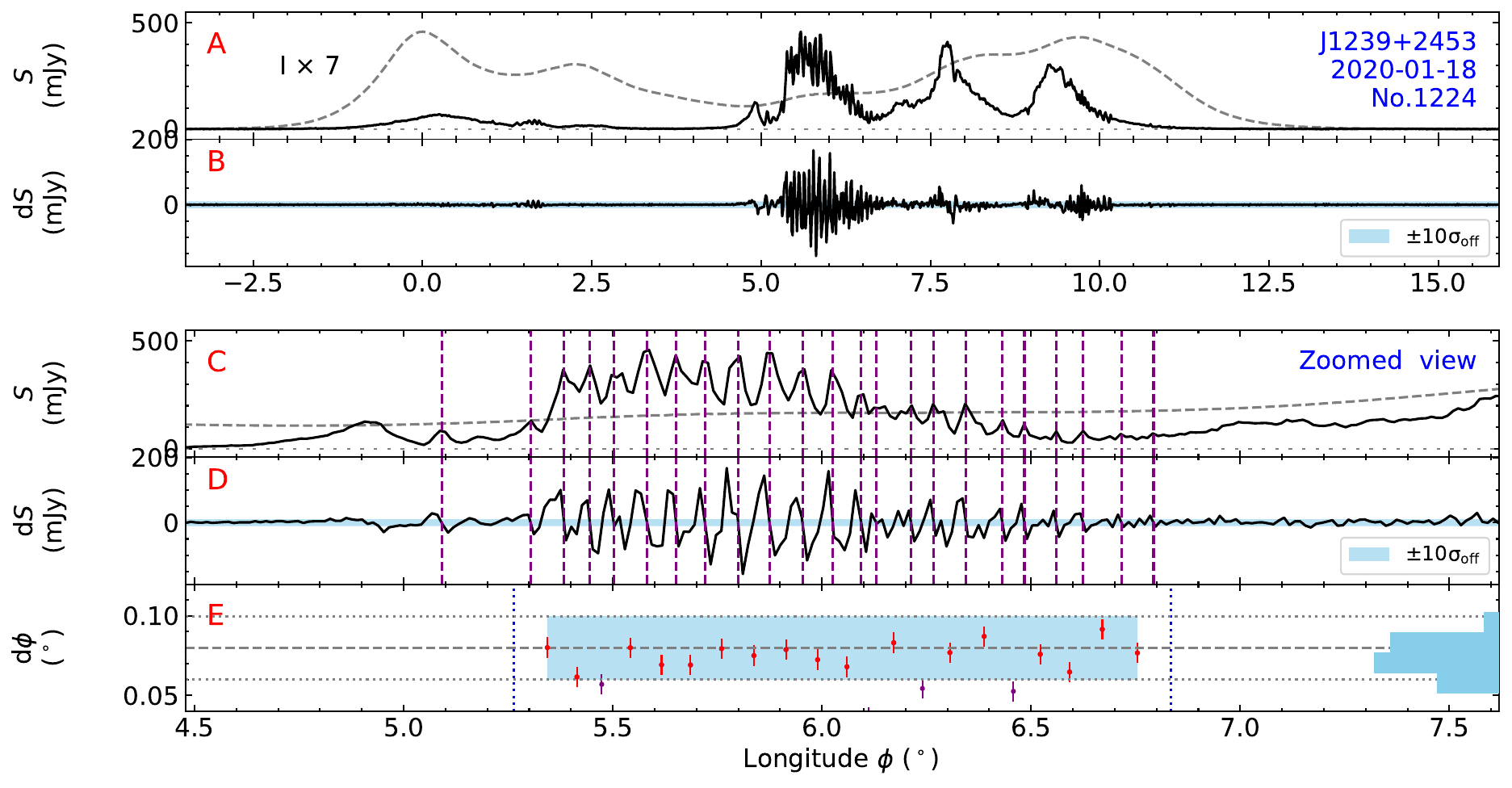}
    \caption{Similar to Figure~\ref{fig:QPSJ1136J2219}, but for the quasi-periodic subpulses in the period number 1224 of PSR J1239+2453 observed on 2020-01-18.}
    \label{fig:QPSJ1239}
    \end{figure}
    
    \begin{figure*}[tb!]
    \centering
    \includegraphics[width=0.600\textwidth]{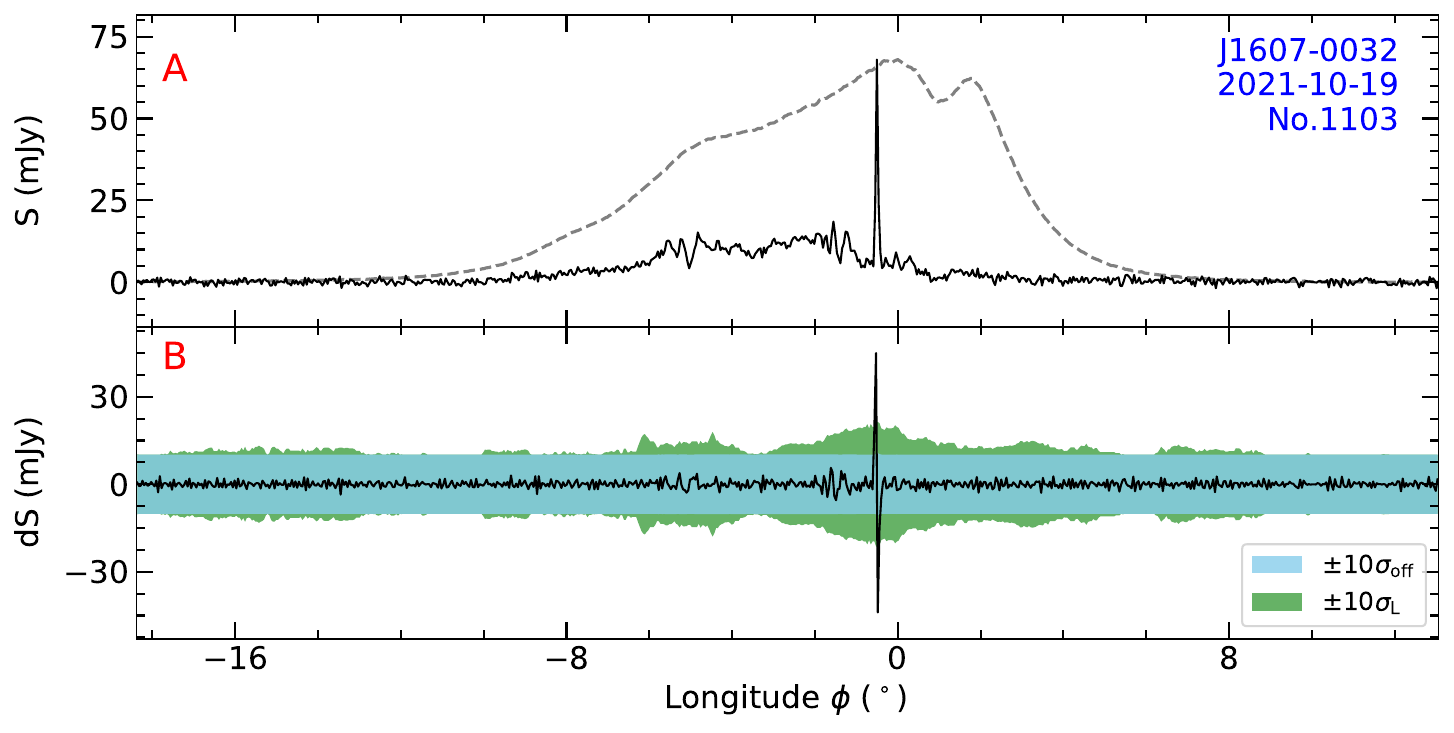}
    \includegraphics[width=0.303\textwidth]{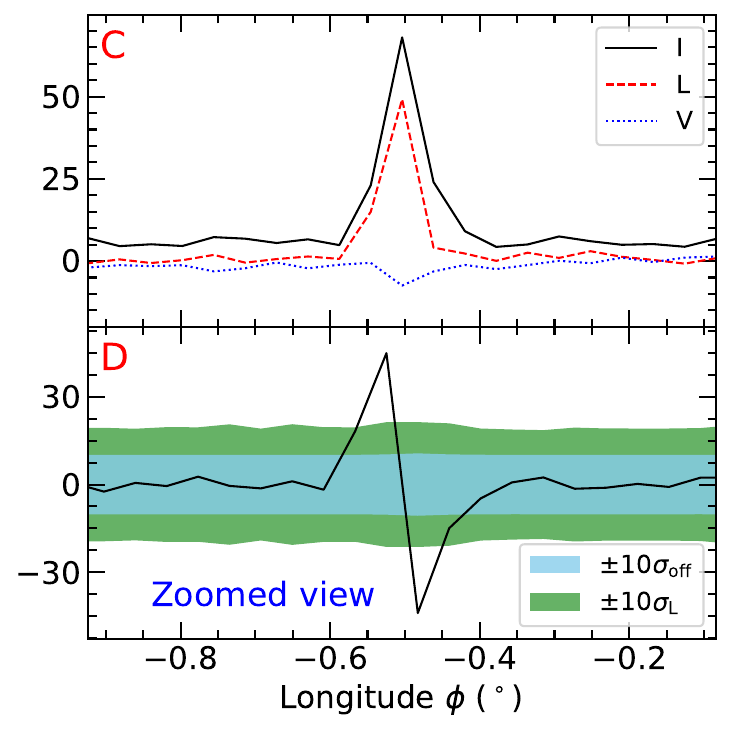}
    \caption{Similar to figure~\ref{fig:J1136N1749}, but for a spike subpulse appearing in the period number 1103 of PSR J1607-0032 observed on 2021-10-19.}
    \label{fig:J1607N1103}
%
    \centering
    \includegraphics[width=0.600\textwidth]{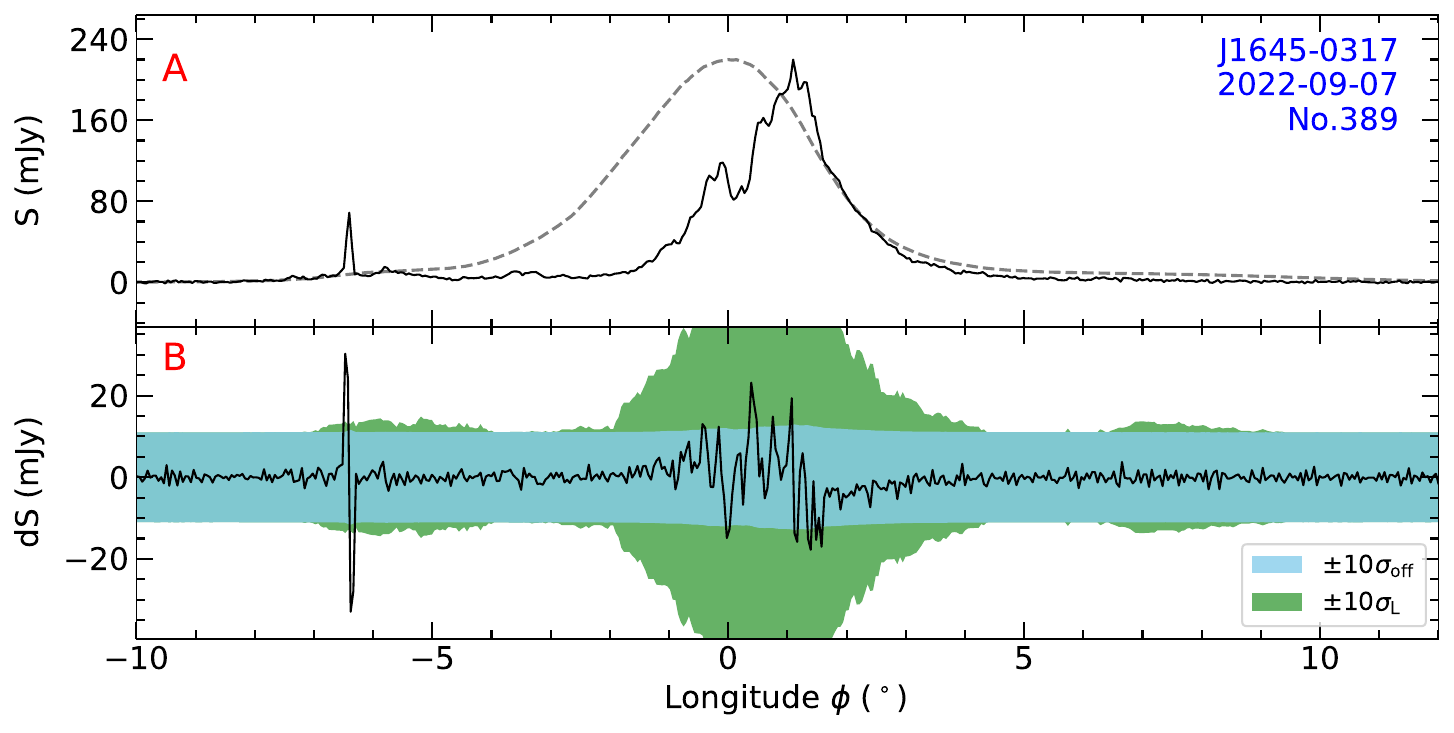}
    \includegraphics[width=0.303\textwidth]{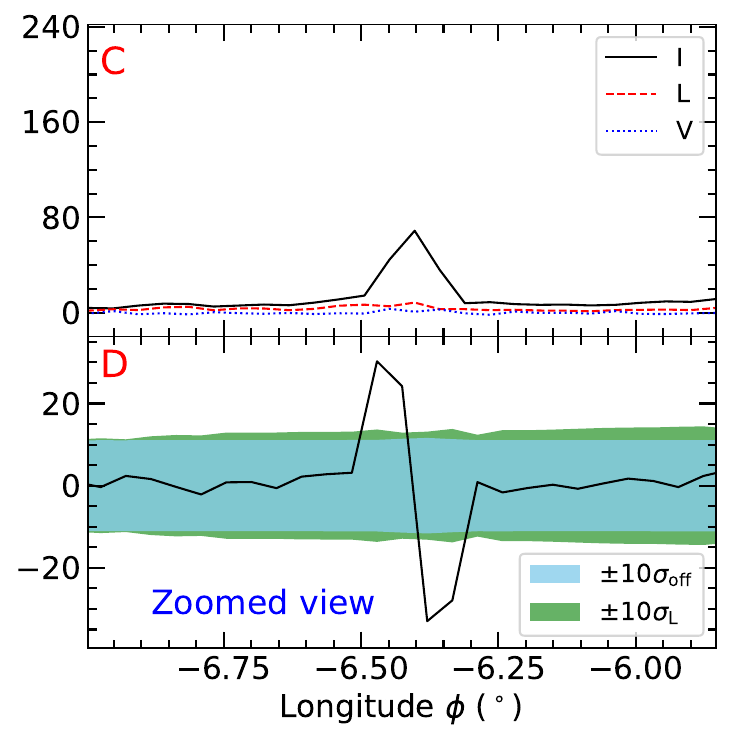}
    \caption{Similar to figure~\ref{fig:J1136N1749}, but for a spike subpulse appearing in the period number 389 of PSR J1645-0317 observed on 2022-09-07.}
    \label{fig:J1645N389}
    \end{figure*}

    \subsubsection{J1136+1551}
    \label{subsec:J1136+1551}
    
   PSR J1136+1551 is one of the first pulsars discovered at 81.5 MHz with a 1 MHz bandwidth \citep{phb+1968}. It shows bright emission at 322 and 4850 MHz \citep{tww+2024, kkg+2003} and periodic nulling at 327 MHz \citep{bgk+2007}. The nulling state is not strictly simultaneous across frequencies according to observations at 341, 626, 1412, and 4850 MHz \citep{kkg+2003}. This pulsar is well known for microstructures first revealed by the Arecibo telescope \citep{Hankins+1972}. Representative widths measured at multiple frequencies are $574~\mu{\rm s}$ at 111 MHz, $651~\mu{\rm s}$ at 196 MHz, and $525~\mu{\rm s}$ at 318 MHz. Microstructures were subsequently confirmed at 430 MHz \citep{ch+1977}, 606 MHz \citep{cwh+1990}, 1.2 GHz \citep{mar+2015}, 1.7 and 2.6 GHz \citep{fs+1978}, and 4.85 GHz \citep{lkw+1998}. 
   Micro-pulses were detected from both main components of the mean pulse profile, though with different properties \citep{ch+1977, pbc+2002, Popov2024}. High-sensitivity Arecibo observations at 1.2 GHz with a time resolution of $59.5~\mu{\rm s}$ revealed many quasi-periodic microstructures, with typical quasi-periods of $830 \pm 440~\mu{\rm s}$ in the leading component and $950 \pm 620~\mu{\rm s}$ in the trailing component. Individual solitons within a quasi-periodic train display diverse polarization properties \citep{fs+1978}. More recently, \citet{kld+2024} estimated a typical microstructure width of $340^{+204}_{-128}~\mu{\rm s}$ and a quasi-period of $780^{+312}_{-223}~\mu{\rm s}$.

    We use PSR J1136+1551 as a representative example, 
    as presented in the previous subsections. FAST observed PSR J1136+1551 on 2019-11-21 for 60.0 minutes, covering 3032 periods. We folded data with the period of ${\rm P_0}=1.187~{\rm s}$ with 24128 phase bins per period, and identified 78 spike subpulses, listed in Table~\ref{table:78spikeJ1136}, and also found 24 sets of quasi-periodic subpulses, as listed in Table~\ref{table:QPSJ1136J2219}.

    \begin{figure}[tbh]
    \centering
    \includegraphics[width=0.98\columnwidth]{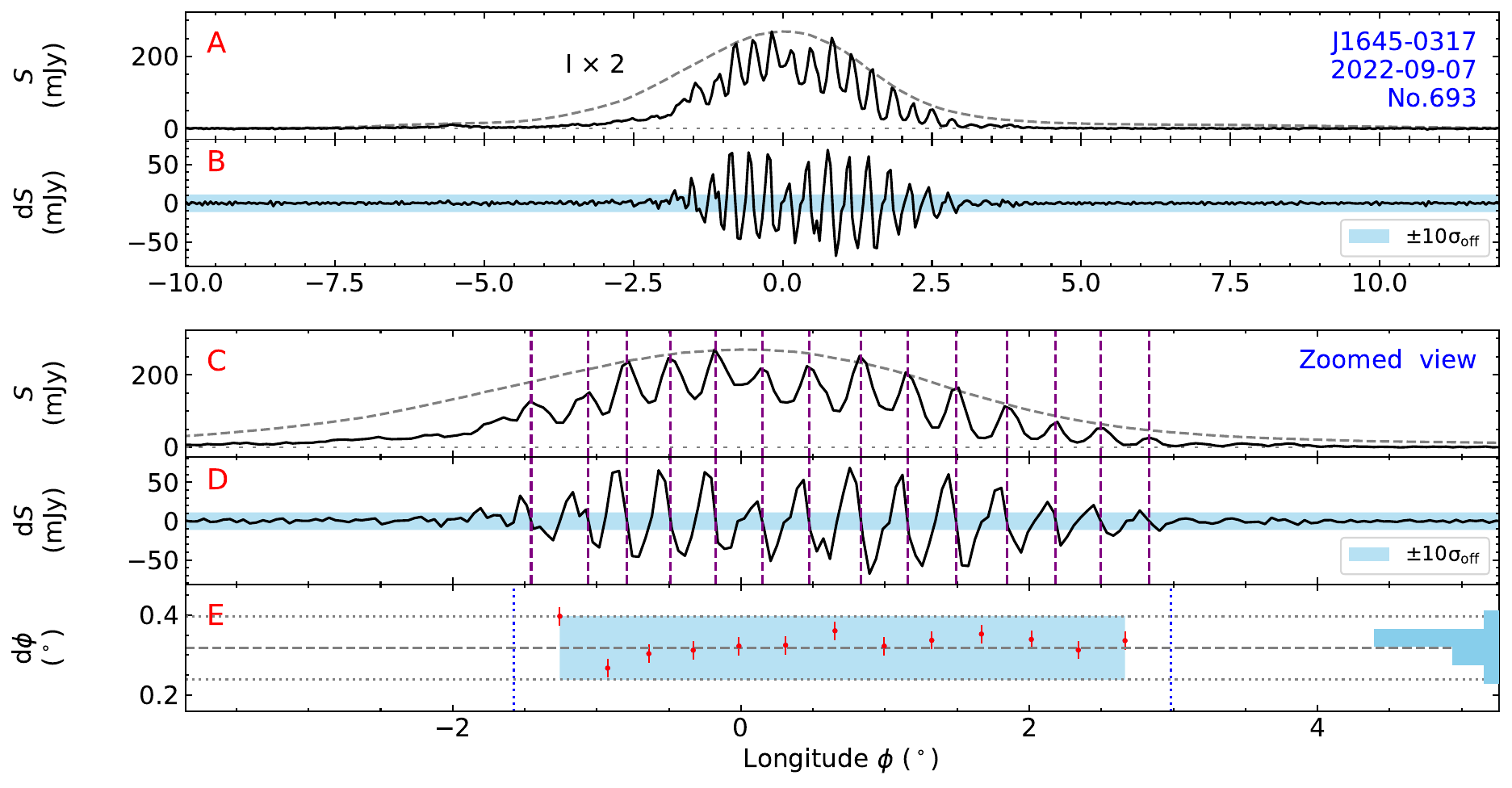}
    \caption{Similar to Figure~\ref{fig:QPSJ1136J2219}, but for the quasi-periodic subpulses in the period number 693 of PSR J1645-0317 observed on 2022-09-07.}
    \label{fig:QPSJ1645}
    \end{figure}

    \begin{figure*}[tbh]
    \centering
    \includegraphics[width=0.600\textwidth]{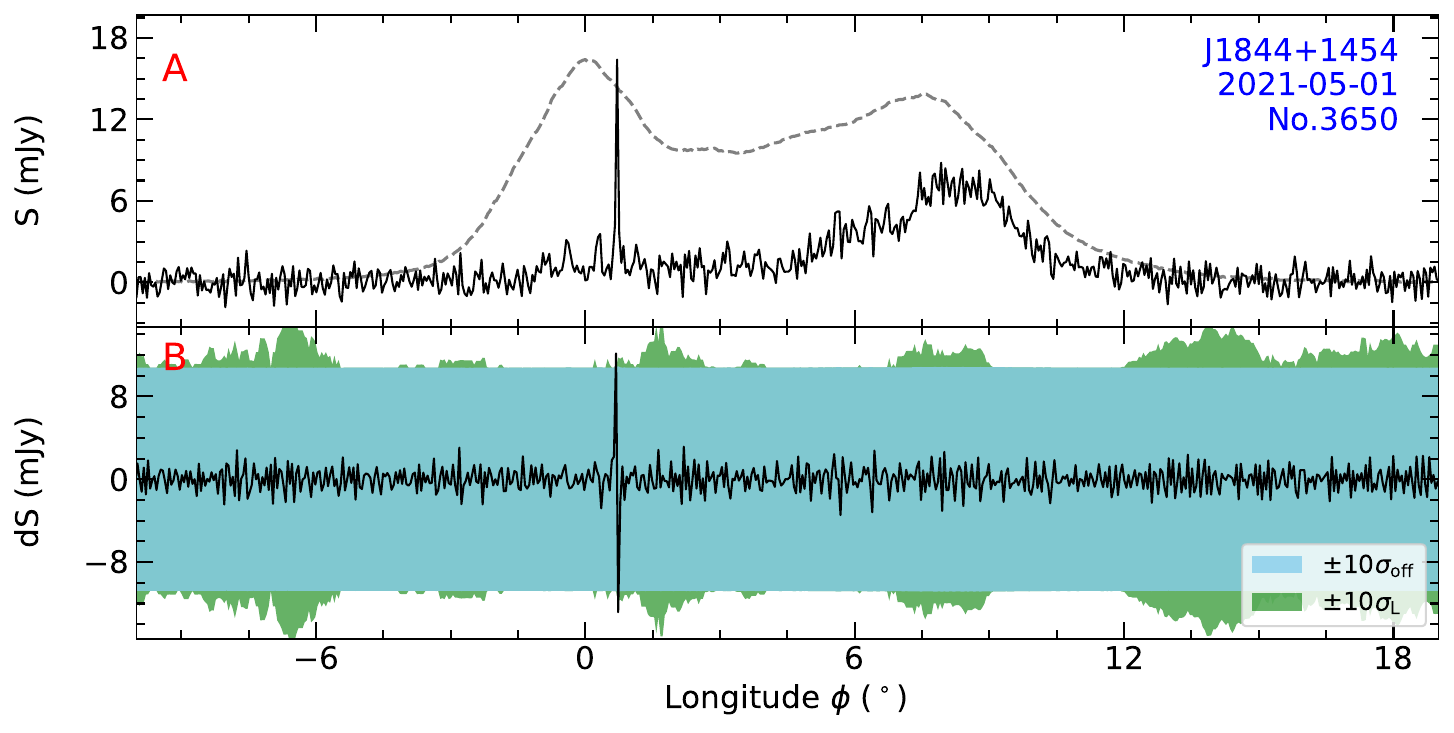}
    \includegraphics[width=0.303\textwidth]{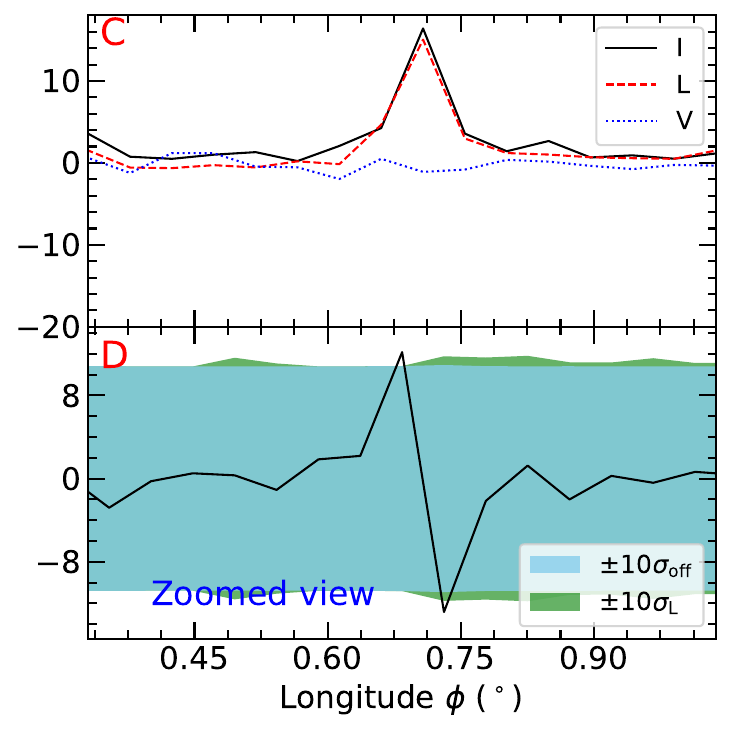}
    \caption{Similar to figure~\ref{fig:J1136N1749}, but for a spike subpulse appearing in the period number 3650 of PSR J1844+1454 observed on 2021-05-01.}
    \label{fig:J1844N3650}
    \end{figure*}

    \begin{figure}[tbh]
    \centering
    \includegraphics[width=0.98\columnwidth]{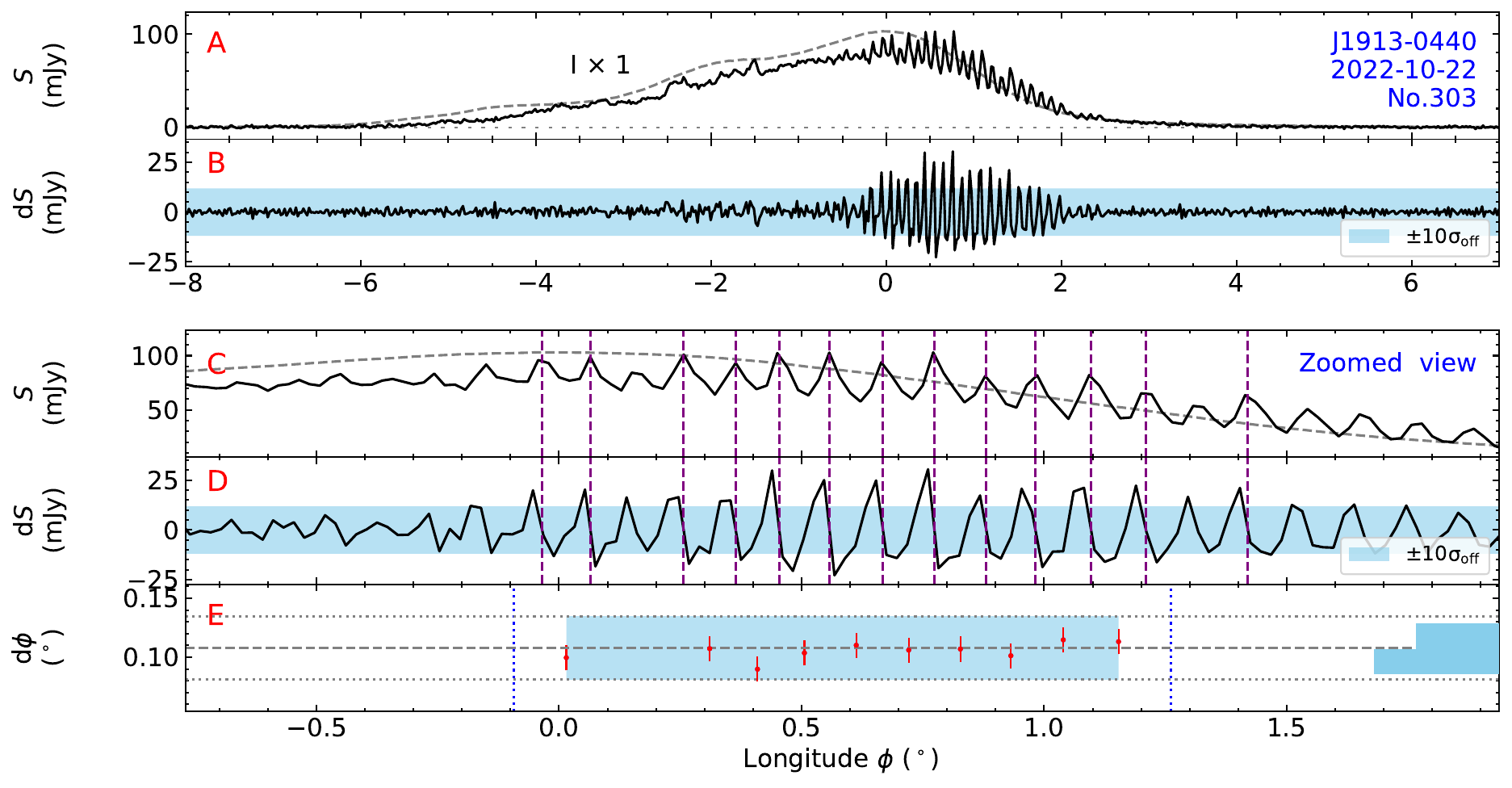}
    \caption{Similar to Figure~\ref{fig:QPSJ1136J2219}, but for the quasi-periodic subpulses in the period number 303 of PSR J1913-0440 observed on 2022-10-22.}
    \label{fig:QPSJ1913}
    \end{figure}
    
    \subsubsection{J1239+2453}
    \label{subsec:J1239+2453}

    

    PSR J1239+2453 was discovered at 318 MHz with the Arecibo telescope \citep{lang+1969}. It shows mode changing, subpulse drifting \citep{sr+2005, njm+2017}, and pulse nulling with a fraction of a few percent of all periods \citep{Backer+1970, r+1976, wes+2006, njm+2017, bmm+2017, whh+2020}. With the high sensitivity of FAST, numerous dwarf pulses have been identified during null states \citep{yhz24}. Arecibo observations with $69.5~\mu{\rm s}$ sampling revealed quasi-periodic microstructures \citep{mar+2015}, with different quasi-periods of $2.16(1.16)$~ms and $1.26(0.45)$~ms for two dominant profile components. More recently, \citet{kld+2024} got the quasi-period of ${\rm P_\mu}=1.63^{+0.65}_{-0.46}$~ms.
    
    We analyzed the FAST data obtained on 2020-01-18. The observation lasted for  120 minutes, covering 5210 single pulses. We folded data with the period of ${\rm P_0}=1.38~{\rm s}$ with 28096 phase bins per period (Table~\ref{tab:PSRlist}), and identified 11 spike subpulses. One example from period No.~1300 is shown in Figure~\ref{fig:J1239N1300}. The measured parameters of 11 spike subpulses are listed in Table~\ref{table:SpikeAllPSR}. 
    We also detected 3 sets of quasi-periodic subpulses. One example is shown in Figure~\ref{fig:QPSJ1239}. The quasi-periodic subpulses in period No.~1224 have a phase quasi-period of $0.0768^\circ$, corresponding to 0.295 ms. The parameters of other sets are listed in Table~\ref{table:QPSallPSR}. 

    \subsubsection{J1607-0032}
    \label{subsec:J1607-0032}


    
    PSR J1607-0032 was discovered by the Molonglo radio telescope \citep{vl+1970}. It exhibits an ``emission-shift'' behaviour that is distinct from subpulse drifting and mode changing \citep{Rankin1988, rrw+2006}.
    
    We analyzed the FAST data obtained on 2021-10-19. The observation lasted for  58.0 minutes, covering 8246 periods. We folded data with the period of ${\rm P_0}=0.42~{\rm s}$ with 8576 phase bins per period (Table~\ref{tab:PSRlist}), and
   identified 29 spike subpulses. One example is shown in Figure~\ref{fig:J1607N1103}. The measured parameters of 29 spike subpulses are listed in Table~\ref{table:SpikeAllPSR}. Their peak flux densities follow a log-normal distribution with $ {\rm \mu} (\log S_{\rm peak}) =  1.55$ and $\sigma({\log S_{\rm peak})=0.21}$, corresponding to a flux density of $35^{+22}_{-13}$ mJy. 

    \subsubsection{J1645-0317}
    \label{subsec:J1645-0317}
        
    PSR J1645-0317 was discovered in 1969 by G.R. Huguenin \& J.H. Taylor \citep{tml+1993}. Using the Giant Metrewave Radio Telescope with a time resolution of $\sim15~\mu{\rm s}$, \citet{sgd+2024} identified quasi-periodic modulations in some subpulses and estimated a typical quasi-period of $\sim180~\mu{\rm s}$.
    
    We analyzed the FAST data obtained on 2022-09-07. The observation lasted for 10.2 minutes, covering 1577 periods. We folded data with the period of ${\rm P_0}=0.388~{\rm s}$ with 7872 phase bins per period (Table~\ref{tab:PSRlist}), and identified one spike subpulse, shown in Figure~\ref{fig:J1645N389}, with its parameters listed in Table~\ref{table:SpikeAllPSR}, and 3 sets of quasi-periodic subpulses. One example from the period No.~693 is shown in Figure~\ref{fig:QPSJ1645}, which has a phase quasi-period of $0.319^\circ$, corresponding to 0.344 ms. The parameters of 3 sets of quasi-periodic subpulses are listed in Table~\ref{table:QPSallPSR}. 

    \subsubsection{J1844+1454}
    \label{subsec:J1844+1454}
         
    PSR J1844+1454 was discovered in the second Molonglo pulsar survey 
    \citep{mlt+1978}. 
    
    We analyzed the FAST data obtained on 2021-05-01. The observation lasted for 30.0 minutes, covering 4797 periods. We folded data with the period of ${\rm P_0}=0.375~{\rm s}$ with 7632 phase bins per period (Table~\ref{tab:PSRlist}), and identified two spike subpulses. One example, from period No.~3650, is shown in Figure~\ref{fig:J1844N3650}. The measured parameters of the two spike subpulses are listed in Table~\ref{table:SpikeAllPSR}. 

    \begin{figure*}[tbh]
    \centering
    \includegraphics[width=0.600\textwidth]{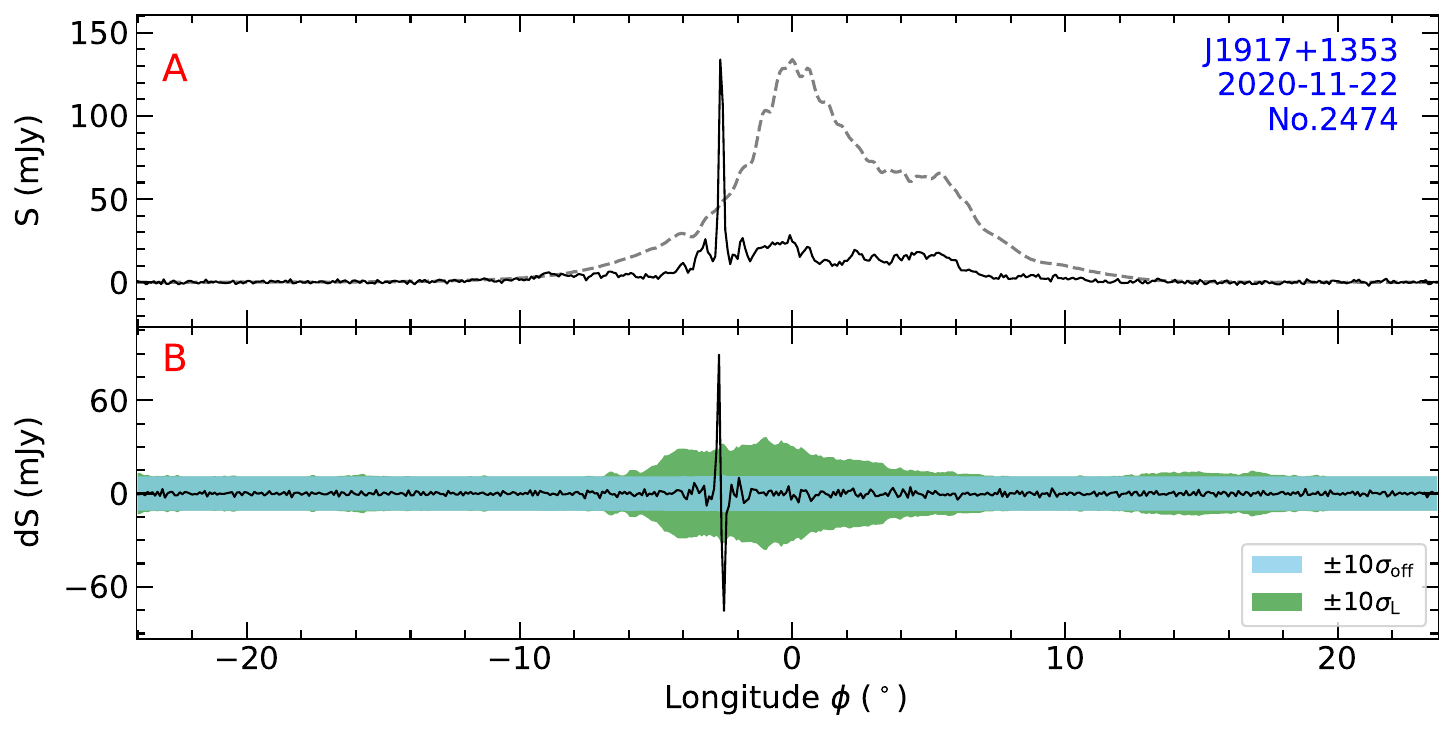}
    \includegraphics[width=0.303\textwidth]{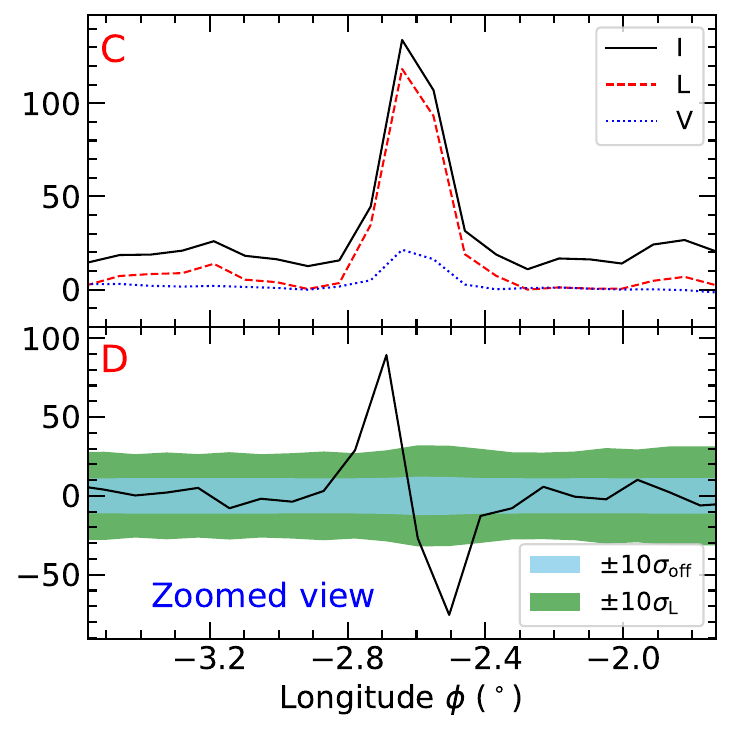}
    \caption{Similar to figure~\ref{fig:J1136N1749}, but for a spike subpulse appearing in the period number 2474 of PSR J1917+1353 observed on 2020-11-22.}
    \label{fig:J1917N2474}
    \end{figure*}
    
    \begin{figure}[tbh]
    \centering
    \includegraphics[width=0.98\columnwidth]{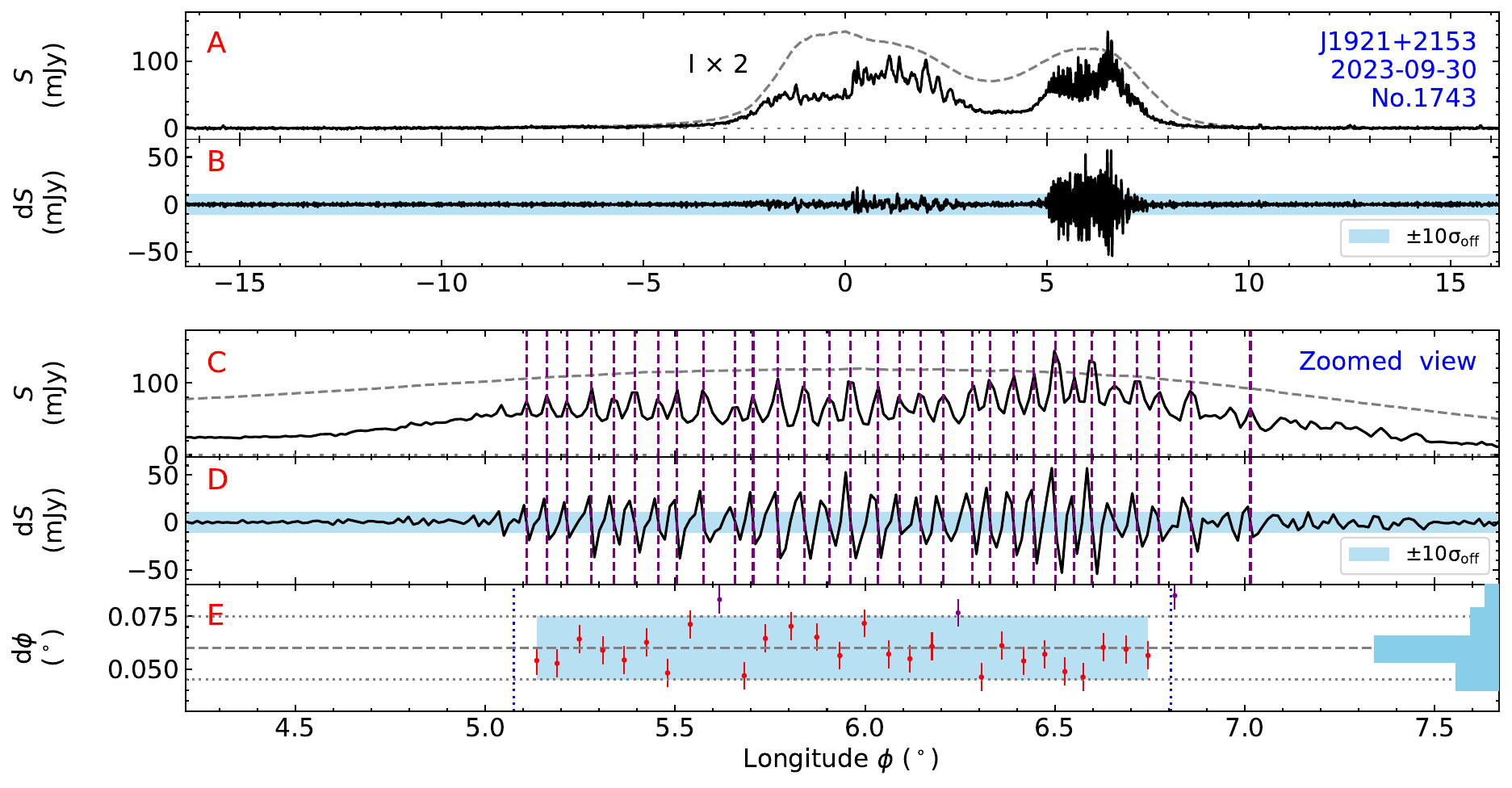}
    \caption{Similar to Figure~\ref{fig:QPSJ1136J2219}, but for the quasi-periodic subpulses in the period number 1743 of PSR J1921+2153 observed on 2023-09-30.}
    \label{fig:QPSJ1921+2153}
    \end{figure}
    
    \subsubsection{J1913-0440}
    \label{subsec:J1913-0440}
        
    PSR J1913-0440 was discovered in a Molonglo pulsar search at 408 MHz \citep{lvw+1969b}. 
    
    We analyzed the FAST data obtained on 2022-10-22. The observation lasted for 23.8 minutes, covering 1729 periods. We folded data with the period of ${\rm P_0}=0.826~{\rm s}$ with 16800 phase bins per period (Table~\ref{tab:PSRlist}).
    No spike subpulse was identified, but we detected one set of quasi-periodic subpulses in period No.~303, shown in Figure~\ref{fig:QPSJ1913}, with a phase quasi-period of $0.107^\circ$, corresponding to 0.246 ms. The measured parameters are listed in Table~\ref{table:QPSallPSR}.

    \subsubsection{J1917+1353}
    \label{subsec:J1917+1353}
        
    PSR J1917+1353 is a bright pulsar discovered by \citet{smb+1971}. 
    
    We analyzed the FAST data obtained on 2020-11-22. The observation lasted for 10.0 minutes, covering 3075 periods. We folded data with the period of ${\rm P_0}=0.194~{\rm s}$ with 3952 phase bins per period (Table~\ref{tab:PSRlist}), and identified only one spike subpulse from period No.2474, shown in Figure~\ref{fig:J1917N2474}, and its parameters are listed in Table~\ref{table:SpikeAllPSR}. Its peak flux density reaches 110.3 mJy.
        
    \subsubsection{J1921+2153}
    \label{subsec:J1921+2153}

    
    It is the first pulsar discovered by \citet{Hewish1968}. It exhibits subpulse drifting \citep{cordes1975,pw+1986,ptv+2022}. Using the Arecibo observations at 1.2 GHz with a time resolution of 59.5~$\mu$s, \citet{mar+2015} reported quasi-periodic microstructures, distributed across the two dominant emission components of the mean pulse profile. The leading and trailing components showed different typical microstructure periods of $1.31\pm0.62$~ms and $1.55\pm1.15$~ms, respectively. More recently, \citet{kld+2024} estimated a typical microstructure width of $1.30^{+1.20}_{-0.65}$~ms and a quasi-period of $1.54^{+0.92}_{-0.58}$~ms.
    
    We analyzed the FAST data obtained on 2023-09-30. The observation lasted for 48.9 minutes, covering 2195 periods. We folded data with the period  of ${\rm P_0}=1.3373~{\rm s}$ with 27200 phase bins per period (Table~\ref{tab:PSRlist}).
    No spike subpulse is identified in these data. However, we detected 8 sets of quasi-periodic subpulses. One example is shown in Figure~\ref{fig:QPSJ1921+2153} with a phase quasi-period of $0.053^\circ$, corresponding to 0.197 ms. The parameters of these 8 sets of quasi-periodic subpulses are listed in Table~\ref{table:QPSallPSR}. 

    \begin{figure*}[tb!]
    \centering
    \includegraphics[width=0.658\textwidth]{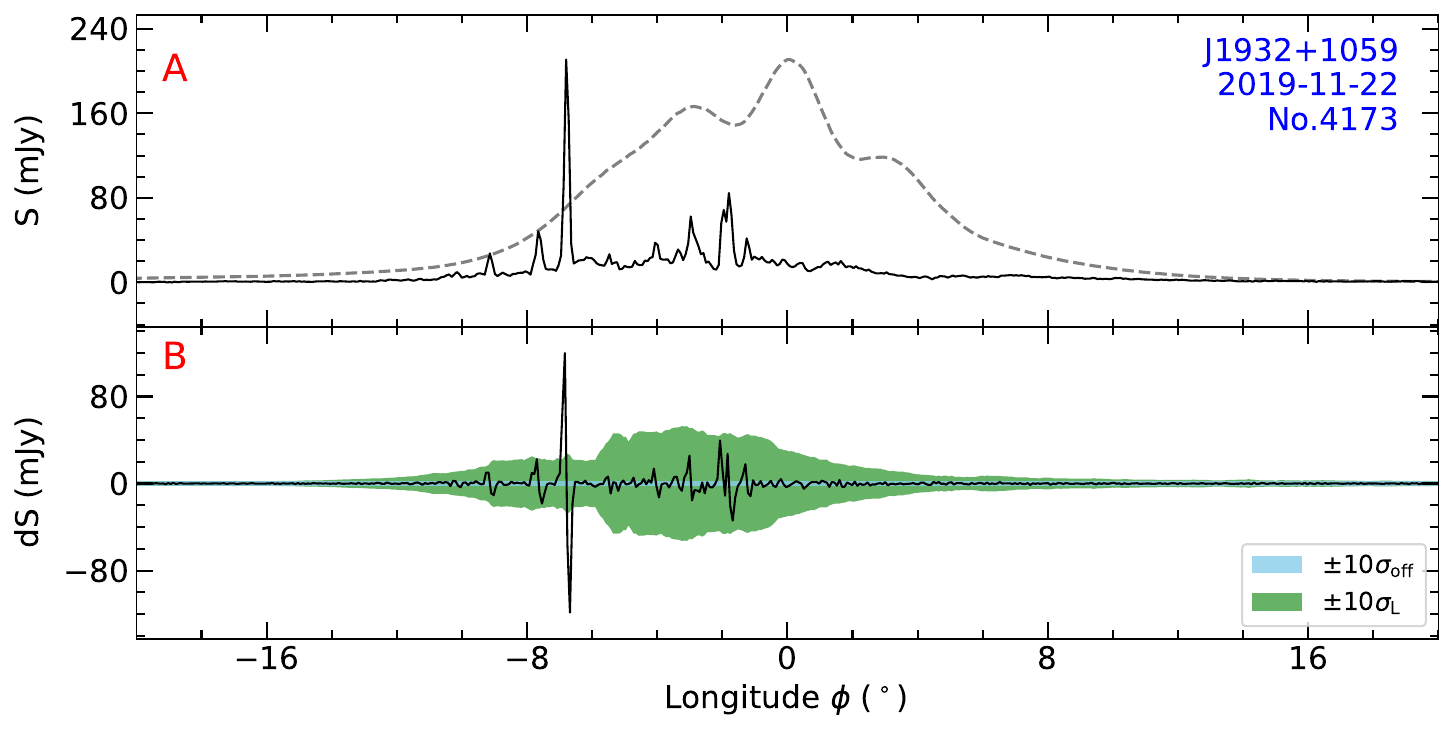}
    \includegraphics[width=0.330\textwidth]{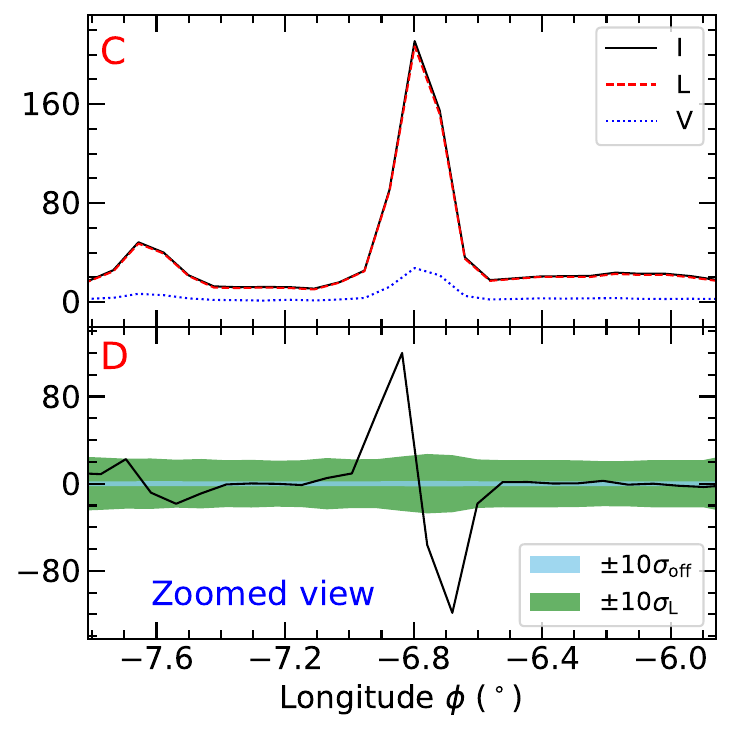}
    \caption{Similar to figure~\ref{fig:J1136N1749}, but for a spike subpulse appearing in the period number 4173 of PSR J1932+1059 observed on 2019-11-22.}
    \label{fig:J1932N4173}
    \end{figure*}
    
    
    \begin{figure}[tb!]
    \centering
    \includegraphics[width=0.98\columnwidth]{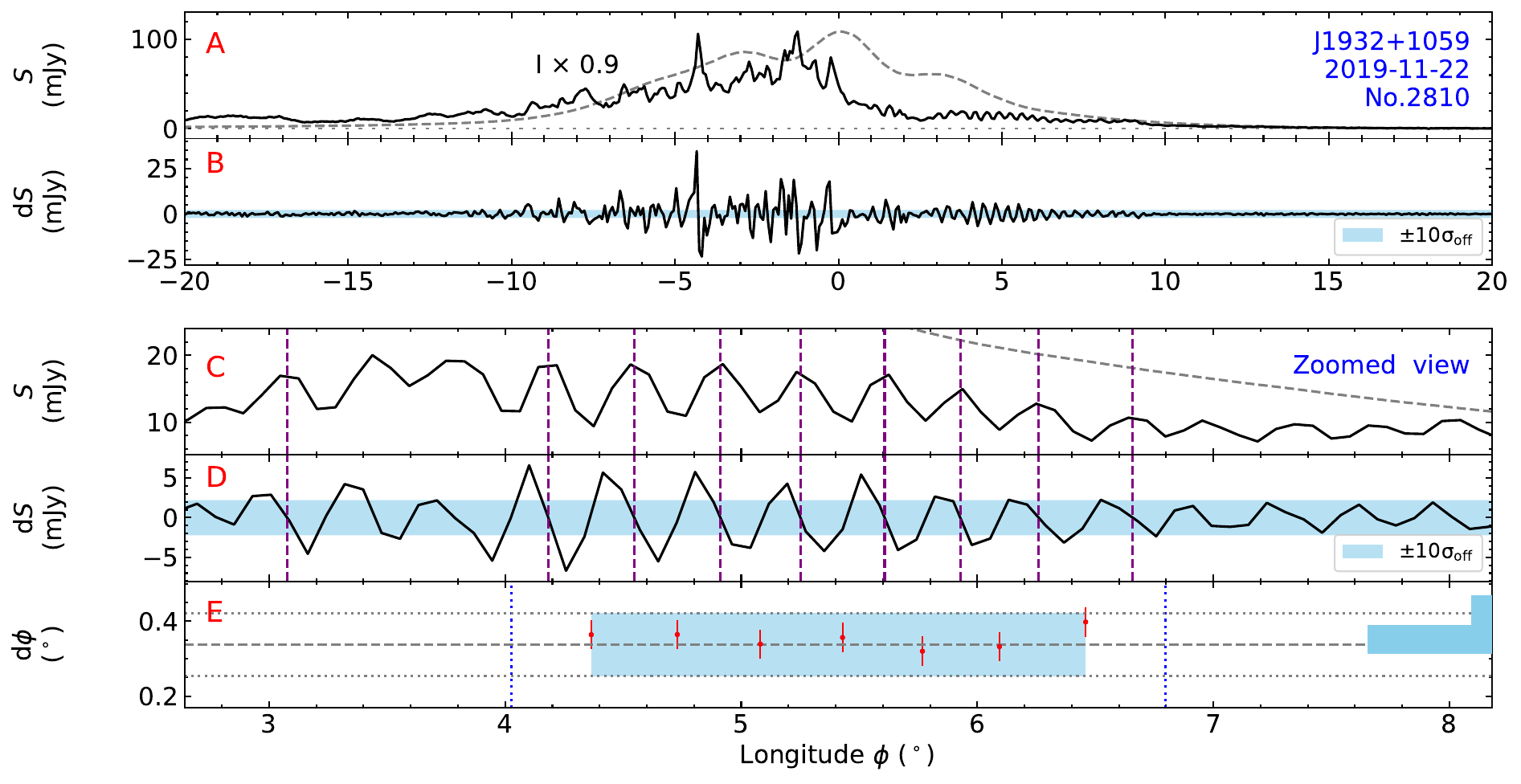}
    \caption{Similar to Figure~\ref{fig:QPSJ1136J2219}, but for the quasi-periodic subpulses in the period number 2810 of PSR J1932+1059 observed on 2019-11-22.}
    \label{fig:QPSJ1932}
    \end{figure}
    
    \begin{figure}[tb!]
    \centering
    \includegraphics[width=0.8\columnwidth]{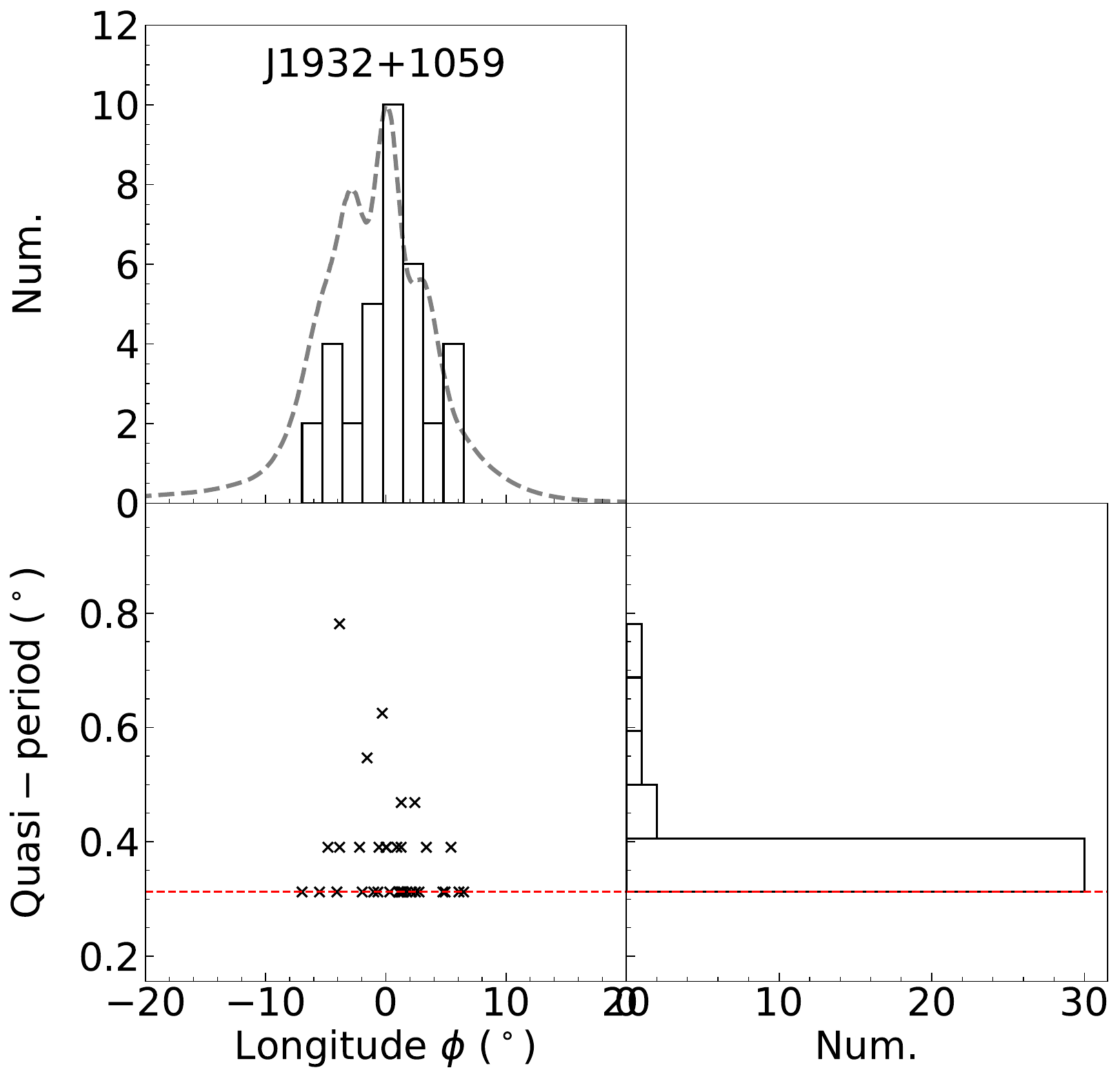}
    \caption{Similar to figure~\ref{fig:QPSJ1136distr}, but for the 35 sets of quasi-periodic subpulses of PSR J1932+1059 observed on 2019-11-22.}
    \label{fig:QPSJ1932distr}
    \end{figure}
    
    \begin{figure}[tb!]
    \centering
    \includegraphics[width=0.98\columnwidth]{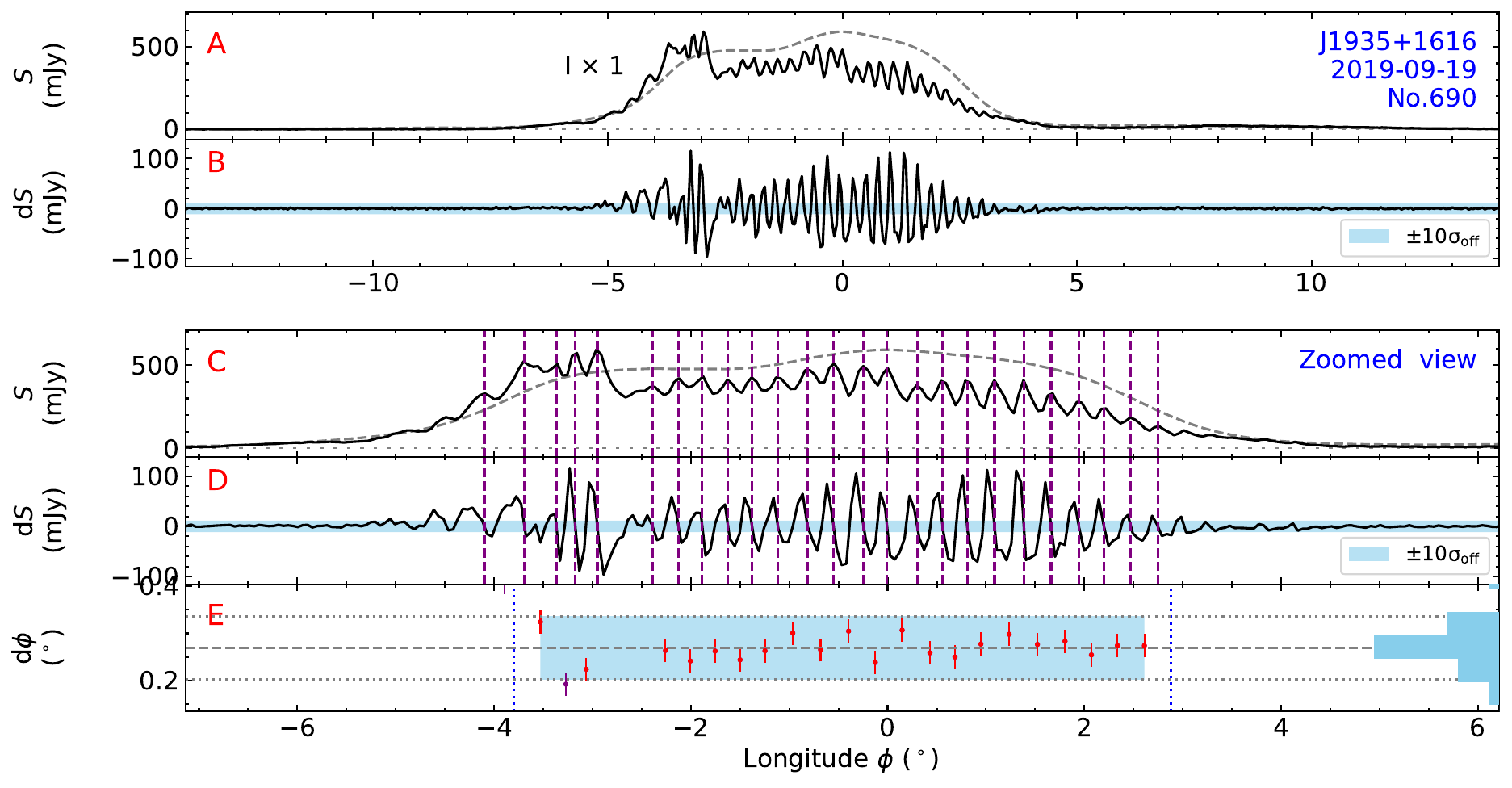}
    \caption{Similar to Figure~\ref{fig:QPSJ1136J2219}, but for the quasi-periodic subpulses in the period number 690 of PSR J1935+1616 observed on 2019-09-19.}
    \label{fig:QPSJ1935}
    \end{figure}
          
    \begin{figure}[tb!]
    \centering
    \includegraphics[width=0.8\columnwidth]{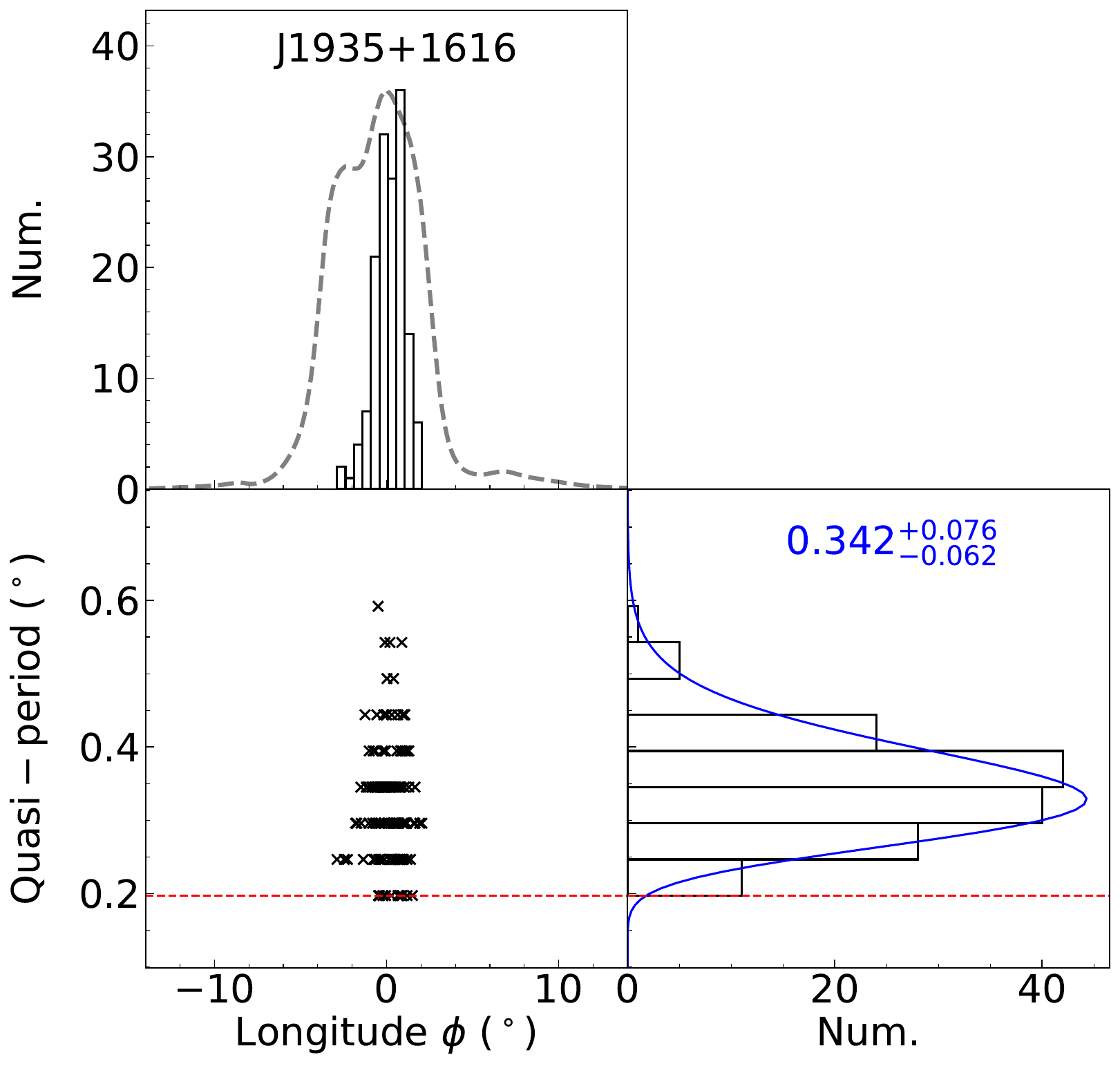}
    \caption{Similar to figure~\ref{fig:QPSJ1136distr}, but for the 151 sets of quasi-periodic subpulses of PSR J1935+1616 observed on 2019-09-19.}
    \label{fig:QPSJ1935distr}
    \end{figure}
    
    \subsubsection{J1932+1059}
    \label{subsec:J1932+1059}
    
    PSR J1932+1059 is a very bright pulsar discovered by \citep{Large1968}. It exhibits nulling with a low null fraction of less than 1\% \citep{r+1976,whh+2020} and diverse subpulse-drifting behaviour, including two modulation periods of ${\rm 11P_0}$ and ${\rm 5.6P_0}$ \citep{backer+1973,rr+1997}. 
    Sensitive FAST observations revealed periodic and phase-locked flux modulation for both the main pulses and interpulses with a period of ${\rm 12P_0}$ \citep{kyp+2021}. About 50\% of single pulses exhibit microstructures in Effelsberg 4.85~GHz data \citep{lkw+1998}, and 
    97\% at 1.65 GHz \citep{pbc+2002}. Reported typical microstructure width and quasi-period are $95\pm10~\mu{\rm s}$ \citep{pbc+2002} and $480\pm180~\mu{\rm s}$ \citep{mar+2015}. \citet{kld+2024} estimated a typical width of microstructures of $120^{+36}_{-28}~\mu{\rm s}$ and a quasi-period of $420^{+294}_{-173}~\mu{\rm s}$.
    
    We analyzed the FAST data obtained on 2019-11-22. The observation lasted for 60 minutes, covering 15896 periods. We folded data with the period  of ${\rm P_0}=0.2265~{\rm s}$ with 4608 phase bins per period (Table~\ref{tab:PSRlist}), and identified 70 spike subpulses. One example, from period No.~4173, is shown in Figure~\ref{fig:J1932N4173}. The measured parameters of 70 spike subpulses are listed in Table~\ref{table:SpikeAllPSR}. 
    Although these spike subpulses are superimposed on strong and gradual pulsed emission, careful examinations of the individual pulses revealed that most of them are nearly completely linearly polarized. In addition, we detected 35 sets of quasi-periodic subpulses. One example from the period No.~2810 is shown in Figure~\ref{fig:QPSJ1932}, with a phase quasi-period of $0.391^\circ$, corresponding to 0.246 ms. The parameters of 35 sets of  quasi-periodic subpulses are listed in Table~\ref{table:QPSallPSR}. As shown in Figure~\ref{fig:QPSJ1932distr}, the mean pulse profile of this pulsar contains multiple emission components, and the quasi-periodic subpulses emerge at any phase across the profile. 
    The mean phase quasi-period is $0.373\pm0.101^\circ$, corresponding to $0.235\pm0.064$ ms.

    \begin{figure*}[tb!]
    \centering
    \includegraphics[width=0.600\textwidth]{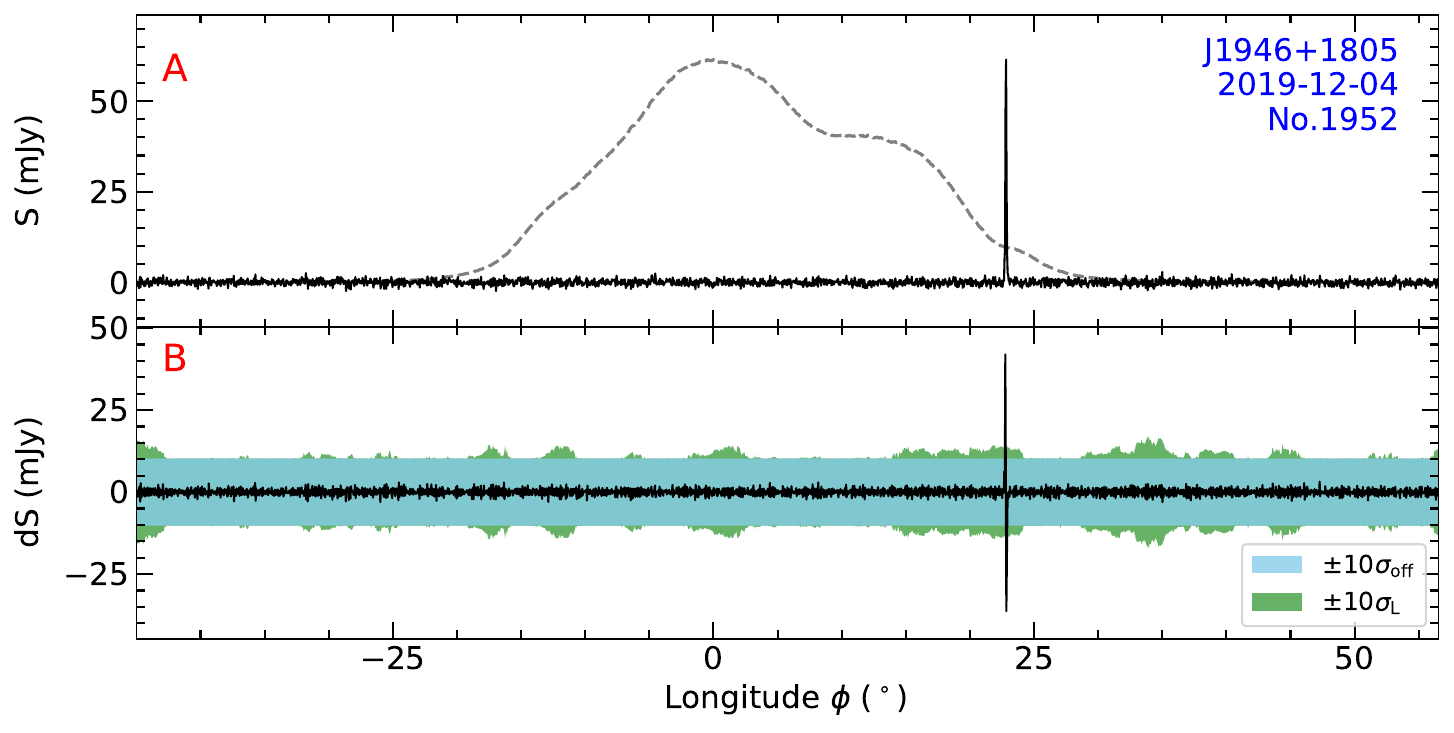}
    \includegraphics[width=0.303\textwidth]{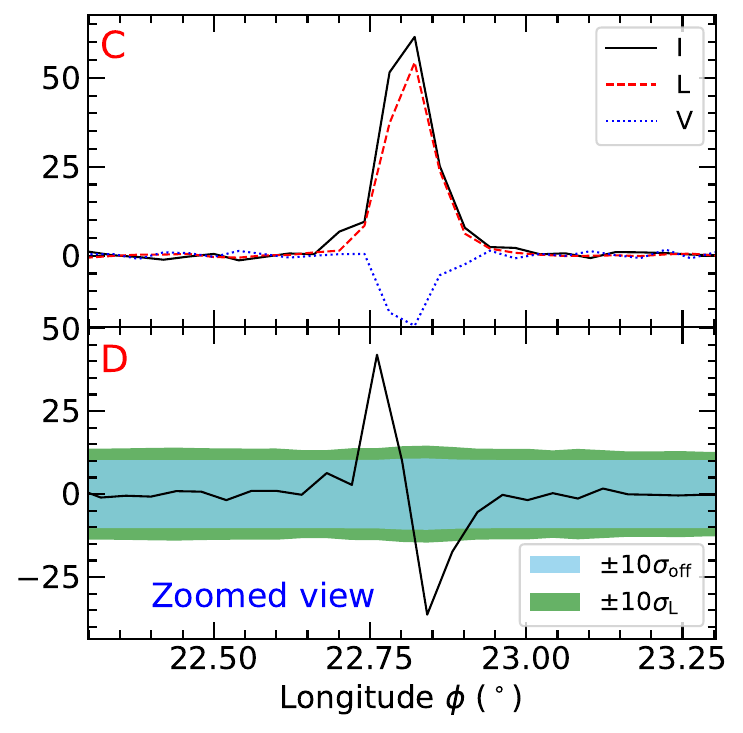}
    \caption{Similar to figure~\ref{fig:J1136N1749}, but for a spike  subpulse from the period No.~1952 of PSR J1946+1805 observed on 2019-12-04.}
    \label{fig:J1946N1952}
    \end{figure*}

    \begin{figure}[tb!]
    \centering
    \includegraphics[width=0.98\columnwidth]{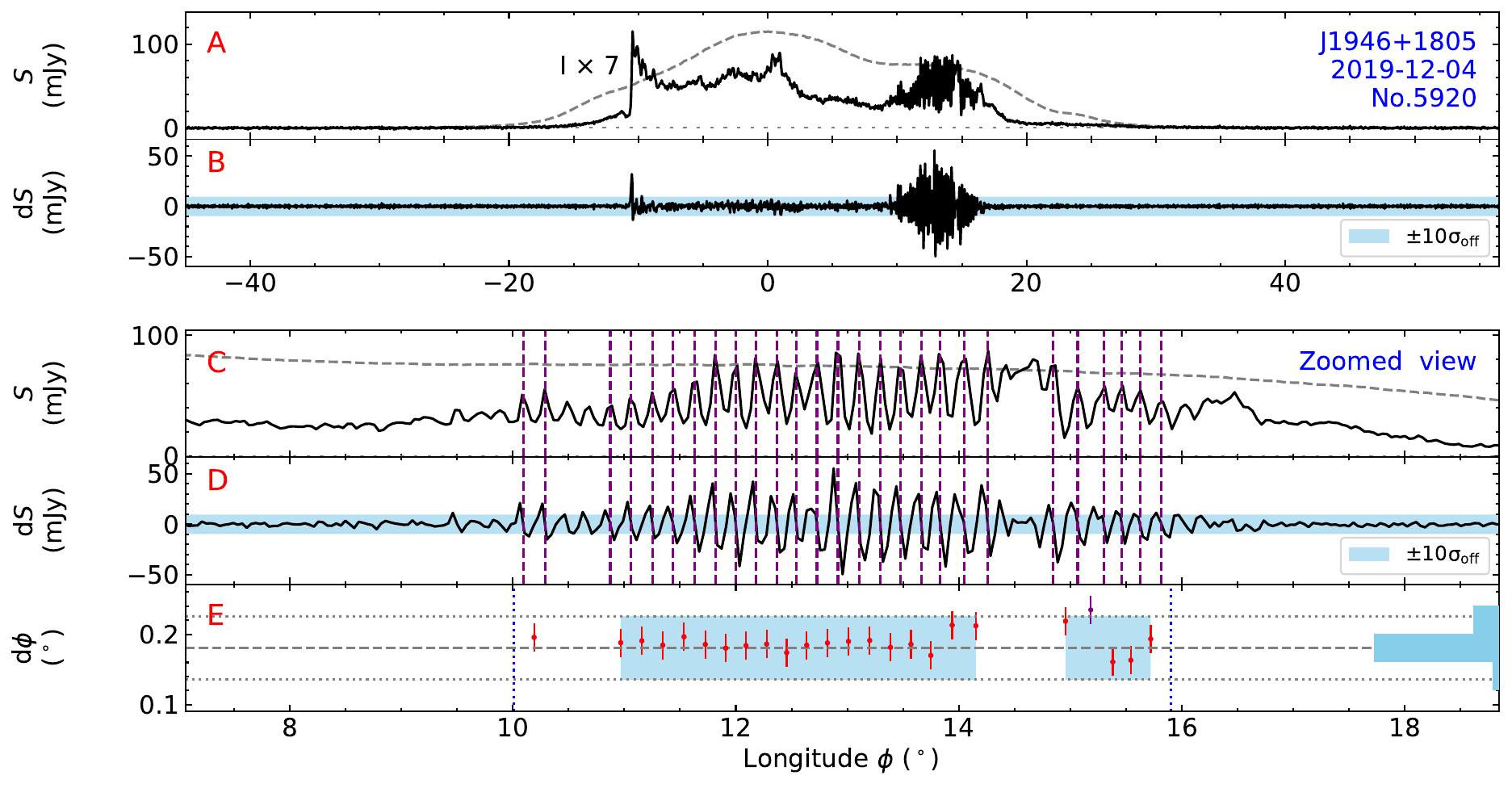}
    \caption{Similar to Figure~\ref{fig:QPSJ1136J2219}, but for the quasi-periodic subpulses in the period number 5920 of PSR J1946+1805 observed on 2019-12-04.}
    \label{fig:QPSJ1946}
    \end{figure}

    \begin{figure}[tb!]
    \centering
    \includegraphics[width=0.9\columnwidth]{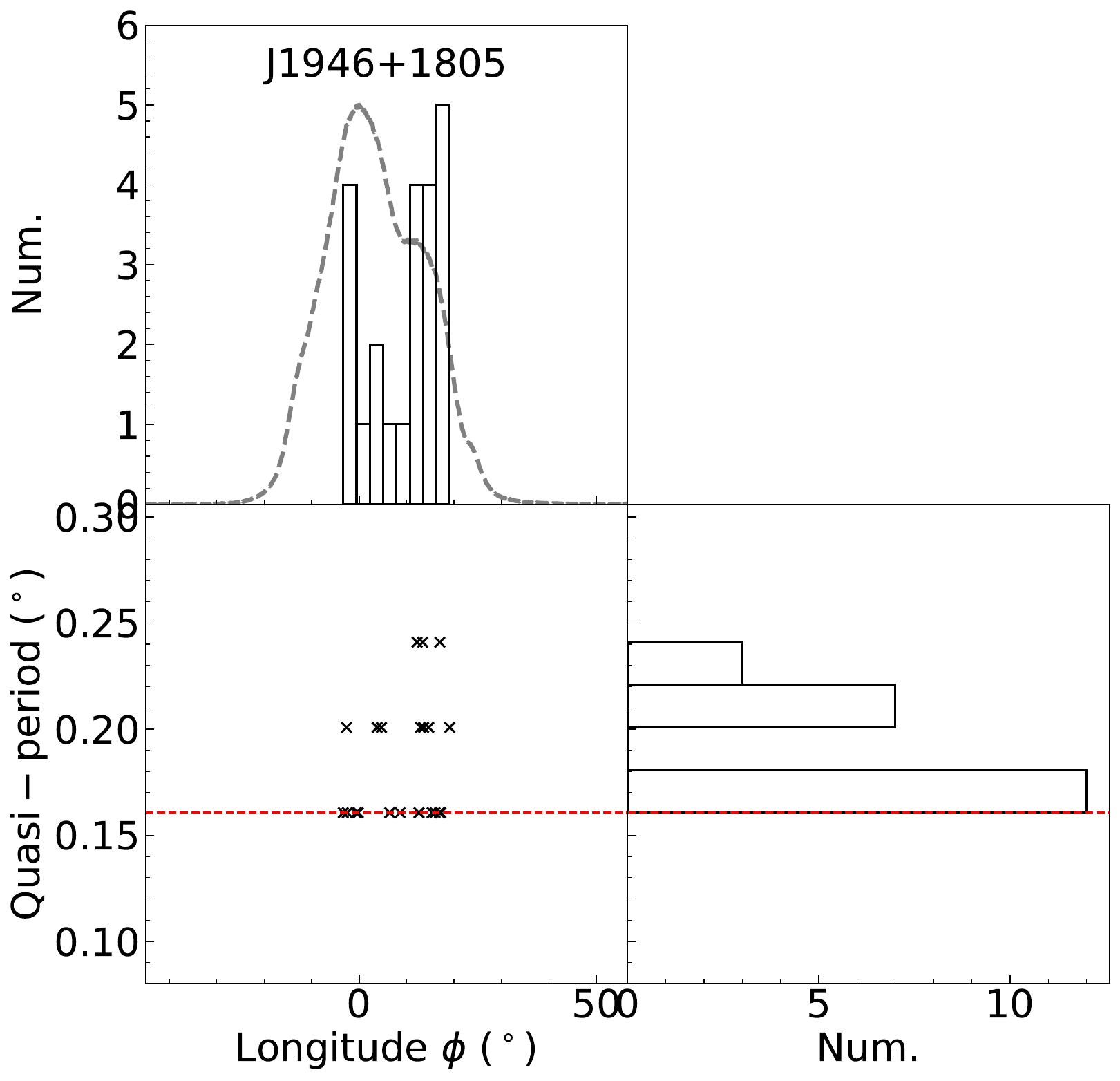}
    \caption{Similar to figure~\ref{fig:QPSJ1136distr}, but for the 22 sets of quasi-periodic subpulses of PSR J1946+1805 observed on 2019-12-04.}
    \label{fig:QPSJ1946distr}
    \end{figure}

    \subsubsection{J1935+1616}
    \label{subsec:J1935+1616}

    PSR J1935+1616 was first discovered in a single-pulse search with the Jodrell Bank Mark I telescope \citep{dl+1970}. 
    \citet{pbc+2002b} reported micropulses with a width of $150\pm10~\mu{\rm s}$. Sensitive observations at 1.5 and 4.5~GHz revealed quasi-periodic microstructures with quasi-periods of $400\pm200$~ms and $350\pm160$~ms, respectively \citep{mra+2016}. 
    From FAST data, \citet{tzc+2025} got the 
    typical width of microstructures and quasi-periods 
    to be $128^{+71}_{-46}~\mu{\rm s}$ and $232\pm9.9~\mu{\rm s}$, respectively.
    
    We analyzed the FAST data obtained on 2019-09-19. The observation lasted for 52.8 minutes, covering 8829 periods. We folded data with the period  of ${\rm P_0}=0.3587~{\rm s}$ with 7296 phase bins per period (Table~\ref{tab:PSRlist}).
    No spike subpulse is identified in these data. We detected 151 sets of quasi-periodic subpulses. One example from the period No.~690 is shown in Figure~\ref{fig:QPSJ1935}, which has a phase quasi-period of $0.247^\circ$, corresponding to 0.246 ms. The parameters of these 151 sets of quasi-periodic subpulses are listed in Table~\ref{table:QPSallPSR}. As shown in Figure~\ref{fig:QPSJ1935distr}, the mean pulse profile of this pulsar consists of multiple emission components, and the quasi-periodic subpulse trains preferentially appear in the central component. The fitted distribution gives a typical phase quasi-period of $0.342^{+0.076}_{-0.062}$ degree, corresponding to $0.341^{+0.076}_{-0.062}$ ms.

    \begin{figure*}[tbh]
    \centering
    \includegraphics[width=0.600\textwidth]{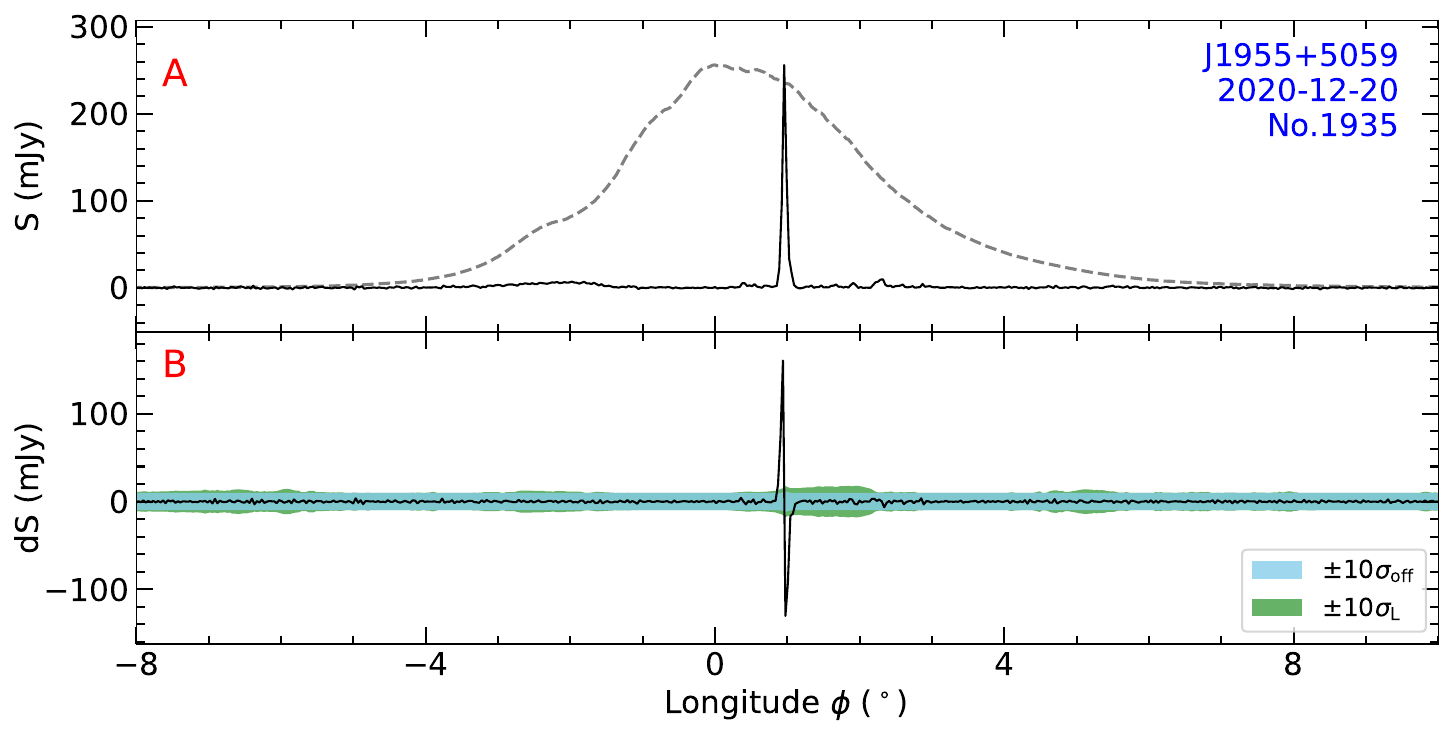}
    \includegraphics[width=0.303\textwidth]{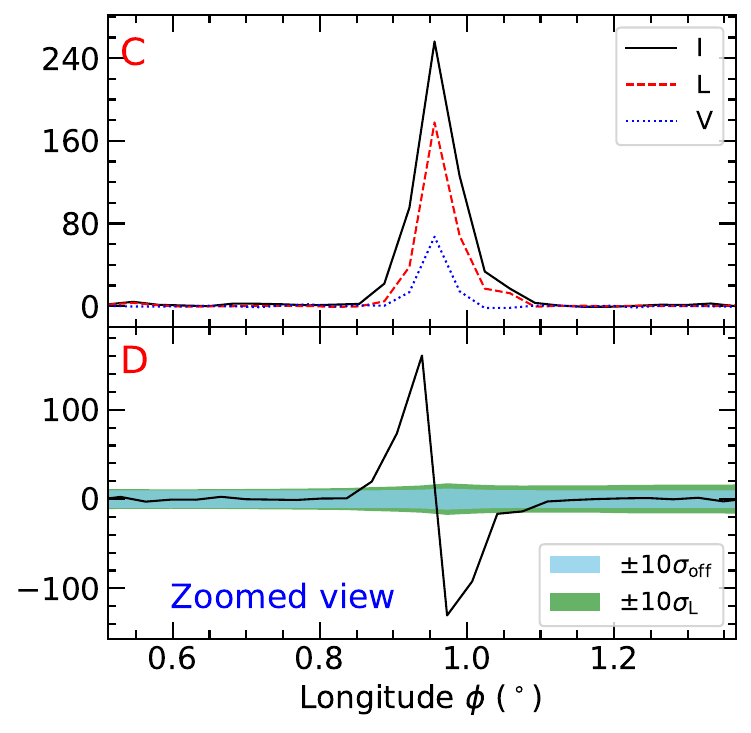}
    \caption{Similar to figure~\ref{fig:J1136N1749}, but for a spike subpulse appearing in the period number 1935 of PSR J1955+5059 observed on 2020-12-20.}
    \label{fig:J1955N1935}
    \end{figure*}
    

    \begin{figure}[tbh]
    \centering
    \includegraphics[width=0.9\columnwidth]{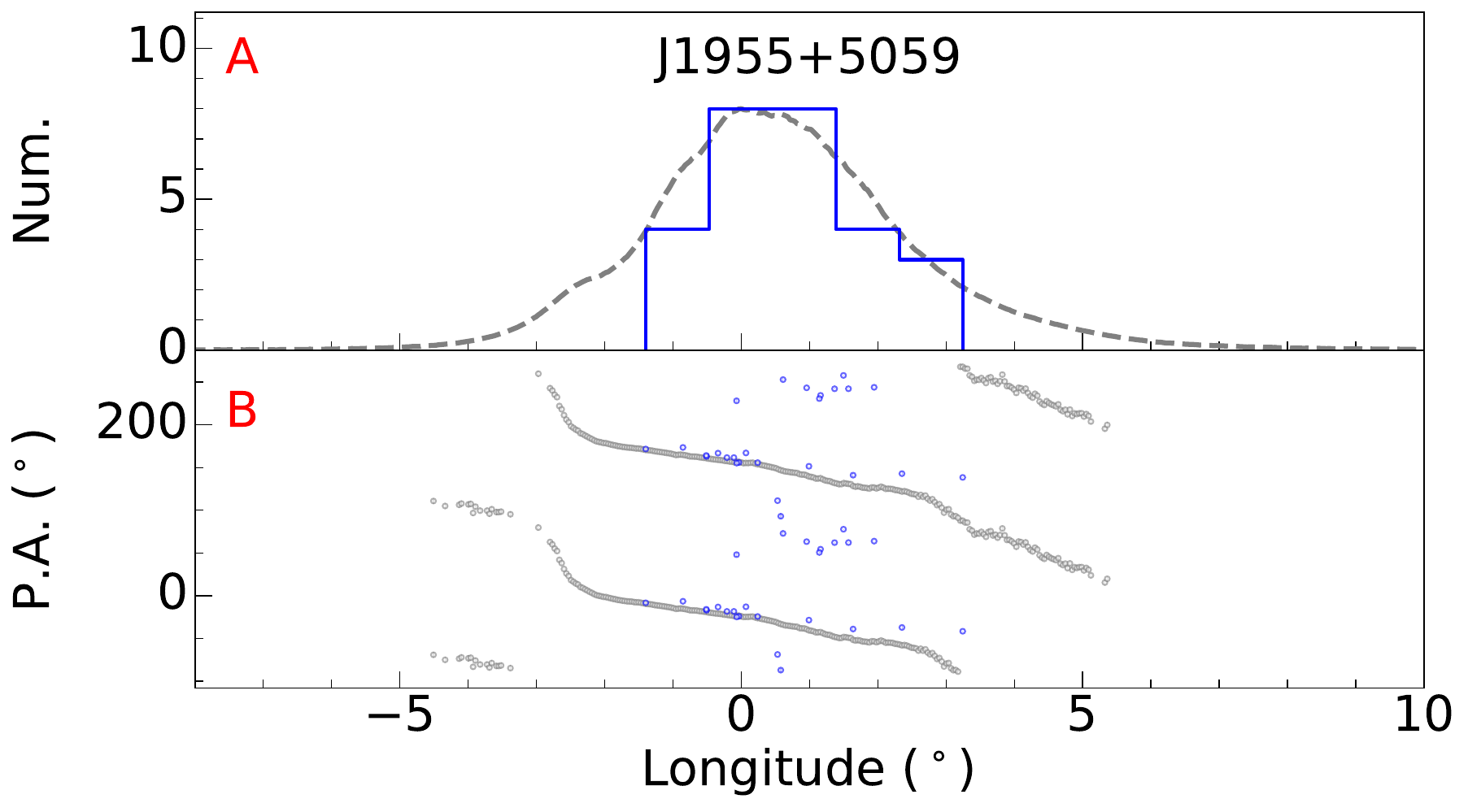}
    \includegraphics[width=0.9\columnwidth]{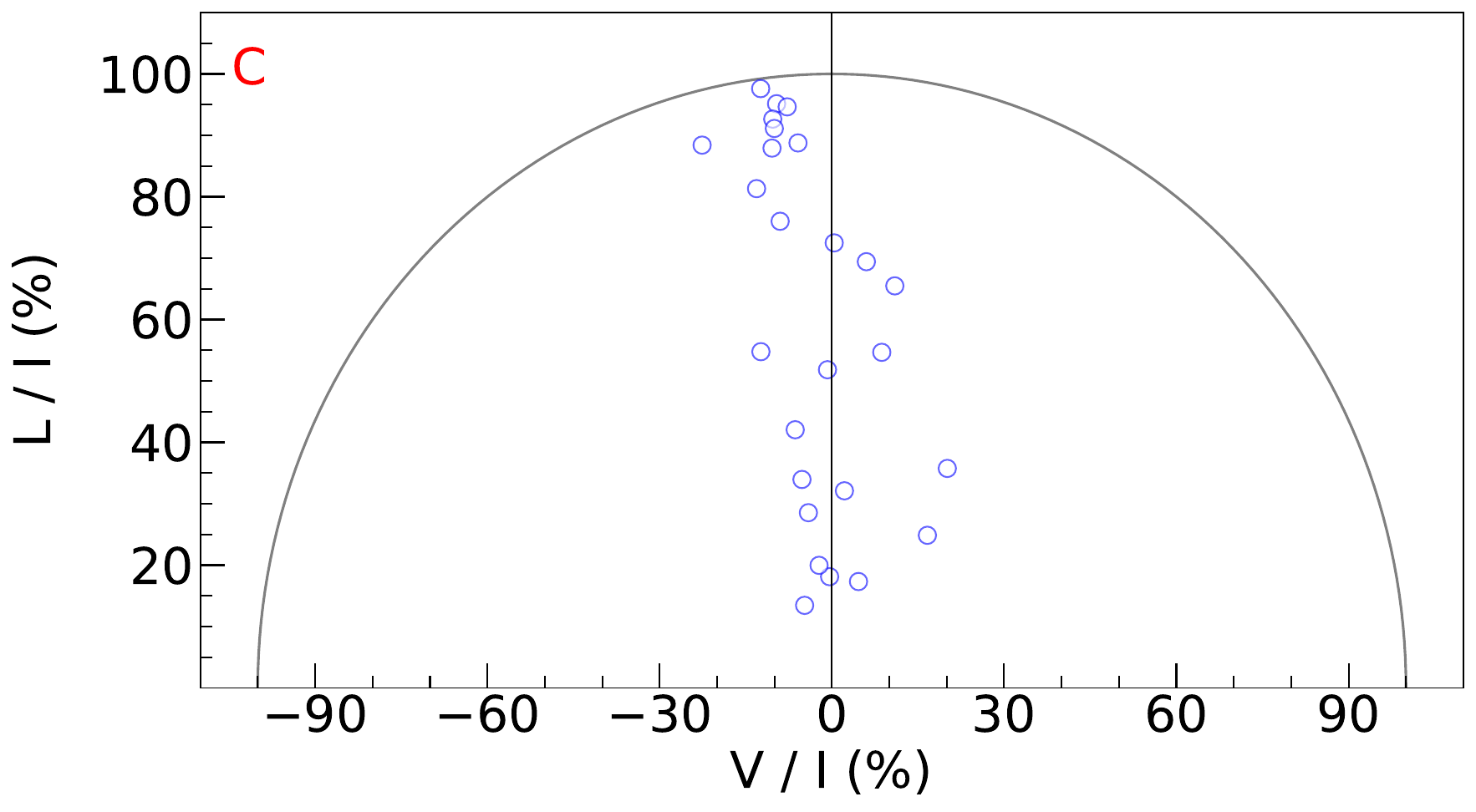}
    \caption{Similar to figure~\ref{fig:J1136spikePoln}, but for the longitude distribution and polarization properties of 27 isolated spike subpulses of PSR J1955+5059.}
    \label{fig:J1955spikePoln}
    \end{figure}
    
    \begin{figure}[tbh]
    \centering
    \includegraphics[width=0.98\columnwidth]{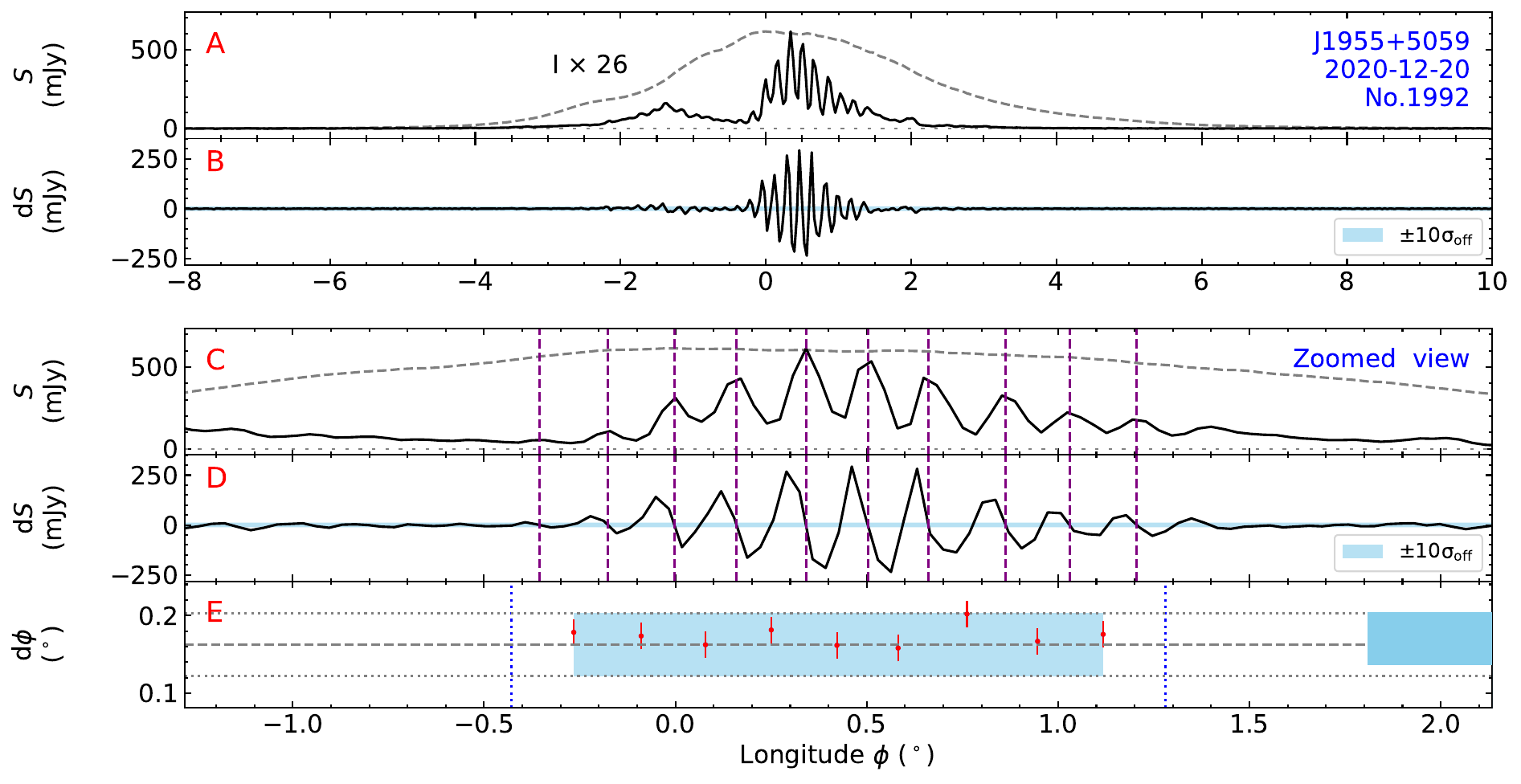}
    \caption{Similar to Figure~\ref{fig:QPSJ1136J2219}, but for the quasi-periodic subpulses in the period number 1992 of PSR J1955+5059 observed on 2020-12-20.}
    \label{fig:QPSJ1955}
    \end{figure}

    \subsubsection{J1946+1805}
    \label{subsec:J1946+1805}
    
    PSR J1946+1805 was discovered with the Molonglo radio telescope \citep{vl+1970}. It exhibits diverse emission phenomena, including nulling, subpulse drifting, and mode changing \citep{dchr+1986}, and four distinct emission modes \citep{kr+2010}. During its nulling state, FAST revealed 234 dwarf pulses \citep{yhz24}. Microstructures in this pulsar were first reported at 430 MHz with an 8~$\mu{\rm s}$ time resolution using the Arecibo telescope \citep{cwh+1990}, with typical widths of about $800~\mu{\rm s}$. More recently, Arecibo observations at 1.2~GHz measured a quasi-period of $710\pm180~\mu{\rm s}$. 
    \citet{kld+2024} estimated a typical microstructure width of $240^{+240}_{-120}~\mu{\rm s}$ and a quasi-period of $750^{+75}_{-68}~\mu{\rm s}$. Using FAST data with a time resolution of about $115~\mu{\rm s}$, \citet{LDW+2025} detected numerous quasi-periodic microstructures in the main pulse, with a typical width of $\tau_\mu=430~\mu{\rm s}$ and a quasi-period of $P_\mu=750^{+160}_{-160}~\mu{\rm s}$, but not from the interpulses. 
    
    We analyzed the FAST data obtained on 2019-12-04. The observation lasted for  60.0 minutes, covering 8172 periods. We folded data with the period  of ${\rm P_0}=0.4406~{\rm s}$ with 8960 phase bins per period (Table~\ref{tab:PSRlist}),  
    and identified two spike subpulses. One example, from period No.~1952, is shown in Figure~\ref{fig:J1946N1952}. The measured parameters of two spike subpulses are listed in Table~\ref{table:SpikeAllPSR}. We also detected 22 sets of quasi-periodic subpulses. One example from the period No.~5920 is shown in Figure~\ref{fig:QPSJ1946}. It has a phase quasi-period of $0.201^\circ$, corresponding to 0.246 ms. The measurement parameters of 22 sets of quasi-periodic subpulses are listed in Table~\ref{table:QPSallPSR}. As shown in Figure~\ref{fig:QPSJ1946distr}, the mean pulse profile of this pulsar contains multiple emission components, and the quasi-periodic subpulse trains tend to appear in the trailing components. Most of them have narrow quasi-periods near the detection limit, suggesting finer emission structures probably detectable at higher time resolutions. The mean phase quasi-period of these quasi-periodic subpulse trains is $0.184\pm0.029^\circ$, corresponding to $0.225\pm0.035$ ms.

    \begin{figure*}[tb!]
    \centering
    \includegraphics[width=0.600\textwidth]{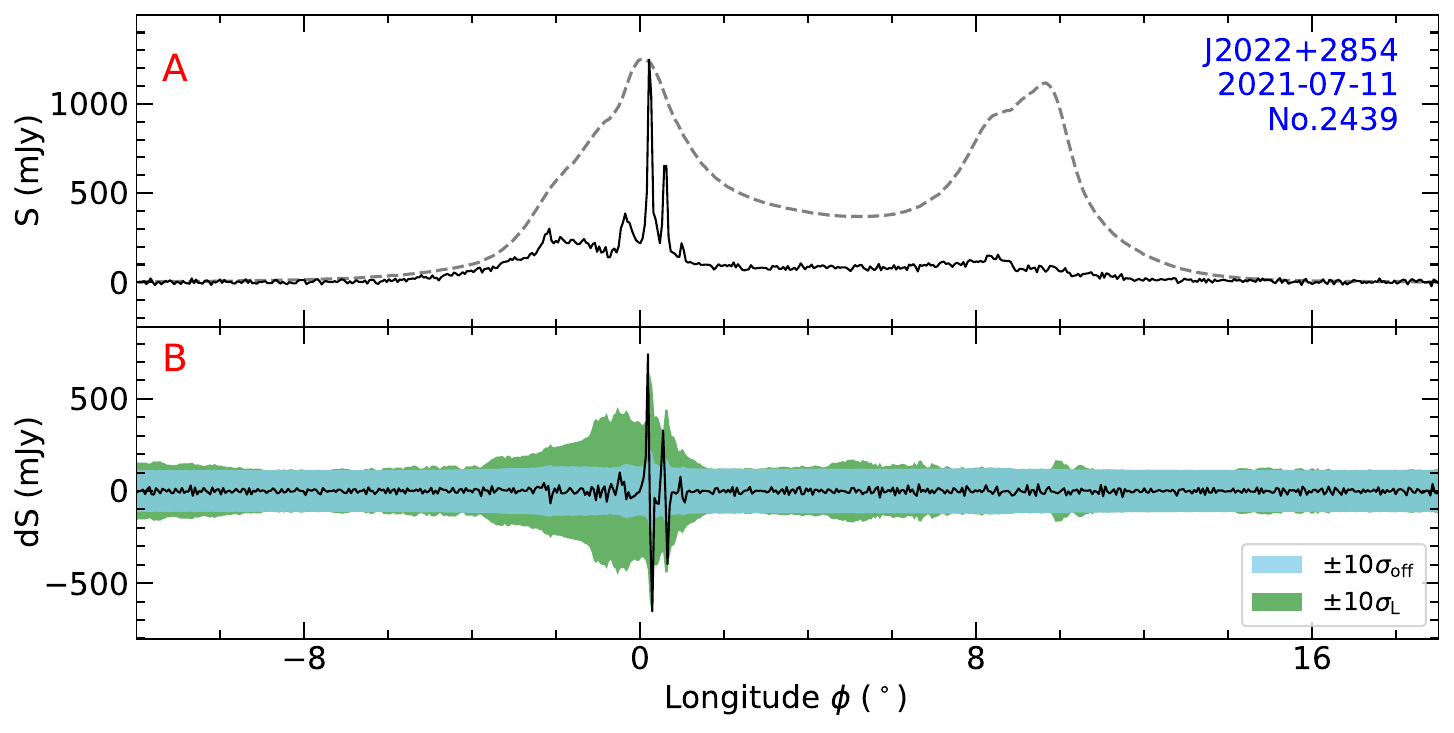}
    \includegraphics[width=0.303\textwidth]{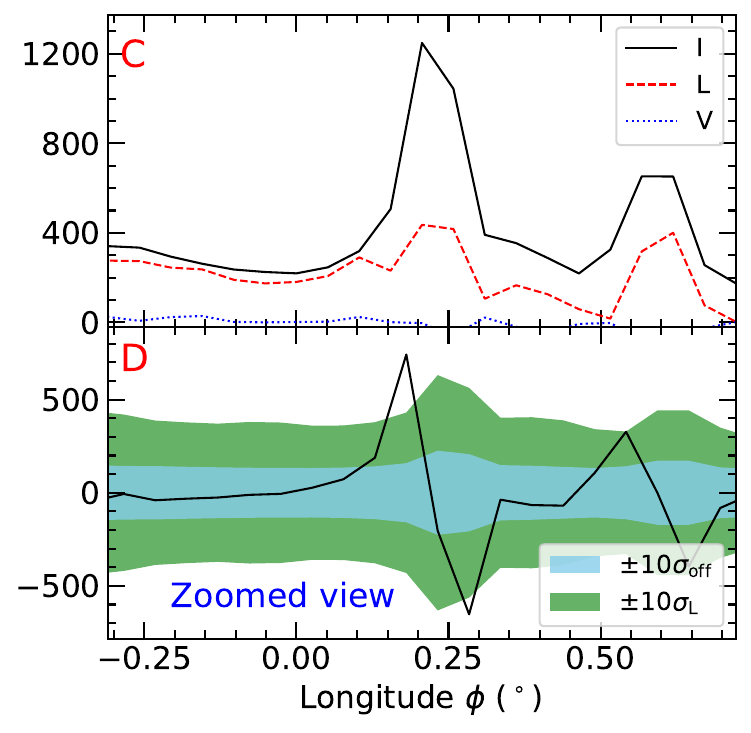}
    \caption{Similar to figure~\ref{fig:J1136N1749}, but for a spike subpulse appearing in the period number 2439 of PSR J2022+2854 observed on 2021-07-11.}
    \label{fig:J2022N2439}
    %
    \centering
    \includegraphics[width=0.600\textwidth]{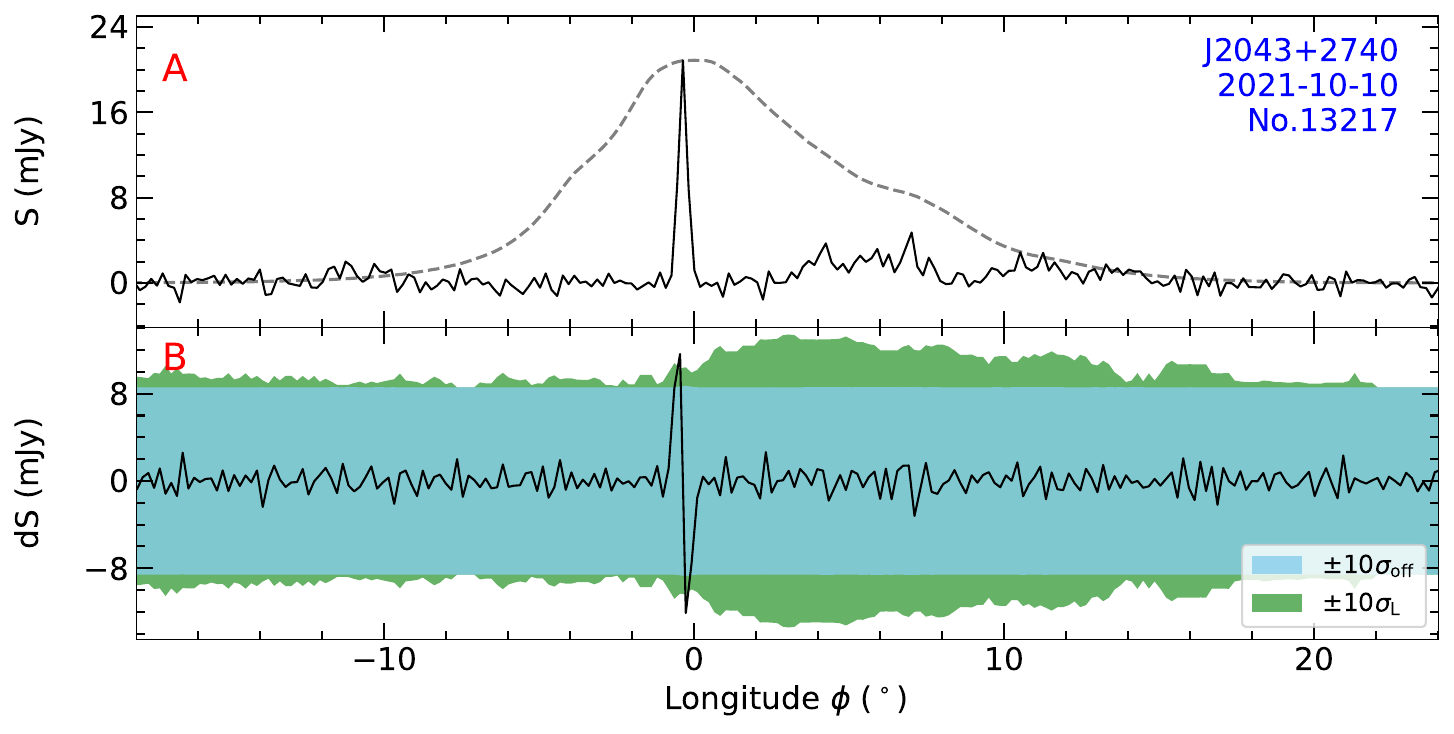}
    \includegraphics[width=0.303\textwidth]{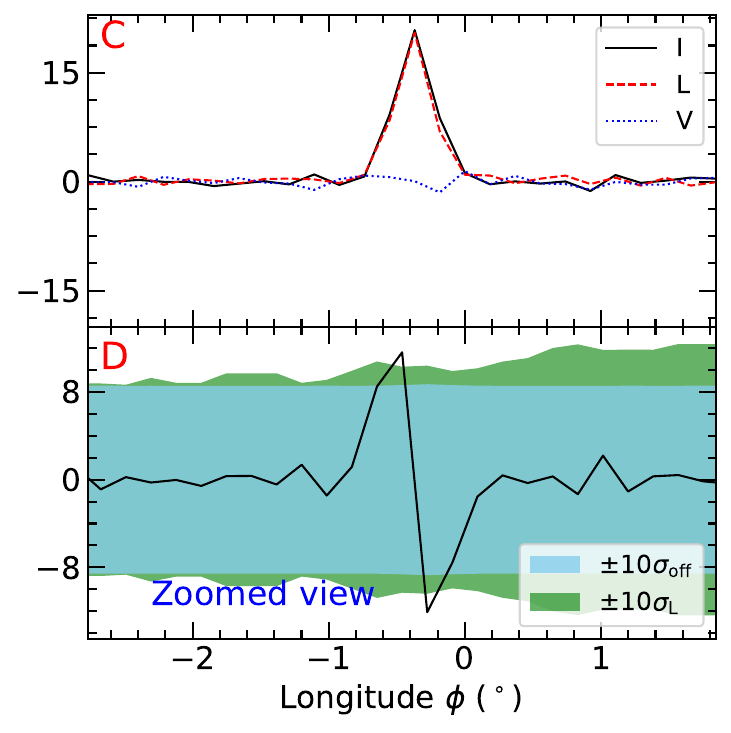}
    \caption{Similar to figure~\ref{fig:J1136N1749}, but for a spike subpulse appearing in the period number 13217 of PSR J2043+2740 observed on 2021-10-10.}
    \label{fig:J2043N13217}
    %
    %
    \centering
    \includegraphics[width=0.600\textwidth]{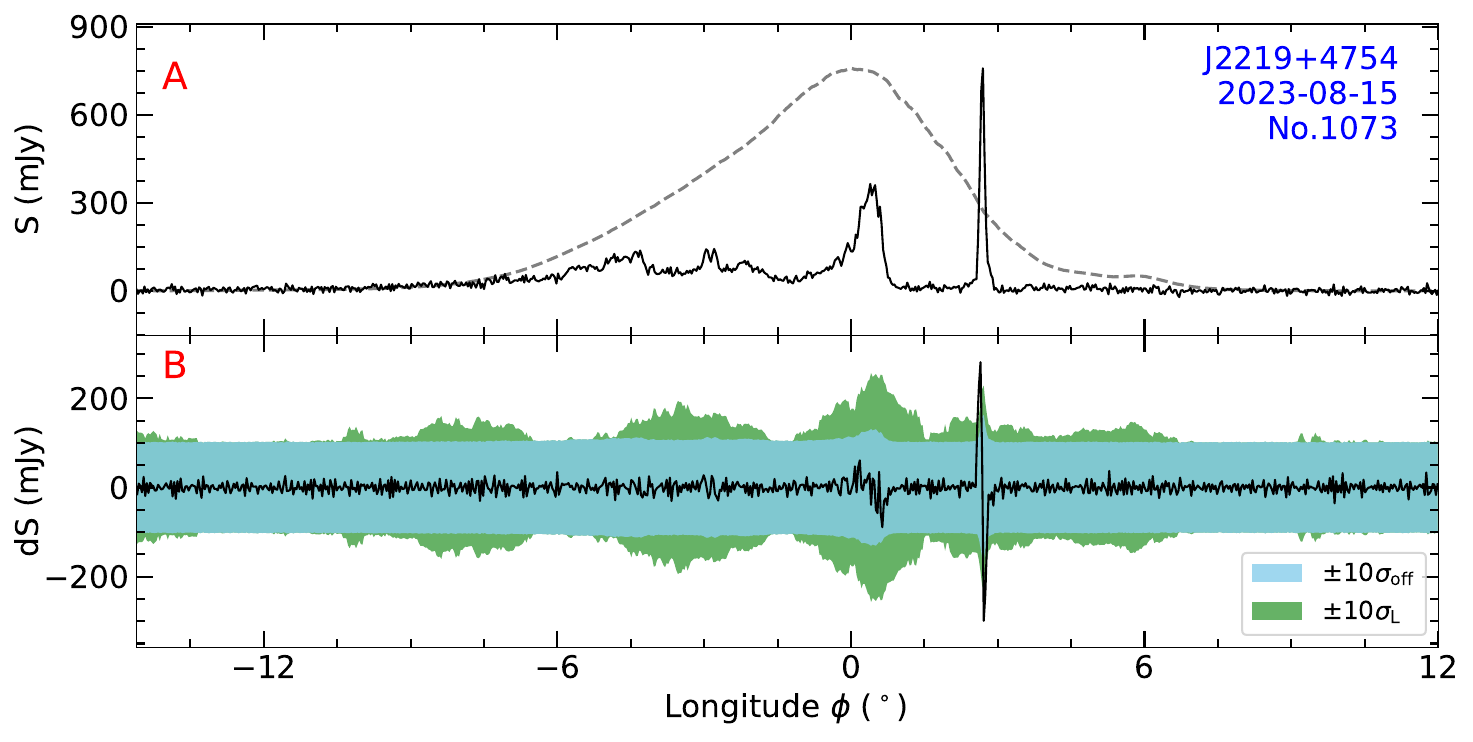}
    \includegraphics[width=0.303\textwidth]{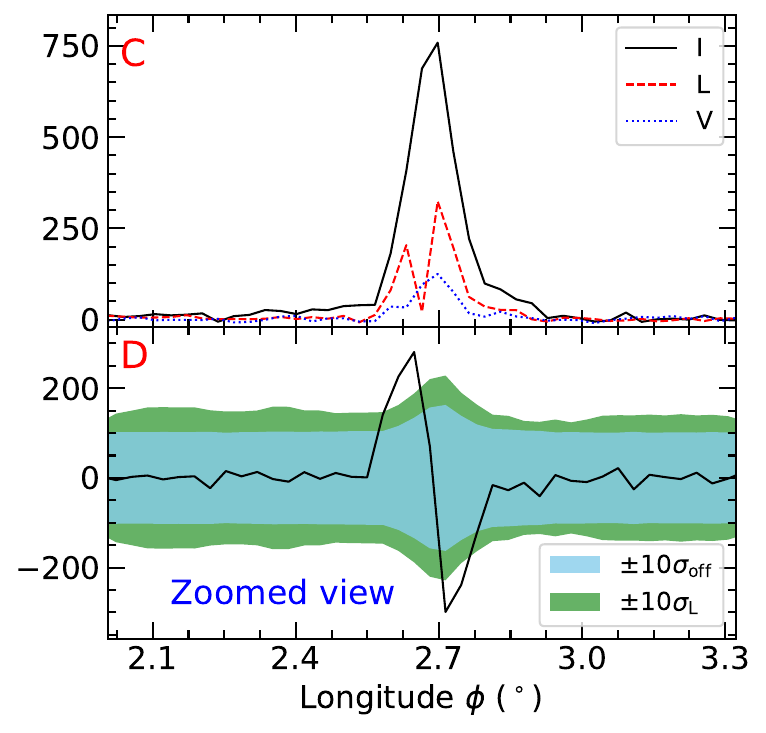}
    \caption{Similar to figure~\ref{fig:J1136N1749}, but for a spike subpulse appearing in the period number 1073 of PSR J2219+4754 observed on 2023-08-15.}
    \label{fig:J2219N1073}
    \end{figure*}

    
    \subsubsection{J1955+5059}
    \label{subsec:J1955+5059}

    PSR J1955+5059 was discovered by 
    \citet{dth+1978}. 
    We analyzed the FAST data obtained on 2020-12-20. The observation lasted for 58.0 minutes, covering 6703 periods. We folded data with the period  of ${\rm P_0}=0.5189~{\rm s}$ with 10544 phase bins per period (Table~\ref{tab:PSRlist}), and identified  67 spike subpulses. One example, from period No.~1935, is shown in Figure~\ref{fig:J1955N1935}. The measured parameters of 67 spike subpulses are listed in Table~\ref{table:SpikeAllPSR}. Their peak flux densities follow a log-normal distribution with $ {\rm \mu} (\log S_{\rm peak}) =  1.75$ and $\sigma({\log S_{\rm peak})=0.29}$, corresponding to a flux density of $56^{+53}_{-27}$ mJy. 
    Some of these isolated spike subpulses are nearly completely linearly polarized, and the overall distribution of polarized emission of these spike subpulses suggests a depolarization trend that may arise from the overlap of coherent radiation cells. We also detect two sets of quasi-periodic subpulses. One example from the period  No.~1992 is shown in Figure~\ref{fig:QPSJ1955}, with a phase quasi-period of $0.171^\circ$, corresponding to 0.247 ms. The parameters of these two sets of  quasi-periodic subpulses are listed in Table~\ref{table:QPSallPSR}. 

    \subsubsection{J2022+2854}
    \label{subsec:J2022+2854}
 
    PSR J2022+2854 was discovered with the Bologna Cross telescope at 408 MHz \citep{fss+1973}, and a mode-switching phenomenon was later reported by \citet{wwy+2016}. Its microstructures were detected at 102.5 MHz with a time resolution of 10~$\mu$s \citep{kkn+1978}. \citet{cordes+1979} reported a typical micropulse width of about 100~$\mu$s. Later, more sensitive Arecibo observations revealed quasi-periodic microstructures in the two dominant components of the mean pulse profile at 1.2~GHz \citep{mar+2015}, with typical quasi-periods of $600\pm270~\mu$s for the leading component and $480\pm180~\mu$s for the trailing component. 
    \citet{kld+2024} estimated a typical microstructure width of $110^{+110}_{-55}~\mu$s and a quasi-period of $500^{+50}_{-45}~\mu$s.
    
    We analyzed the FAST data obtained on 2021-07-11. The observation lasted for 30.0 minutes, covering 5238 periods. We folded data with the period  of ${\rm P_0}=0.343~{\rm s}$ with 6976 phase bins per period (Table~\ref{tab:PSRlist}), and identified 7 spike subpulses. One example, from period No.~2439, is shown in Figure~\ref{fig:J2022N2439}. The measured parameters of 7 spike subpulses are listed in Table~\ref{table:SpikeAllPSR}. 
    No quasi-periodic subpulses satisfying our conditions above were found from the FAST data.

    \subsubsection{J2043+2740}
    \label{subsec:J2043+2740}
    
    PSR J2043+2740 was discovered during the Arecibo millisecond pulsar survey \citep{rtj+1996}. 
    We analyzed the FAST data obtained on 2021-10-10. The observation lasted for  38.0 minutes, covering 23702 periods. We folded data with the period  of ${\rm P_0}=0.0961~{\rm s}$ with 1952 phase bins per period (Table~\ref{tab:PSRlist}), and identified  10 spike subpulses. One example, from period No.~13217, is shown in Figure~\ref{fig:J2043N13217}. The measured parameters of 10 spike subpulses are listed in Table~\ref{table:SpikeAllPSR}. 

    \subsubsection{J2219+4754}
    \label{subsec:J2219+4754}
        
    PSR J2219+4754 was first discovered by \citet{th+1969}. 
    We analyzed the FAST data obtained on 2023-08-15. The observation lasted for 49.0 minutes, covering 5465 periods. We folded data with the period  of ${\rm P_0}=0.5385~{\rm s}$ with 10944 phase bins per period (Table~\ref{tab:PSRlist}), and identified 49 spike subpulses. One example, from period No.~1073, is shown in Figure~\ref{fig:J2219N1073}. The measured parameters of 49 spike subpulses are listed in Table~\ref{table:SpikeAllPSR}. Their peak flux densities follow a log-normal distribution with $ {\rm \mu} (\log S_{\rm peak}) =  1.75$ and $\sigma({\log S_{\rm peak})=0.29}$, corresponding to a flux density of $56^{+53}_{-27}$ mJy. 
    The detected spike subpulses occur within the single emission component of the mean pulse profile. We also detect one train of quasi-periodic subpulses in period No.~1920 (see figure~\ref{fig:QPSJ1136J2219}). Its parameters are listed in Table~\ref{table:QPSJ1136J2219}. The phase quasi-period is $0.164^\circ$, corresponding to 0.245 ms.

\section{Discussion and conclusions}
\label{discussion}

%
We identified spike subpulses and quasi-periodic subpulses based on the strict procedures to the differential profiles. 
Relying on the high sensitivity of FAST, we identified spike subpulses and quasi-periodic subpulses of 25 pulsars. We understand that the analyses in this paper are limited by the time resolution of FAST observations; therefore, they do not cover a wide range of pulsar periods and quasi-periodicity in subpulses.  

\begin{figure}
    \centering
    \includegraphics[width=0.85\columnwidth]{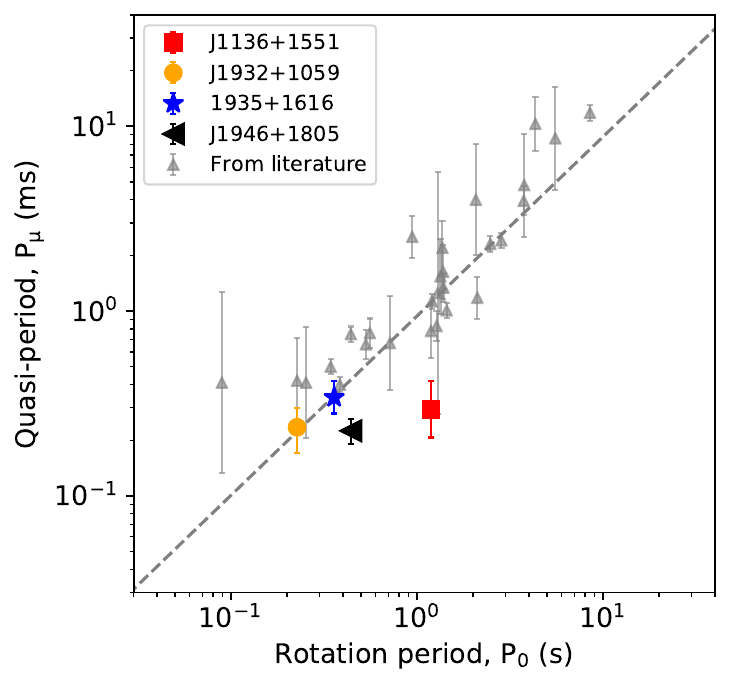}
    \caption{Quasi-period ${\rm P_\mu}$ versus rotation period ${\rm P_0}$ of pulsars. The colored symbols denote the four pulsars in this work, for the average of more than 20 detected quasi-periodic subpulses, namely PSRs J1136+1551, J1932+1059, J1935+1616, and J1946+1805. The gray triangles are measurements collected from the literature by \citet{kld+2024}.
    }
    \label{fig:Pu_P0}
\end{figure}

Using the same standard, we measured the sharpness $R$, peak flux density $S_{\rm peak}$, subpulse width $\Delta \phi$ of spike subpulses, and their linear and circular polarization percentages, as presented in Section~\ref{result} and \ref{otherPSRs}. Most isolated spike subpulses are too narrow to be resolved, with a width $\Delta \phi$ of only one or a few phase bins, each bin corresponding to about 49 $\mu$s.
Their polarization properties show that these narrow structures are often strongly linearly polarized, suggesting that they represent highly localized emission elements embedded in broader single-pulse structures. 
%
%
Because these identified spike subpulses are extremely narrow and remain unresolved or only marginally resolved with the FAST sampling time of $49.152~\mu{\rm s}$, it is hard to do further analyses. 

We showed that quasi-periodic subpulses can be robustly identified similarly as 
narrow spike-like structures. From currently available FAST data, none or few sets of quasi-periodic subpulses were found from most pulsars in Table~\ref{tab:PSRlist}. Nevertheless, 
more than 20 sets of quasi-periodic subpulses have been identified from four pulsars, PSRs J1136+1551, J1932+1059, J1935+1616, and J1946+1805. 

With a similarly strict procedure, we got the quasi-periodicity for these 
repetitive subpulses.
Previously, pulse oscillatory features have been studied through the autocorrelation function together with visual inspection 
\citep{kjs+2002, mar+2015, dgs+2016, lab+2022, kld+2024}.
As shown in Figure~\ref{fig:Pu_P0} for comparison between the characteristic quasi-period ${\rm P_\mu}$ and the rotation period ${\rm P_0}$, their measured quasi-periods are concentrated in a relatively narrow range of $\sim 0.2$--$0.35$ ms. Two of them are very consistent with the correlation between the quasi-periods and pulsar spin periods \citep{kld+2024}, but the other two are lower than the expected values, especially PSR J1136+1551. 

Several classes of models have been proposed for the origin of quasi-periodic microstructures, including geometric effects, temporal modulation, propagation effects in the magnetospheric plasma, and intrinsic plasma instabilities in the emission region \citep{Benford+1977, van+1980, Gil2003, ht+1981, cjd+2004, bmm+2016, pts+2020, t+2022b, bbl+2023, Jones+2024, kld+2024}. In this context, our results are better interpreted as empirical constraints on these theoretical pictures. The quasi-periodic structures identified here are built on narrow spike-like subpulses, which suggests that quasi-periodicity can be associated with narrow and localized emission structures. 
A notable example is PSR J0139+3336, in which quasi-periodic narrow structures can reappear after a long nulling interval, as shown in Section~\ref{sec:J0139+3336} and Figure~\ref{fig:J0139exam}. For models that rely on persistent plasma structures along the propagation path, or on a relatively stable external modulation environment, this behavior provides an observational constraint that deserves careful consideration \citep{ht+1981, cjd+2004}. More generally, the present results are more consistent with localized plasma dynamics in the emission region, 
%
rather than with a purely geometric or propagation-induced modulation. 

\section*{acknowledgments}
    
\add{We thank the referee for his careful reading and thoughtful comments.} 
The authors were supported by National Key R\&D Program of China No. 2025YFA161400, the National Natural Science Foundation of China (NSFC), grant numbers 12588202 and 12041303, the Chinese Academy of Sciences via project JZHKYPT-2021-06, C. Wang and P.~F. Wang are also supported by NSFC No. 12133004.
This work made use of the data from FAST (https://cstr.cn/31116.02.FAST).  FAST is a Chinese national mega-science facility, operated by National Astronomical Observatories, Chinese Academy of Sciences. 
%
%
%
%

%

\bibliography{citation}{}
\bibliographystyle{aasjournal}

%








\clearpage




 \renewcommand{\thetable}{\arabic{table}}
 \setcounter{table}{1} 


\begin{table*}[tb]
  \caption{Parameters of spike subpulses.}
  \label{table:SpikeAllPSR}
  \centering
  \footnotesize
  \renewcommand{\arraystretch}{0.8}
\begin{tabular}{crrcccrrr}
  \hline
  No. of & Longitude $\phi$ & $S_{\rm peak}$ & $\sigma_{\rm Loc}$ & $\Delta \phi$ & R & L/I & V/I & PA\\
  Period & ($^\circ$) & (mJy) & (mJy) & (bin) & & (\%) & (\%) & ($^\circ$)\\
  \hline
    \multicolumn{9}{c}{1 spike subpulse of PSR J0139+3336 observed on 2020-01-17} \\
  \hline
      9492      &     -0.84&           46.8&      1.20&         2&      24.2&      28.8(0.7)&      -8.2(0.9)&      49.1(1.0)\\
  \hline
    \multicolumn{9}{c}{17 spike subpulses of PSR J0426+4933 observed on 2021-11-17} \\
  \hline
      142       &     -0.05&           33.2&      0.80&         2&      34.4&      62.5(1.0)&      42.9(1.4)&      56.1(0.4)\\
      196       &      0.13&           87.8&      1.20&         2&      33.0&      58.1(0.5)&      46.4(0.7)&      64.0(0.2)\\
      262       &     -0.66&           46.2&      1.10&         2&      31.0&      73.8(0.9)&      36.5(1.0)&      36.1(0.3)\\
\hline
  \end{tabular}
\end{table*}

\addtocounter{table}{-1} \begin{table*}[tb]
  \caption{--- Continued.}
  \label{table:SpikeAll2}
  \centering
  \footnotesize
  \renewcommand{\arraystretch}{0.8}
  \begin{tabular}{crrcccrrr}
  \hline
  No. of & Longitude $\phi$ & $S_{\rm peak}$ & $\sigma_{\rm Loc}$ & $\Delta \phi$ & R & L/I & V/I & PA\\
  Period & ($^\circ$) & (mJy) & (mJy) & (bin) & & (\%) & (\%) & ($^\circ$)\\
  \hline
      11109     &     -3.38&           51.0&      1.20&         2&      21.7&       2.9(0.5)&       0.6(0.8)&              -\\
      11195     &      1.49&          238.9&      3.20&         2&      41.1&      23.7(0.2)&       1.0(0.2)&     -30.3(0.3)\\
      11205     &     -3.01&          165.9&      1.20&         2&      77.2&      19.9(0.2)&      -2.4(0.3)&     -45.1(0.5)\\
      11307     &      0.53&           36.8&      1.60&         1&      23.8&      96.7(2.6)&       8.1(1.1)&      62.2(0.3)\\
      11403     &     -3.80&           40.4&      0.80&         2&      26.0&       5.6(0.8)&      -0.3(0.9)&              -\\
      11502     &     -2.27&           70.5&      1.50&         2&      20.9&      91.4(1.1)&      -2.7(0.6)&      54.7(0.2)\\
\hline
  \end{tabular}
\end{table*}

\addtocounter{table}{-1} \begin{table*}[tb]
  \caption{--- Continued.}
  \label{table:SpikeAll3}
  \centering
  \footnotesize
  \renewcommand{\arraystretch}{0.8}
  \begin{tabular}{crrcccrrr}
  \hline
  No. of & Longitude $\phi$ & $S_{\rm peak}$ & $\sigma_{\rm Loc}$ & $\Delta \phi$ & R & L/I & V/I & PA\\
  Period & ($^\circ$) & (mJy) & (mJy) & (bin) & & (\%) & (\%) & ($^\circ$)\\
  \hline
    \multicolumn{9}{c}{53 spike subpulses of PSR J0826+2637 observed on 2019-11-26} \\
  \hline
      281       &     -2.57&           26.0&      1.10&         1&      42.5&      20.3(1.3)&      -1.7(1.6)&     -86.6(2.1)\\
      477       &     -0.07&          186.3&      11.2&         1&      21.7&      44.3(0.3)&      10.4(0.3)&      47.5(0.1)\\
      570       &      0.03&           95.8&      2.80&         1&      39.2&      95.8(1.1)&      -3.5(0.5)&      71.4(0.1)\\
      572       &      0.65&          124.1&      4.00&         2&      28.5&      58.2(0.2)&      23.9(0.3)&      12.9(0.1)\\
      712       &      1.27&           16.4&      1.00&         1&      21.3&      61.6(2.7)&      -9.5(2.4)&     -72.2(1.0)\\
\hline
  \end{tabular}
\end{table*}

\addtocounter{table}{-1} \begin{table*}[tb]
  \caption{--- Continued.}
  \label{table:SpikeAll4}
  \centering
  \footnotesize
  \renewcommand{\arraystretch}{0.8}
  \begin{tabular}{crrcccrrr}
  \hline
  No. of & Longitude $\phi$ & $S_{\rm peak}$ & $\sigma_{\rm Loc}$ & $\Delta \phi$ & R & L/I & V/I & PA\\
  Period & ($^\circ$) & (mJy) & (mJy) & (bin) & & (\%) & (\%) & ($^\circ$)\\
  \hline
      305       &     -1.76&           12.6&      1.00&         1&      25.9&      35.5(3.1)&       3.2(4.4)&     -15.6(3.0)\\
      322       &      7.37&           26.1&      1.60&         1&      31.5&      96.8(2.2)&      11.6(1.3)&      66.8(0.3)\\
      323       &      6.97&           24.7&      1.30&         1&      35.0&      86.9(2.7)&      14.1(2.3)&      57.0(0.7)\\
      338       &      0.37&          285.7&      7.80&         1&      69.4&      96.6(0.3)&       2.1(0.2)&       1.1(0.1)\\
      368       &      4.22&           32.3&      1.00&         1&      61.4&      90.9(2.2)&      12.2(1.9)&      40.4(0.5)\\
      444       &      4.82&           16.0&      1.00&         1&      31.6&      81.9(3.2)&      13.2(3.0)&      40.1(0.9)\\
\hline
 \end{tabular}
\end{table*}

\addtocounter{table}{-1} \begin{table*}[tb]
  \caption{--- Continued.}
  \label{table:SpikeAll5}
  \centering
  \footnotesize
  \renewcommand{\arraystretch}{0.8}
  \begin{tabular}{crrcccrrr}
  \hline
  No. of & Longitude $\phi$ & $S_{\rm peak}$ & $\sigma_{\rm Loc}$ & $\Delta \phi$ & R & L/I & V/I & PA\\
  Period & ($^\circ$) & (mJy) & (mJy) & (bin) & & (\%) & (\%) & ($^\circ$)\\
  \hline
      2873      &     -0.10&           87.4&      4.50&         1&      25.4&      95.2(1.2)&       0.2(0.4)&      78.2(0.1)\\
      3342      &      6.09&           61.9&      4.10&         1&      30.5&      46.3(0.7)&      41.2(0.9)&     -75.3(0.5)\\
      4138      &      2.13&           22.7&      1.00&         1&      41.6&      67.2(2.6)&       7.5(2.7)&      73.2(1.0)\\
      4389      &      5.92&           36.4&      3.60&         1&      20.5&      30.3(0.9)&      17.6(1.3)&     -86.8(1.1)\\
      4772      &      6.33&           23.6&      1.80&         1&      25.7&      59.3(1.6)&       7.2(1.8)&      34.1(0.8)\\
  \hline
    \multicolumn{9}{c}{29 spike subpulses of PSR J1607-0032 observed on 2021-10-19} \\
\hline
  \end{tabular}
\end{table*}

\addtocounter{table}{-1} \begin{table*}[tb]
  \caption{--- Continued.}
  \label{table:SpikeAll6}
  \centering
  \footnotesize
  \renewcommand{\arraystretch}{0.8}
  \begin{tabular}{crrcccrrr}
  \hline
  No. of & Longitude $\phi$ & $S_{\rm peak}$ & $\sigma_{\rm Loc}$ & $\Delta \phi$ & R & L/I & V/I & PA\\
  Period & ($^\circ$) & (mJy) & (mJy) & (bin) & & (\%) & (\%) & ($^\circ$)\\
  \hline
      6941      &     -5.08&          104.5&      3.20&         2&      20.2&      97.1(0.2)&      10.9(0.1)&      44.8(0.0)\\
      7454      &     -4.85&           54.5&      4.10&         1&      25.5&      94.0(0.3)&       4.7(0.1)&      40.1(0.0)\\
      7716      &     -6.72&           78.4&      3.40&         1&      30.4&      95.5(0.3)&       5.7(0.2)&      48.2(0.0)\\
      7784      &     -6.87&           22.2&      1.00&         1&      29.1&      97.1(1.2)&       5.7(0.5)&      48.2(0.1)\\
      8220      &     -5.00&           66.4&      4.80&         1&      26.2&      96.2(0.3)&       7.9(0.2)&      45.2(0.1)\\
      8600      &     -0.08&          243.8&      12.8&         1&      21.2&      98.7(0.1)&      -0.3(0.0)&      39.9(0.0)\\
\hline
  \end{tabular}
\end{table*}

\addtocounter{table}{-1} \begin{table*}[tb]
  \caption{--- Continued.}
  \label{table:SpikeAll7}
  \centering
  \footnotesize
  \renewcommand{\arraystretch}{0.8}
  \begin{tabular}{crrcccrrr}
  \hline
  No. of & Longitude $\phi$ & $S_{\rm peak}$ & $\sigma_{\rm Loc}$ & $\Delta \phi$ & R & L/I & V/I & PA\\
  Period & ($^\circ$) & (mJy) & (mJy) & (bin) & & (\%) & (\%) & ($^\circ$)\\
  \hline
      2821      &     -0.07&           37.6&      2.10&         1&      23.7&      94.6(3.2)&      -7.8(1.9)&     -24.8(0.6)\\
      2982      &      0.61&           21.7&      1.10&         1&      26.2&      42.1(3.3)&      -6.4(3.6)&      72.9(2.1)\\
      3042      &      3.72&           19.1&      1.20&         1&      28.8&      37.7(2.2)&     -66.3(3.5)&     -40.9(2.0)\\
      3115      &      0.58&           40.6&      1.80&         1&      30.2&      17.3(1.3)&       4.6(1.8)&     -87.0(2.6)\\
      3156      &      0.07&           28.7&      1.50&         1&      26.3&      95.1(4.1)&      -9.6(2.5)&     -12.9(0.7)\\
      3189      &      1.50&          171.4&      3.20&         2&      34.4&      24.9(0.2)&      16.6(0.4)&      77.8(0.5)\\
\hline
  \end{tabular}
\end{table*}

\addtocounter{table}{-1} \begin{table*}[tb]
  \caption{--- Continued.}
  \label{table:SpikeAll8}
  \centering
  \footnotesize
  \renewcommand{\arraystretch}{0.8}
  \begin{tabular}{crrcccrrr}
  \hline
  No. of & Longitude $\phi$ & $S_{\rm peak}$ & $\sigma_{\rm Loc}$ & $\Delta \phi$ & R & L/I & V/I & PA\\
  Period & ($^\circ$) & (mJy) & (mJy) & (bin) & & (\%) & (\%) & ($^\circ$)\\
  \hline
      2774      &      2.80&         1564.2&      25.8&         2&      20.8&      20.8(0.2)&      -5.7(0.3)&     -23.0(0.4)\\
      2787      &      1.46&         2133.4&      19.2&         3&      26.6&      21.9(0.2)&      -1.4(0.2)&     -18.7(0.3)\\
      2941      &      0.72&         1980.8&      18.3&         2&      40.1&      22.0(0.2)&       1.7(0.2)&       2.2(0.3)\\
      3048      &      0.52&         1381.2&      17.5&         2&      42.1&      28.8(0.4)&       0.1(0.3)&     -39.1(0.3)\\
      3051      &     -0.01&         2200.4&      23.2&         2&      36.8&      18.4(0.2)&       5.9(0.2)&     -22.4(0.3)\\
      3161      &      2.19&          399.4&      11.6&         2&      24.4&      98.7(2.0)&       6.4(1.2)&      12.3(0.3)\\
\hline
  \end{tabular}
\end{table*}



\clearpage

\clearpage


\begin{table}[th]
  \caption{Parameters of quasi-periodic subpulse sets.}
  \label{table:QPSallPSR}
  \centering
  \footnotesize
  \renewcommand{\arraystretch}{0.8}
  \begin{tabular}{crcc}
  \hline
  Period & Longitude range & Num. of & Quasi-period \\
  Number & ($^\circ$) & subpulses & ($^\circ$) \\
  \hline
  \multicolumn{4}{c}{1 QPS of PSR J0139+3336 observed on 2020-01-17}\\
  \hline
  5789&     (0.17$^\circ$~,~1.05$^\circ$)&         8&     0.113\\
  \hline
  \multicolumn{4}{c}{1 QPS of PSR J0304+1932 observed on 2021-09-13}\\
  \hline
  779&     (-1.01$^\circ$~,~-0.05$^\circ$)&         8&     0.115\\
  \hline
  \multicolumn{4}{c}{5 QPS of PSR J0826+2637 observed on 2019-11-26}\\
  \hline
  360&     (-1.83$^\circ$~,~0.19$^\circ$)&         9&     0.233\\
\hline
  \end{tabular}
\end{table}

\addtocounter{table}{-1} 
\begin{table}[th]
  \caption{--- Continued.}
  \label{table:QPSall2}
  \centering
  \footnotesize
  \renewcommand{\arraystretch}{0.8}
  \begin{tabular}{crcc}
  \hline
  Period & Longitude range & Num. of & Quasi-period \\
  Number & ($^\circ$) & subpulses & ($^\circ$) \\
  \hline
  6326&     (-2.85$^\circ$~,~0.83$^\circ$)&        12&     0.312\\
  6410&     (-1.87$^\circ$~,~2.52$^\circ$)&        14&     0.312\\
  6605&     (-2.03$^\circ$~,~1.43$^\circ$)&         6&     0.625\\
  7250&     (4.32$^\circ$~,~8.62$^\circ$)&        14&     0.312\\
  7262&     (0.72$^\circ$~,~4.8$^\circ$)&        13&     0.312\\
  8982&     (3.54$^\circ$~,~8.62$^\circ$)&        16&     0.312\\
\hline
  \end{tabular}
\end{table}

\addtocounter{table}{-1} 
\begin{table}[th]
  \caption{--- Continued.}
  \label{table:QPSall3}
  \centering
  \footnotesize
  \renewcommand{\arraystretch}{0.8}
  \begin{tabular}{crcc}
  \hline
  Period & Longitude range & Num. of & Quasi-period \\
  Number & ($^\circ$) & subpulses & ($^\circ$) \\
  \hline
  2421&     (0.4$^\circ$~,~2.77$^\circ$)&         8&     0.296\\
  2446&     (0.18$^\circ$~,~2.6$^\circ$)&        10&     0.247\\
  2469&     (-2.06$^\circ$~,~2.18$^\circ$)&        14&     0.296\\
  2486&     (-1.05$^\circ$~,~2.27$^\circ$)&        11&     0.296\\
  2562&     (-0.25$^\circ$~,~1.8$^\circ$)&         8&     0.247\\
  2563&     (-1.47$^\circ$~,~1.23$^\circ$)&        11&     0.247\\
\hline
  \end{tabular}
\end{table}

\addtocounter{table}{-1} 
\begin{table}[th]
  \caption{ --- Continued.}
  \label{table:QPSall5}
  \centering
  \footnotesize
  \renewcommand{\arraystretch}{0.8}
  \begin{tabular}{crcc}
  \hline
  Period & Longitude range & Num. of & Quasi-period \\
  Number & ($^\circ$) & subpulses & ($^\circ$) \\
  \hline
  6576&     (0.7$^\circ$~,~3.39$^\circ$)&         9&     0.296\\
  6627&     (-2.1$^\circ$~,~2.78$^\circ$)&        16&     0.296\\
  6788&     (-2.11$^\circ$~,~2.51$^\circ$)&        13&     0.345\\
  6883&     (-2.41$^\circ$~,~2.85$^\circ$)&        15&     0.345\\
  6910&     (-3.27$^\circ$~,~2.82$^\circ$)&        18&     0.345\\
  6937&     (-1.1$^\circ$~,~2.57$^\circ$)&        11&     0.345\\
\hline
  \end{tabular}
\end{table}

\end{document}